\documentclass[aps,
prd,
reprint,
nolongbibliography,
preprintnumbers,
nobibnotes,
nofootinbib,
]{revtex4-2}

\usepackage[absolute,overlay]{textpos}

\usepackage{graphicx}
\usepackage{bm}
\usepackage{xspace}
\usepackage[dvipsnames, usenames]{xcolor}
\usepackage{amsmath}
\usepackage{amssymb}
\usepackage{booktabs}
\usepackage{multirow}
\usepackage[normalem]{ulem}
\usepackage{hyperref}

\newcommand{\be}{\begin{equation}}
\newcommand{\ee}{\end{equation}}
\newcommand{\bea}{\begin{eqnarray}}
\newcommand{\eea}{\end{eqnarray}}

\newcommand{\bi}{\begin{itemize}}
\newcommand{\ei}{\end{itemize}}
\newcommand{\bn}{\begin{enumerate}}
\newcommand{\en}{\end{enumerate}}
\newcommand{\La}{\ensuremath{\Lambda}\xspace}
\renewcommand{\th}{\ensuremath{\theta}\xspace}
\newcommand{\fa}{\ensuremath{f_a}\xspace}

\newcommand{\dLa}{\ensuremath{\mathrm{d}\La\;}\xspace}

\newcommand{\dt}{\ensuremath{\mathrm{d}\th\;}\xspace}
\newcommand{\dDup}{\ensuremath{\mathrm{d}\Dup\;}\xspace}

\newcommand{\rhoup}{\ensuremath{\rho_{\uparrow}}\xspace}
\newcommand{\Dup}{\ensuremath{D_{\uparrow}}\xspace}
\newcommand{\msun}{\ensuremath{M_{\odot}}\xspace}
\newcommand{\ct}{\ensuremath{\cos\tau}\xspace}
\newcommand{\cto}{\ensuremath{\cos\tau_1} \xspace}
\newcommand{\ctt}{\ensuremath{\cos\tau_2} \xspace}
\newcommand{\sct}{\ensuremath{\sigma_{c\tau}}\xspace}
\newcommand{\mct}{\ensuremath{\mu_{c\tau}}\xspace}
\newcommand{\ab}{\ensuremath{\alpha_{c\tau}} \xspace}
\newcommand{\bb}{\ensuremath{\beta_{c\tau}} \xspace}

\newcommand{\lvk}{\texttt{LVK}\xspace}
\newcommand{\lvkc}{\texttt{LVK corr}\xspace}

\newcommand{\isobeta}{\texttt{Isotropic + Beta}\xspace}
\newcommand{\isogaus}{\texttt{Isotropic + Gaussian}\xspace}
\newcommand{\isotuk}{\texttt{Isotropic + Tukey}\xspace}

\newcommand{\isocbeta}{\texttt{Isotropic + Beta corr}\xspace}
\newcommand{\isocgaus}{\texttt{Isotropic + Gaussian corr}\xspace}
\newcommand{\isoctuk}{\texttt{Isotropic + Tukey corr}\xspace}

\newcommand{\fatone}{\ensuremath{f_{q=1}}\xspace}
\newcommand{\enn}{\ensuremath{n}\xspace}
\newcommand{\faq}{\ensuremath{f(q)}\xspace}
\newcommand{\truthcolor}{yellow\xspace}

\graphicspath{{Plots/}}

\newcommand{\comment}[1]{}

\newcommand{\sv}[1]{{\color{black}#1}}

\newcommand{\fractionBelowLvkOneFifty}{\ensuremath{36.5^{+6.8}_{-7.9}}\%\xspace}
\newcommand{\fractionBelowLvkFiveHundred}{\ensuremath{41.4^{+3.7}_{-4.2}}\%\xspace}
\newcommand{\fractionBelowLvkFifteenHundred}{\ensuremath{40.0^{+2.8}_{-1.6}\%}\xspace}
\newcommand{\fractionBelowIsogaussOneFifty}{\ensuremath{37.1^{+7.4}_{-7.8}\%}\xspace}
\newcommand{\fractionBelowIsogaussFiveHundred}{\ensuremath{41.8^{+3.9}_{-4.6}\%}\xspace}
\newcommand{\fractionBelowIsogaussFifteenHundred}{\ensuremath{40.7^{+2.8}_{-2.2}\%}\xspace}
\newcommand{\truenegativectfrac}{\ensuremath{38.9\%}\xspace}
\newcommand{\truepositivectfrac}{\ensuremath{1.0\%}\xspace}
\newcommand{\fractionAboveLvkOneFifty}{\ensuremath{1.1^{+0.4}_{-0.2}\%}\xspace}
\newcommand{\fractionAboveLvkFiveHundred}{\ensuremath{1.0^{+0.2}_{-0.1}\%}\xspace}
\newcommand{\fractionAboveLvkFifteenHundred}{\ensuremath{1.0^{+0.1}_{-0.1}\%}\xspace}
\newcommand{\fractionAboveIsogaussOneFifty}{\ensuremath{1.0^{+0.3}_{-0.3}\%}\xspace}
\newcommand{\fractionAboveIsogaussFiveHundred}{\ensuremath{0.9^{+0.2}_{-0.2}\%}\xspace}
\newcommand{\fractionAboveIsogaussFifteenHundred}{\ensuremath{0.9^{+0.1}_{-0.1}\%}\xspace}

\newcommand{\fractionBelowIsoBetaOneFifty}{\ensuremath{35.0^{+8.0}_{-7.9}\%}\xspace}
\newcommand{\fractionBelowIsoBetaFiveHundred}{\ensuremath{40.9^{+4.6}_{-5.0}\%}\xspace}
\newcommand{\fractionBelowIsoBetaFifteenHundred}{\ensuremath{40.2^{+1.7}_{-1.8}\%}\xspace}
\newcommand{\fractionBelowIsoTukeyOneFifty}{\ensuremath{37.4^{+12.0}_{-9.8}}\%\xspace}
\newcommand{\fractionBelowIsoTukeyFiveHundred}{\ensuremath{41.8^{+4.6}_{-5.0}}\%\xspace}
\newcommand{\fractionBelowIsoTukeyFifteenHundred}{\ensuremath{40.8^{+2.9}_{-2.1}\%}\xspace}

\newcommand{\fractionAboveIsoBetaOneFifty}{\ensuremath{0.9^{+3.2}_{-0.6}\%}\xspace}
\newcommand{\fractionAboveIsoBetaFiveHundred}{\ensuremath{0.7^{+1.5}_{-0.2}\%}\xspace}
\newcommand{\fractionAboveIsoBetaFifteenHundred}{\ensuremath{0.6^{+0.2}_{-0.2}\%}\xspace}

\newcommand{\fractionAboveIsoTukeyOneFifty}{\ensuremath{0.9^{+0.6}_{-0.4}}\%\xspace}
\newcommand{\fractionAboveIsoTukeyFiveHundred}{\ensuremath{0.9^{+0.2}_{-0.2}}\%\xspace}
\newcommand{\fractionAboveIsoTukeyFifteenHundred}{\ensuremath{0.9^{+0.1}_{-0.1}\%}\xspace}

\newcommand{\gwtcthree}{GWTC-3\xspace}
\newcommand{\LIGO}{\affiliation{LIGO Laboratory, Massachusetts Institute of Technology, Cambridge, MA 02139, USA}}
\newcommand{\MKI}{\affiliation{Kavli Institute for Astrophysics and Space Research, Massachusetts Institute of Technology, Cambridge, MA 02139, USA}}
\newcommand{\MIT}{\affiliation{Department of Physics, Massachusetts Institute of Technology, Cambridge, MA 02139, USA}}

\begin{document}
\begin{textblock*}{4cm}(0.5pt,5pt)
    \includegraphics[width=4cm]{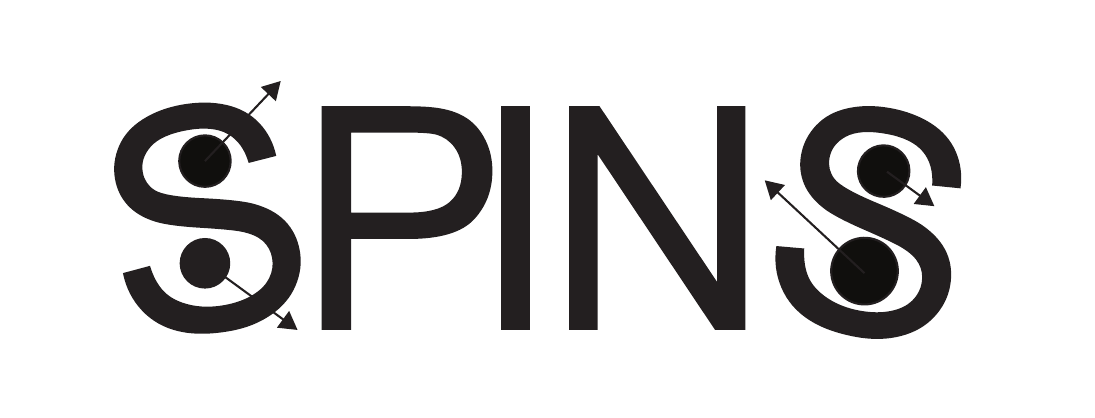}
\end{textblock*}
\title{The Long Road to Alignment: Measuring Black Hole Spin Orientation with Expanding Gravitational-Wave Datasets}

\author{Salvatore Vitale}
\email{salvo@mit.edu}
\LIGO\MKI\MIT

\author{Matthew Mould}
\LIGO\MKI\MIT

\collaboration{Society of Physicists Interested in Non-aligned Spins, SPINS}
\thanks{\url{sites.mit.edu/spins}}
\date{\today}

\begin{abstract}
Measuring the distribution of spin tilts—the angles between the spin vectors and the binary orbital angular momentum—in stellar-mass binary black holes detected by LIGO-Virgo-KAGRA would provide valuable insight into their astrophysical origins. Analyses of the 69 binary black holes detected through LIGO-Virgo-KAGRA's third observing run yielded model-dependent conclusions, particularly regarding whether the spin tilt distribution exhibits a peak near alignment, as expected for binaries formed in galactic fields. In this work, we simulate populations of up to 1500 binary black hole systems with parameters consistent with the default \gwtcthree analysis, while introducing a correlation that favors small spin tilts for binaries with mass ratios near unity. We find that: (a) spurious peaks away from perfect alignment are possible even with catalogs of up to 300 sources; (b) establishing a definitive peak at alignment remains difficult even with 1500 detections; (c) integrated measurements -- such as the fraction of events with tilt angles smaller than $10^\circ$ or greater than $90^\circ$ -- are more robust and should be preferred, achieving relative 90\% credible uncertainties of $\sim20\%-80\%$ with 1500 sources; and (d) even with the largest simulated catalogs, evidence for a mass ratio–tilt correlation remains inconclusive. Our results suggest that identifying the formation channels of merging black holes using spin tilts will remain challenging, but that model-independent measurements may yield more informative insights over model parameters themselves.
\end{abstract}
\maketitle
\preprint{INT-PUB-25-016}

\section{\label{sec:intro} Introduction}

The field of gravitational-wave (GW) astrophysics will become ten years old in September 2025. After the momentous discovery of GW150914~\cite{LIGOScientific:2016aoc}, the LIGO~\cite{TheLIGOScientific:2014jea}-Virgo~\cite{TheVirgo:2014hva}-KAGRA~\cite{Aso:2013eba} (LVK) collaboration and others have published on the detection of nearly 100 binary black hole (BBH) binaries~\cite{KAGRA:2021vkt,Nitz:2021uxj,Olsen:2022pin}. With another $\sim 200$ significant compact binary mergers~\cite{LVK:gracedbO4} - most of which are BBHs - reported in low-latency in GCNs during the fourth observing run (O4) and yet to be published~\cite{LVK:ObservingPlan}, population analyses are likely to become more and more constraining. These studies aim at characterizing the properties of the underlying astrophysical population - or populations - of the stellar-mass BBHs being detected in LVK data~\cite{Mapelli:2021taw,Callister:2024cdx}. The ultimate goal would be to understand how many such populations exist {across the universe}, their merger rates over redshift, and the distribution of BBH parameters (masses, spins, eccentricity, though eccentricity is much harder to measure~\cite{Favata:2021vhw}) arising from each population. In practice, one deals with the inverse problem: given a set of $N\gg 1$ detected BBHs, for each of which a noisy measurement of masses, spins, etc. is obtained, what can be said about the astrophysical populations that produced them?

To attempt answering this question, predictions about the distributions of some or all parameters from each population (also known as formation channel) should be available. Those can be either partial reasonable expectations on only one or a subset of the parameters or very detailed distributions of all parameters and their correlations. These latter are generated by so called population synthesis suites: end-to-end simulations that may yield the distribution of all intrinsic parameters and their correlations~\cite{Belczynski:2005mr,2017PASA...34...58E,2023ApJS..264...45F,COMPASTeam:2021tbl,Spera:2018wnw,2022ApJS..258...22R}. 
The analyst must thus decide what level of expectations to fold in when setting up their models for the astrophysical distribution of BBH parameters. 

The most conservative approach is to use flexible models~\cite{Vitale:2018yhm,Ray:2023upk,Callister:2023tgi,Farah:2024xub,Edelman:2021zkw,Edelman:2022ydv,Golomb:2022bon,Heinzel:2024hva,Mandel:2016prl,Heinzel:2024jlc,Heinzel:2024hva,Ray:2023upk,Toubiana:2023egi,Tiwari:2020otp,Godfrey:2023oxb,Sadiq:2023zee}. This approach has the benefit of minimizing the risk of biases but comes at the cost of larger statistical uncertainties driven by the large number of model's parameters. It also usually yields constraints about the properties of the \emph{overall} set of sources, without characterizing eventual subpopulations. The next most conservative approach is to use parametric families of functions to model some or all of the source properties (e.g., Refs.~\cite{Vitale:2015tea,Talbot:2018cva,Talbot:2017yur,Fishbach:2018edt,Callister:2021fpo,Biscoveanu:2022qac,LIGOScientific:2020kqk,LIGOScientific:2021psn}). This is the approach that was historically followed first, as it is simple to implement and has the advantage that at least some of the model’s parameters can approximate astrophysical quantities of interest. More recently there have been proposals to use directly the output of populations synthesis codes as the model that enters the population analysis~\cite{Zevin:2020gbd,Wong:2020ise,Wong:2019uni,Colloms:2025hib,Plunkett:2025mjr}. The prospects of using population synthesis results as models for the analysis of GW data is intriguing, as they could constrain the very parameters that affect binary evolution (e.g. the properties of the progenitor stars initial mass function). At the same time, limitations exist in the complexity and availability of such population synthesis simulations everywhere in the relevant parameter space. Using limited information can naturally lead to biases~\cite{Franciolini:2021tla,Cheng:2023ddt}. In practice, one might use a combination of all the methods above in various steps of the analysis, or for different subset of parameters~\cite{Heinzel:2023hlb,Plunkett:2025mjr}.

Among the tracers of BBHs formation channels there are BH spins~\cite{Gerosa:2013laa,Gerosa:2014kta,Vitale:2015tea,Farr:2017uvj,Talbot:2017yur,Mould:2021xst,Varma:2021xbh,Varma:2021csh,Biscoveanu:2025jpc,Roulet:2021hcu,Stevenson:2017dlk}. While predictions on the expected spin magnitude have changed over the years and are hard to make in a solid fashion \sv{(see for example the introduction of Ref.~\cite{Perigois:2023ihi} and references therein)}, a rather simple prediction can be made about their direction. That is that BBHs formed dynamically should have spins that are randomly oriented, since no direction is preferred, whereas black holes formed in galactic field stellar binaries should have spins preferentially aligned with the binary orbital angular momentum~\cite{Rodriguez:2016vmx,Marchant:2021hiv}. The exact degree of alignment is not known, as it depends on the poorly-understood details of the supernovae explosions that generate the two BHs, for example, the geometry and amount of the fallback material, and on the orbital separation at the time of the supernovae~\cite{Fryer:1999ht,Kalogera:1999tq,2012ApJ...749...91F}. 

This expectation can be folded in a simple model for the spin angles (called tilts \sv{and indicated with the letter $\tau$ in this paper}), that includes a mixture of two components: one isotropic component and a preferentially-aligned component. In Ref.~\cite{Vitale:2015tea} this preferentially-aligned component was a Gaussian in cosine tilt space centered at $\ct=1$ and with a fixed width. Ref.~\cite{Talbot:2017yur} later extended the model by making the width of the Gaussian a parameter measured from the data. This is the model that was used in the default spin analyses of the LVK up to their catalog \gwtcthree based on the end of the third observing run (O3b). In their \gwtcthree paper~\cite{LIGOScientific:2021psn}, the LVK finds that the data are consistent with the tilt distribution having a peak at aligned spins but also with a broad distribution (they find that $44^{+6}_{-11}\%$ of BHs have $\ct\leq0$\footnote{As shown in our Fig~\ref{Fig.FractionCtBelow} and Tab.~\ref{Tab.FractionsUncorr}, the prior for this fraction peaks in the same region.}). Ref~\cite{Vitale:2022dpa} extended the spin tilt model by allowing the location of the preferentially-aligned component to be measured from the data and re-analyzed \gwtcthree with it. They found that while the data is not inconsistent with a peak at $\ct=1$ it does not require it either. In fact, with this extended model they obtain a mild preference for a broad peak in the \ct distribution \emph{away} from unity. This unexpected result was corroborated in that same paper using different parametric models~\cite{Vitale:2022dpa} and by other independent analyses~\cite{Li:2023yyt,Callister:2023tgi,Edelman:2022ydv,Golomb:2022bon,Heinzel:2024hva}.

The challenges of measuring the magnitude and tilt of individual BHs with GWs has been known for a long time~\cite{vanderSluys:2007st,vanderSluys:2009bf,Vitale:2014mka,Purrer:2015nkh,Vitale:2016avz}, and naturally these uncertain measurements on an event-by-event basis propagate at the population level, making the choices of models, priors and analyses details more important~\cite{Vitale:2022dpa,Dhani:2024jja}. Indeed, Ref.~\cite{Miller:2024sui} investigated the extent to which GW observations constrain the full spin distributions of BBHs beyond the commonly used effective spin parameters. Using simulated populations with identical effective spin distributions but differing component spin magnitudes and tilt angles, they found that while gravitational waves do encode information about full spin vectors, this information is extremely difficult to extract in practice. 

In this work we focus on the measurability of the astrophysical spin tilt distribution using simulated BBH populations. Our ``true'' population is chosen to be consistent with what measured in \gwtcthree but we also endow the population with a correlation between BBHs mass ratio and spins (which does leave it consistent with \gwtcthree). We consider catalog sizes of up to $1500$ sources. That is, we attempt to make predictions about the evolution of this measurements for the next several years. These sources are analyzed with several of the models proposed by Ref.~\cite{Vitale:2022dpa}, in order to assess model-dependence and enable model selection. Our main findings are that:

\begin{itemize}
\item Even if the true \ct distribution peaks at $1$, spurious peaks away from unity are not impossible with catalogs including as many as $300$ sources. However, they become increasingly less likely \sv{as the catalog size increases}. Should a peak away from $\ct=1$ still be observed in O4 data (which could include $\simeq 300$ BBHs, based on public alerts~\cite{LVK:gracedbO4}) it might still go away as more sources are added. 
\item Even with $1500$ BBH sources, some of the measurements of key parameters --- such as the fraction of events in the non-isotropic component --- are very uncertain, and may depend on the exact model being used, especially for our most elastic model. 
\item While it is tempting to focus on marginalized posteriors of the model parameters, as some of those can be directly connected to interesting astrophysical quantities, it is better to work directly with the posterior predictive distribution (PPD) of the relevant source parameter or with integrated quantities obtained from the PPD. Given the complexity of these models, looking at marginalized posteriors of individual parameters can be challenging and might belie the whole story. 
\item For our simulated population, the true fraction of sources with $\ct\leq 0$ is \truenegativectfrac and the true fraction of sources with tilts within $10^\circ$ from perfect alignment is \truepositivectfrac. Using a mixture model with an isotropic spin component and a Gaussian component with both location and width measured from the data, we measure these two fractions to be \fractionBelowIsogaussOneFifty, \fractionBelowIsogaussFiveHundred, \fractionBelowIsogaussFifteenHundred and \fractionAboveIsogaussOneFifty, \fractionAboveIsogaussFiveHundred, \fractionAboveIsogaussFifteenHundred with catalogs including $N=150$, $500$ and $1500$ sources, respectively. 
\item Irrespective of the catalog size, we cannot reveal in a definitive way the existence of the mass ratio--spin tilt correlation. Both posterior-based analyses and evidence-based analyses are inconclusive. In fact, the simplest model--that does not allow for the existence of this correlation--is slightly preferred for all catalog sizes, with most other models achieving a similar evidence.
\end{itemize}

The rest of the paper is structured as follows: in Sec~\ref{Sec.SimulatedPop} we discuss in detail the parameters of the simulated BBH populations; in Sec.~\ref{Sec.BayesianAnalysis} we review the technical details of hierarchical Bayesian analysis; in Sec.~\ref{Sec.HyperModels} we list the models used in the paper. Results are reported in Sec.~\ref{Sec.Results}. Sec.~\ref{Sec.Conclusions} discusses our findings and future prospects.

\section{Simulated population}\label{Sec.SimulatedPop}

We simulate BBH mergers from a population that is consistent with \gwtcthree data~\cite{LIGOScientific:2021psn}. 
Specifically, the joint distribution for the source-frame primary mass ($m_1$) and mass ratio ($q$) is the \texttt{Power Law + Peak} model~\cite{Talbot:2018cva} with hyper parameters: $\alpha_{m_1}=3.4, m_{\rm{min}}=5\msun, m_{\rm{max}}=87\msun, m_{\rm{lam}}=0.04, m_{\mu}=34\msun,m_{\sigma}=3.6\msun, \delta\!m=4.8\msun, \beta_q=1.1$.
The spin magnitudes are assumed to be independent and identically distributed from a beta distribution~\cite{Wysocki:2018}, with parameters $\alpha_\chi=1.67, \beta_\chi=4.43$. The redshifts are draws from the  \texttt{power law} redshift model~\cite{Fishbach:2018edt} with slope $\lambda_z=2.73$. Fig.~\ref{Fig.TrueCorner} in App.~\ref{App.Truth} shows the hyper posteriors for the default LVK model~\cite{LIGOScientific:2021psn} for \gwtcthree (as run by Ref.~\cite{Vitale:2022dpa}) in blue, and mark the values of the hyper parameters we used to simulate our sources with \truthcolor lines. 

The cosine of the black hole spin tilt angles are drawn from a distribution that is a mixture of an isotropic distribution, and a Gaussian peaking at $\ct=1$ (i.e., spin vector aligned with the angular momentum). In order to verify if and when it will be possible to reveal eventual correlations between the distribution of the spins tilts and other parameters, we make the branching ratio between the preferentially aligned component and the isotropic component a function of the mass ratio. Specifically:

\begin{widetext}
\be
p(\cos\tau_1, \cos\tau_2 | q, f_{q=1}, \sigma_{c\tau}, \mu_{c\tau}, n) = \frac{1-f_a(q,f_{q=1},n)}{4} +f_a(q,f_{q=1},n) \mathcal{N}(\cos\tau_1 | \mu_{c\tau}, \sigma_{c\tau}) \mathcal{N}(\cos\tau_2| \mu_{c\tau}, \sigma_{c\tau}),
\ee
\end{widetext}

\noindent where $\mathcal{N}$ indicates a normal distribution. We have defined the mass-ratio-dependent fraction of preferentially aligned systems as

\bea
f_a(q,f_{q=1},n) &\equiv&  f_{q=1} \frac{g(q,n)-g(0.1,n)}{g(1,n)-g(0.1,n)}\nonumber\\
g(q,n) &\equiv& \exp{\left[(q-0.1)^n - 0.9^n\right]}\label{Eq.Correlatedfa}
\eea
and set $f_{q=1}=1, \sigma_{c\tau}=1.15, \mu_{c\tau}=1$ and $n=2$.
\begin{figure}
\includegraphics[width=\columnwidth]{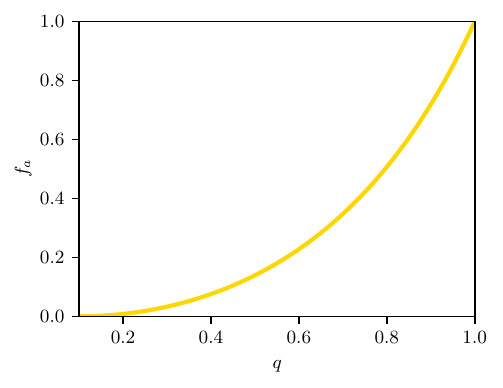}
\caption{Fraction of sources in the preferentially aligned-spin component as a function of the binary mass ratio. This distribution is used when generating the synthetic catalog of BBHs.}\label{Fig.Truefa}
\end{figure}

This non-linear correlation results in a higher fraction of systems with preferentially aligned spins as the mass ratio of the system gets closer to unity, and is shown in Fig.~\ref{Fig.Truefa}. While this functional form is not meant to represent a correlation based on solid astrophysical grounds, it can at least qualitatively capture one possible scenario in which dynamical environments, where isotropic spins are expected, produces BBHs with unequal masses.

\begin{figure}
\includegraphics[width=\columnwidth]{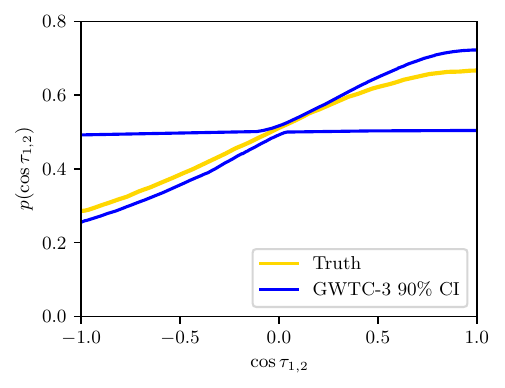}\\
\includegraphics[width=\columnwidth]{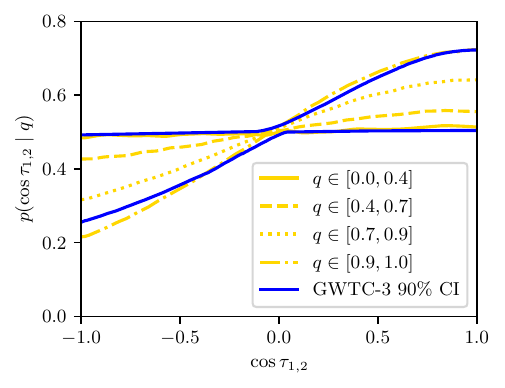}\\
\caption{The distribution of cosine tilts in the simulated universe (yellow), compared with the 90\% credible interval from \gwtcthree (blue). The top panel shows the distributions for the two tilts (which are identical) marginalized over the mass ratio, while the bottom panel shows the conditional distribution for $\cos\tau_{1,2}$ for $q$ in four disjoint intervals.}\label{Fig.CosTau}
\end{figure}

To better visualize how the population we simulate compares with the default LVK results from \gwtcthree, in the top panel of Fig.~\ref{Fig.CosTau} we show the resulting distribution of the cosine tilts (marginalized over the mass ratio), and the 90\% credible interval from the LVK analysis for comparison. The bottom panel shows the distribution of cosine tilts conditional on the mass ratio for $q$ in  4 different intervals together with the \gwtcthree 90\% credible interval. We notice that because the mass-ratio distribution peaks at equal masses (see e.g. Fig.~10 of Ref.~\cite{LIGOScientific:2021psn}), in practice most sources in our simulated population will be drawn from a \ct distribution with a value of the aligned fraction $f_a$ close to 1.

We generate a catalog of {1599} detectable BBHs\footnote{Ask me about this number next time you see me at a conference.} from this distribution, and add them into simulated LIGO-Virgo Gaussian noise, with a sensitivity corresponding to that of the fourth observing run, using the power spectral densities (PSD) provided by~\cite{LIGOT2000012}. To mitigate the problems highlighted by Ref.~\cite{Essick:2023upv}, we define detectability based on the matched filter signal-to-noise ratio (SNR), rather than the optimal SNR. Specifically, a source is detectable if its network matched-filter SNR --- defined as the square root of the sum of the squares of the SNR of each detector --- is greater than 11. We run the parameter estimation algorithm \texttt{Bilby}~\cite{Ashton:2018jfp,Romero-Shaw:2020owr} on the detectable sources, using the \texttt{IMRPhenomXP}~\cite{Pratten:2020fqn} waveform model and the relative binning likelihood~\cite{Krishna:2023bug}. \sv{We do not include higher-order modes to keep the computation cost manageable (See discussion in Sec.~\ref{Sec.Conclusions}).} These sources are used to form catalogs of various size and analyzed using \texttt{GWPopulation}~\cite{Talbot:2024yqw} to measure the hyper parameters, using the models described in Sec.~\ref{Sec.HyperModels}.

\section{Hierarchical Likelihood and Selection effects}\label{Sec.BayesianAnalysis}

Bayesian inference\footnote{The reader familiar with data analysis for gravitational-wave populations can skip this section (though they might find Sec.~\ref{SubSec.SelEffects} interesting).} on catalogs of GW sources can be performed given a model for the underlying astrophysical population, parametrized by hyper parameters \La~\cite{1995ApJS...96..261L,Loredo:1997qf,Loredo:2001rx,Mandel:2018mve,Vitale:2020aaz}:

\be
p(\La | {D}) \propto \pi(\La) \prod_{i=1}^{N_{\rm{events}}} \frac{p(d_i | \La)}{\alpha(\La)},\label{eq.HyperPost}
\ee

\noindent where $\alpha(\La)$ is the population-dependent fraction of detectable events (see Eq.~\ref{Eq.Alpha}); $\pi(\La)$ the hyper prior; $D=\{d_1,\ldots, d_{N_{\rm{events}}}\}$ is the data in the catalog; and $p(d_i | \La)$ are the likelihoods of the individual events. This expression assumes that the merger rate has already been marginalized over using a log-uniform prior.

The choice of the population model affects the individual-event likelihoods which, upon explicit marginalization of the individual-event source parameters \th, can be written as~\cite{2019PASA...36...10T}:

\bea
p(d_i | \La ) &=& \int{\mathrm{d}\th\, p(d_i| \th) \pi(\th | \La)}\nonumber\\
&\propto& \int{\mathrm{d}\th\, p(\th|d_i, \mathrm{PE}) \frac{\pi(\th | \La)}{\pi(\th|\mathrm{PE})}}.\label{Eq.SingleLike}
\eea

\noindent The parameter estimation (PE) label indicates that the posterior $ p(\th|d_i, \mathrm{PE})$ is evaluated using a parameter estimation software, with associated prior $\pi(\th|\rm{PE})$. {We notice that this integral is formally inconsistent with our data selection procedure~\cite{Essick:2023upv}, since we condition detectability on the true values of each event source parameters (see also the discussion around Eq.~(30) of Ref.~\cite{Heinzel:2024hva}) but the systematics this approximation introduces are smaller than statistical uncertainties, see App.~\ref{App.PPD}.}
In practice, the integral in Eq.~\ref{Eq.SingleLike} is evaluated numerically as a Monte Carlo integral:

\be
p(d_i | \La ) \simeq \frac{1}{M} \sum_{j=1}^{M}\left. \frac{\pi(\th_j | \La)}{\pi(\th_j|\mathrm{PE})} \right|_{\th_j \sim p(\theta | d_i, \rm{PE})}.
\ee

\noindent The precision of this estimation depends on the number of posterior samples that are available. We run \texttt{Bilby} with the \texttt{Dynesty}~\cite{Speagle_2020} sampler and 4000 live points. This results in at least 16,000 posterior samples for each source. 

In order to properly account for selection effects, the analyst must be able to calculate $\alpha(\La)$, the fraction of detectable sources (with a definition of detectability that is self-consistent with that used to select the catalog) for all plausible values of the model hyper parameters. Specifically,

\be
\alpha(\La) = \int \dt p(\rhoup | \th) \pi(\th|\La),
\ee

\noindent where the first term is the probability of a source with parameters \th to have detection statistics above threshold.
This can be written as:

\be
p(\rhoup | \th) = \int \dDup p(\Dup| \th),
\ee

\noindent where the integration domain is the data (i.e. noise realization) that would result on the source with parameters \th to be detectable. Plugging this expression into the previous one we have:

\be
\alpha(\La) = \iint \dt \dDup  \pi(\th|\La) p(\Dup| \th),\label{Eq.Alpha}
\ee

\noindent which can be evaluated by introducing a sampling distribution $\zeta$ from which the values of \th can be drawn:

\bea
\alpha(\La) &=& \iint \dt \dDup  \frac{\pi(\th|\La)}{\zeta(\th | \mathcal{H}_\zeta)} p(\Dup| \th)\zeta(\th| \mathcal{H}_\zeta)\nonumber\\
&=&\iint \dt \dDup  \frac{\pi(\th|\La)}{\zeta(\th| \mathcal{H}_\zeta)} p(\Dup \th| \mathcal{H}_\zeta)\nonumber\\
&=& \frac{1}{N_\mathrm{tot}} \sum_{i=1}^{N_\mathrm{detectable}}\left.\frac{\pi(\th_i|\La)}{\zeta(\th_i| \mathcal{H}_\zeta)}\right|_{\th_i \sim \zeta(\th| \mathcal{H}_\zeta)},
\eea

\noindent where we have numerically evaluated the integrals over data and \th by sampling values of \th from the sampling distribution $\zeta(\th| \mathcal{H}_\zeta)$ and sampling values of the data (i.e. adding the signal associated to \th to a random segment of noise, real or simulated) and only keeping sources for which the resulting data stream corresponds to a detectable source~\cite{Tiwari:2017ndi}.  $N_\mathrm{tot}$ sources will have to be drawn from  $\zeta(\th| \mathcal{H}_\zeta)$ before $N_\mathrm{detectable}$ are collected and in general $N_\mathrm{tot}\gg N_\mathrm{detectable}$, with the exact ratio depending on the details of the proposal distribution (e.g. the maximum redshift at which sources can be, and the mass function). A list of detectable sources to be used for evaluating the sum above during the analysis can be prepared in advance and stored to disk.

In general, the approximated estimate of $\alpha(\La)$ will be better with more detectable sources~\cite{2019RNAAS...3...66F,Talbot:2023pex}, and with a proposed distribution that is similar to the (unknown) astrophysical population. We create a catalog of detectable sources in two steps. First, we generate $\sim${8.3} millions detectable sources drawing their parameters from the same population model used to generate the sources in the catalog, i.e., $\zeta(\th| \mathcal{H}_\zeta)= p(\theta | \La_{\rm{true}})$. Then, we generate $\sim${2.4} millions detectable sources drawing their parameters from a population model that has the same mass and redshift distribution as the previous one, but has uniform spin magnitude and cosine tilt distribution. The two sets are then combined using the method described by Ref.~\cite{Essick_2021} and the overall list of {10.7} million sources are used to evaluate $\alpha(\La)$ numerically. In order to obtain $\mathcal{O}(10^7)$ detectable sources we have to generate $\mathcal{O}(10^9)$ sources, i.e., around 1\% of sources are detectable. Unless otherwise stated, we run \texttt{GWPopulation} with a maximum total variance of 2 for the log likelihood~\cite{Talbot:2023pex}, i.e., when sampling the hyper parameter space, samples that corresponds to a variance larger than 2 are rejected. In practice, for most of the runs the actual variance is much lower, and we will enforce a more stringent cut as discussed below.

We will often show the PPD of the astrophysical parameters that characterize individual sources, i.e., spins, masses and redshift. This can be thought of as the expected distribution of those parameters in light of the detected sources. It can be written as:

\bea
p(\th | D) &=& \int \dLa \pi(\th | \La) p(\La | D) \nonumber\\
&=&\frac{1}{M} \sum_{i=1}^{M} \left. \pi(\th | \La_i)\right|_{\La_i \sim p(\La |D)}
\eea
That is, the PPD is the expectation of the population model calculated with draws from the posterior $p(\La|D)$. To make the notation lighter, we'll often use $\mathrm{PPD}(\th)$ to mean $p(\th | D)$.

\subsection{A note about efficient production of detectable sources}\label{SubSec.SelEffects}

We notice that producing a list of detectable signals for the evaluation of $\alpha(\La)$ can be computationally very expensive, as one in general has to calculate the SNR of all sources, even those that won't eventually end up being detectable. This problem is even more severe when estimating the sensitivity of real GW searches, as one needs to also run search algorithms, not just calculate SNRs~\cite{LIGOScientific:2020kqk,Essick:2023toz,Lorenzo-Medina:2024opt}. 
The problem becomes more important if one is sampling many low-mass and/or high-redshift sources, a smaller fraction of which are detectable. For our purposes, the process can be trivially parallelized over as many CPU cores as possible. To further enhance the efficiency of the algorithm, we also pre-build a look-up table as follows:

\begin{itemize}
\item For a given {source-frame} total mass, we assume equal masses, and take both spins magnitudes equal to 0.99 and aligned with the angular momentum. We generate a waveform with these intrinsic parameters.
\item For each detector in the network, we calculate the {optimal} SNR that the source would have if it were overhead and face-on. We calculate the square root of the squared sum of these SNRs. We calculate the redshift at which this SNR would be equal to 11.
\end{itemize}

We notice that -- given our setup -- both steps are conservative: a) for a given total mass, a system with equal masses produces the strongest signal (if higher order modes are neglected, as in our waveforms). Similarly, maximal spins aligned with the binary orbital angular momentum yield higher SNRs;  b) a face-on source produces the strongest signal and obviously a source cannot be overhead to all detectors in the network. Thus, with both steps we are \emph{overestimating} the optimal SNR that the source would produce\footnote{To decide whether any given source is detectable or not, we use the \emph{matched filter} SNR, as mentioned above. The use of the optimal SNR while building this lookup table is not self-consistent, but also not consequential because our setup is so conservative that in practice we are only excluding sources that have an SNR well below our threshold of 11. Given that matched filter SNR and optimal SNR rarely differ by more than $\sim 1$, and often less, this inconsistency is not problematic. An indirect proof that this is indeed the case, is the fact that our redshift PPD--- which is very sensitive to eventual issues with the estimation of the detectors' sensitivity---is not biased; see App.~\ref{App.PPD}.}. 

With this method, we obtain a mapping $z_{\rm{max}}(M_{\rm{tot}})$ that can be quickly interpolated anywhere in the mass range where our population model is defined. This gives a very conservative estimate of the maximum redshift at which a source with given masses could be detectable. Therefore, when we sample $\zeta(\th| \mathcal{H}_\zeta)$, we can compare the proposed redshift with $z_{\rm{max}}(M_{\rm{tot}})$ and only compute the waveform if the proposed redshift is below the maximum. For this waveform, a random noise realization is generated and a matched filter SNR calculated to decide if the source is detectable or not, in a way that is consistent with the way we selected the detectable sources of the simulated population we analyze. 
With our population, this precomputed look-up table with an over-conservative value of the horizon reduces to 20\% (of $\mathcal{O}(10^9)$ draws) the number of sources for which a waveform must be generated. We have not attempted to optimize this process, and it is certainly possible that a further speed-up could be obtained. 

\subsection{A note on variances}

In this work we will consider catalogs comprising up to 1500 sources and a variety of models. It is well known that the total variance on the hyper log likelihood estimator depends on the catalog size~\cite{2019RNAAS...3...66F,Talbot:2023pex}. \sv{As mentioned above, samples with log likelihood variance larger than 2 are rejected already during sampling. In practice, }with our settings, the total variance is usually smaller than 1 for all but the largest catalogs, comprising 1500 sources. \sv{In Fig.~\ref{Fig.Variances} we show the distribution of variances for the models described in Sec.~\ref{Sec.HyperModels} and three exemplary catalog sizes. Each histogram represent the distribution of variances across hyper samples for one run.} For the smaller catalogs, variances are usually as small as $\sim 0.1$, though we do occasionally have hyper samples that yield much larger variances. We treat those as outliers --- samples that explore region of the parameter spaces where our numerical estimation of the likelihood is unusually uncertain and should not be trusted --- and remove them when reporting numerical results and plots henceforth, unless otherwise stated. 
Specifically, we use the following thresholds for the various catalog sizes, informed by the highest 90\% percentile of the total variance across models for any given catalog size:

\begin{figure}[htb]
\includegraphics[width=\columnwidth]{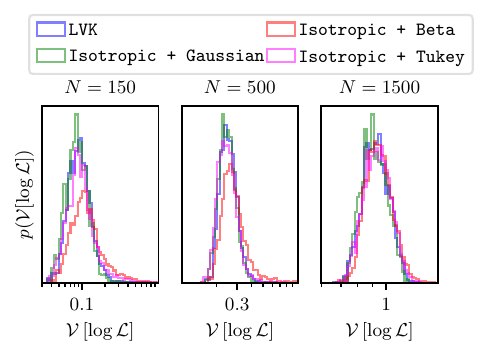}
\caption{Distribution of total variances for all models (described in Sec.~\ref{Sec.HyperModels}) and catalog sizes (we do not show . The variances for correlated models look visually identical. }
\label{Fig.Variances}
\end{figure}
\begin{itemize}
    \item $N=69$, $\mathcal{V}[\log\mathcal{L}]<0.3$
    \item $N=150$, $\mathcal{V}[\log\mathcal{L}]<0.75$
    \item $N=300$, $\mathcal{V}[\log\mathcal{L}]<1.0$
    \item $N=500$, $\mathcal{V}[\log\mathcal{L}]<1.0$
    \item $N=1500$, $\mathcal{V}[\log\mathcal{L}]<1.1$
\end{itemize}

\section{Hyper models}\label{Sec.HyperModels}

In this paper we will use a subset of \ct models from Ref.~\cite{Vitale:2022dpa}:
\begin{widetext}
\begin{flalign}
    &\lvk: && p(\cto,\ctt|\sct,f_a)= \frac{1-f_a}{4} +f_a \prod_{i=1}^{2} \mathcal{N}(\cos\tau_i,\mct=1,\sct)&\\
    &\isogaus: && p(\cto,\ctt|\mct,\sct,f_a)= \frac{1-f_a}{4} +f_a \prod_{i=1}^{2} \mathcal{N}(\cos\tau_i,\mct,\sct)&\\
    &\isobeta: && p(\cto,\ctt|\ab,\bb,f_a)= \frac{1-f_a}{4} +f_a \prod_{i=1}^{2} \mathcal{B}(\cos\tau_i,\ab,\bb)&\\
    &\isotuk: && p(\cto,\ctt|T_{x0},T_k,T_r,f_a)= \frac{1-f_a}{4} +f_a \prod_{i=1}^{2} \mathcal{T}(\cos\tau_i,T_{x0},T_k,T_r).&
\end{flalign}
\end{widetext}

\noindent where $\mathcal{B}$ and $\mathcal{T}$ indicate a beta distribution and Tukey window, respectively, as defined in \sv{Eq. E.1 of} Ref.~\cite{Vitale:2022dpa}. For primary mass, mass ratio, spin magnitude and redshift, we use the same parametric models employed when generating the sources. We set hyper priors equal to Tab.~G2 of Ref.~\cite{Vitale:2022dpa} (the location of the \isogaus's Gaussian component is restricted in the range $[-1,1]$), with the only difference that the two parameters controlling the beta component of \isobeta are uniform in the range $[0.05,7]$. 
These models can be run either assuming that the branching ratio $f_a$ is constant, or that it is correlated with the mass ratio, with the same functional form we used when simulating the sources, Eq.~\ref{Eq.Correlatedfa}. For all models, the priors on \fa is $\mathcal{U}[0,1]$ and -- when enabling correlations -- the prior on \enn is $\mathcal{U}[0.01,12]$. When correlations are allowed we will append ``corr'' to label of the model (e.g., \isocbeta)

We stress that the correlated \isogaus model can exactly match the true astrophysical distribution. This constitutes \emph{an unrealistic best case scenario}, since in general for real data we won't have the luxury of believing that our parametric model is a perfect match to nature, for some values of its parameters. This specific analysis will thus represent the absolute best that can be done given the available data, when all of the possible sources of systematics, namely model mismatch, have been removed. The results obtained with it will constitute a useful quantitative assessment of what one might possible hope to obtain, in optimal conditions.

\section{Results}\label{Sec.Results}

\subsection{Is a peak at $\ct\neq 1$ significative?}

The broad peak or plateau found in the \ct distribution of the \gwtcthree data~\cite{Vitale:2022dpa,Callister:2023tgi,Edelman:2022ydv,Golomb:2022bon} is surprising, as it is not obviously predicted or associated with any of the main astrophysical formation channels. At the same time, \gwtcthree only included 69 BBHs, and \ct is notoriously hard to measure, with most BBH sources producing very broad posteriors~\cite{Vitale:2016avz,Islam:2023zzj}.

The first question we want to address is thus whether -- given a limited catalog size -- one could expect a peak in the PPD of \ct away from $\ct=1$ even if the spins in the true underlying distribution were in fact {born that way}. The answer will clearly depend on what the true distribution is and---for distributions that peak at $\ct=1$---exactly how broad the distribution is. Given that running these analyses is still relatively computationally expensive, we use our fiducial set of sources, described in Sec.~\ref{Sec.SimulatedPop} to attack this question\footnote{In Sec.~\ref{Sec.Conclusions} we will comment on how these results could change given a different true distribution.}.
To that end, we generated 20 catalogs of 69, 150, and 300 BBHs each, drawing from the 1500 sources for which we have produced single-event posteriors. We chose these three sizes to be somewhat representative of the sizes of the \gwtcthree catalog, and what might be available at the end of O4a and O4b. For each of these catalogs we perform the population analysis with the \isogaus model, without allowing for correlations~\footnote{We will show later that even much larger catalogs cannot definitively reveal such a correlation, and thus we are not significantly biasing the analyses in this section by not allowing for it.}. In other words, we consider the minimal extension to the \lvk model introduced in \citet{Vitale:2022dpa}. 

\begin{figure}
 \includegraphics[width=\columnwidth]{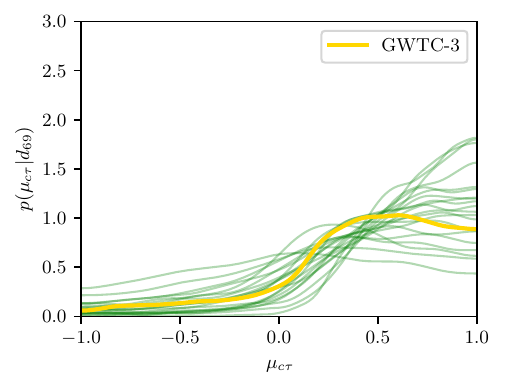}\\
 \includegraphics[width=\columnwidth]{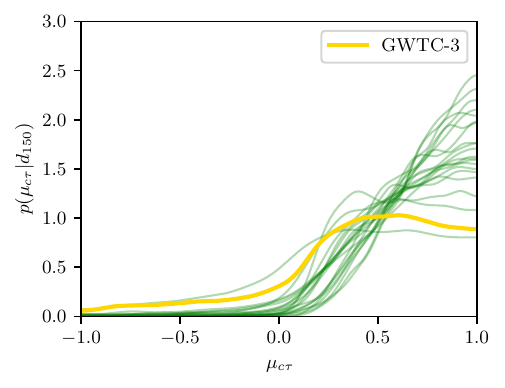}\\
 \includegraphics[width=\columnwidth]{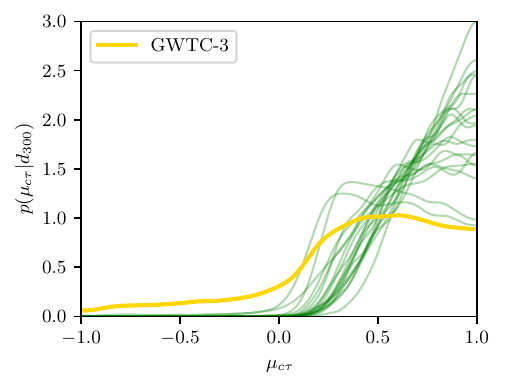}
    \caption{Posteriors of \mct for each of 20 catalogs containing 69 (top panel), 150 (middle panel) and 300 (lower panel) sources. The \truthcolor line shows the posterior obtained by Ref.~\cite{Vitale:2022dpa} using the \isogaus model on \gwtcthree.}
    \label{Fig.MucosRepeaters}
\end{figure}

In the top panel of Fig.~\ref{Fig.MucosRepeaters} we show the hyper posterior measurements for $\mct$ for each of the catalogs of 69 BBH (solid  curves) and, for reference, the posterior obtained in Ref.~\cite{Vitale:2022dpa} using the \isogaus on the \gwtcthree data (\truthcolor curve). 
We find that---depending on the realization of events in each catalog--- \mct can have rather different shapes: some catalogs yield relatively narrow distributions peaked at 1 (i.e. the true value) while others return more shallow distributions or plateau. 
The \gwtcthree hyper posterior is consistent with what we find here, and we conclude that it is not impossible that a true underlying distribution with a broad peak at $\ct=1$ might have yielded a measurement like \gwtcthree's. We notice that for only 10\% of our catalogs of 69 BBH is the 5th percentile for \mct positive. This fraction increases to 85\% (100\%) with catalogs made of 150 (300) BBHs, as seen in the middle and bottom panels of Fig.~\ref{Fig.MucosRepeaters}.

\begin{figure}
 \includegraphics[width=\columnwidth]{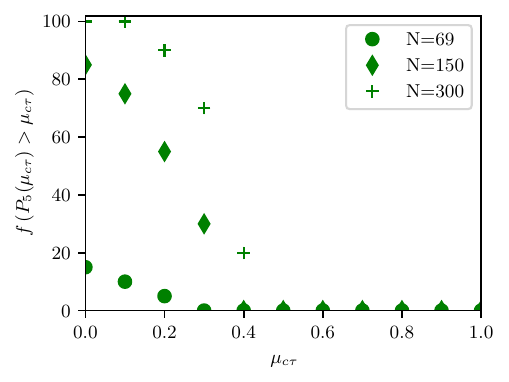}\\
    \caption{For each value \mct in the abscissa, the fraction of catalogs for which the $5^\mathrm{th}$ percentile of $p(\mct|d)$ is larger than \mct. The catalog size is given in the legend and the average is taken over 20 catalogs for each size. The analyses are all performed with the \isogaus model.}
    \label{Fig.FractionCatalogsAbove}
\end{figure}

In Fig.~\ref{Fig.FractionCatalogsAbove} we show for each of the catalog sizes the fraction of catalogs for which the $5^\mathrm{th}$ percentile of the \mct posterior is above the value given in the abscissa. We see that even for our larger catalogs there is a rather sudden drop of this fraction for abscissas in the range $[0,0.4]$. For none of our simulated catalogs can we significantly constrain \mct to be above $0.5$.

Even with larger catalogs comprising 300 sources it is not impossible or even unusual to find posteriors of \mct that peak far from 1, in some cases with peaks in the range $0\leq \mct \leq 0.5$. These posteriors in \mct are usually paired with broad posteriors for \sct, such that the resulting inference in \ct still supports a wide range of possibilities. 
This is a common issue with looking at individual, fully marginalized, posteriors of complicated models with many parameters. We usually still indulge in that exercise because it is often the case that some of these parameters have a clear and useful physical or astrophysical interpretation, and are directly linked to what the analysis is trying to measure. 

\begin{figure}
 \includegraphics[width=\columnwidth]{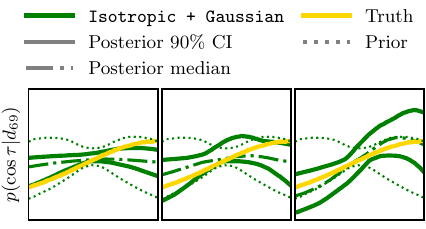}\\
 \includegraphics[width=\columnwidth]{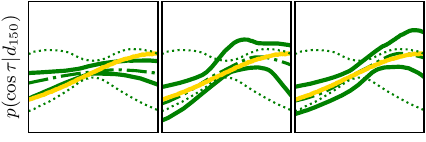}\\
 \includegraphics[width=\columnwidth]{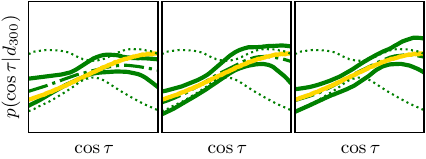}
    \caption{For each of the catalog sizes, $N=69$ (top row) $N=150$ (middle row) and $N=300$ (bottom row) three exemplary PPDs for \ct. In each panel the \truthcolor curve shows the truth, the cyan solid line the median, and the dark band the 90\% credible interval. For the $N=300$ catalogs, one can still get either broad 90\% posterior plateaus (bottom left) or even peaks away from unity (bottom mid) though those are more rare than in smaller catalog sizes.}
    \label{Fig.PPDRepeaters}
\end{figure}

Ultimately, what is being constrained is the high-dimensional shape of the model's parameters, which we can plot through the PPD of the relevant part of the model. It would be unpractical to show 20 sets of \ct PPD for each of the three catalog sizes we have considered in this section. Instead, for each size we chose three exemplary catalogs: one that yields a rather flat posterior, one that yields a peak away from unity, and one that peaks at unity. These are shown in Fig.~\ref{Fig.PPDRepeaters}: in each panel, the green dot-dashed line is the median and the green solid lines enclose the 90\% central credible region. Dotted lines enclose the central 90\% of the prior and the solid \truthcolor line shows the true distribution of \ct. We see that it is not impossible for even a catalog with 300 BBHs to produce a flat PPD in \ct, like the one we show in the bottom left panel. However, it is much more common for catalogs of that size to result in PPDs that peak toward unity, like the bottom right panel. 

We thus conclude that, to the extent that the true distribution of \ct in nature is similar to what simulated here, it would not be surprising to measure spurious peaks in the PPD distribution away from unity after 150 BBHs are collected, and it would not be impossible (though more unlikely) after 300 BBHs are collected.

\subsection{The long road to alignment}

In this section we report on the measurements we obtain with three catalogs, with $N=150$, $N=500$ and $N=1500$ sources. Our data release also includes catalogs with $N=300$ and $N=1000$ sources but we are not showing these in the body of the paper to keep plots less busy. Given the current median merger rate form BBHs, this implies we are able to make projections that span the fourth and the fifth observing run\footnote{Even though we used power spectral density representative of O4, our results can be used to make projections for O5 because the main impact of a more sensitive network is to increase the number of detectable sources.}. 
We'll first focus on the results one obtains using models that do \emph{not} allow for the \ct--$q$ correlation we have introduced in our simulated universe in Sec.~\ref{SubSec.UncorrModels_a} (\lvk and \isogaus models) and \ref{SubSec.UncorrModels_b} (\isobeta and \isotuk models). We allow for that correlation in Sec.~\ref{SubSec.CorrModels}.

\subsubsection{Uncorrelated models: \lvk and \isogaus }\label{SubSec.UncorrModels_a}

{\begin{center}{\small\textbf{Hyper parameters}}\end{center}}

\begin{figure}[t]
    \centering
    \includegraphics[width=\columnwidth]{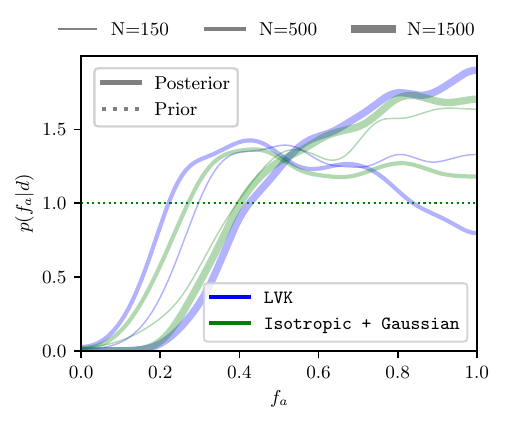}
    \caption{Fraction of sources in the preferentially aligned component for the \lvk and \isogaus models. The prior is marked with a dotted line (in this case, the same for both models).}
    \label{Fig.fa_lvk_isogauss}
\end{figure}

The only thing that all models have in common, in addition to yielding a PPD for the same set of single-event parameters (masses, spins, redshift), is that they all include the fraction of sources in the non-isotropic component - \fa - as one of their parameters. Therefore, this parameter can always be compared across models and catalog sizes. We show the posterior on \fa for the \lvk and \isogaus models in Fig.~\ref{Fig.fa_lvk_isogauss}. The plot shows that the two models perform similarly well, and that while the uncertainty shrinks as the number of sources increases, even the largest catalogs can't exclude small values of \fa. While even with the smallest catalog size we find that $\fa=0$ is excluded at high credibility, values as small as $\fa\gtrsim 0.3$ cannot be ruled out even after $1500$ sources.

\begin{figure}[t]
    \centering
    \includegraphics[width=\columnwidth]{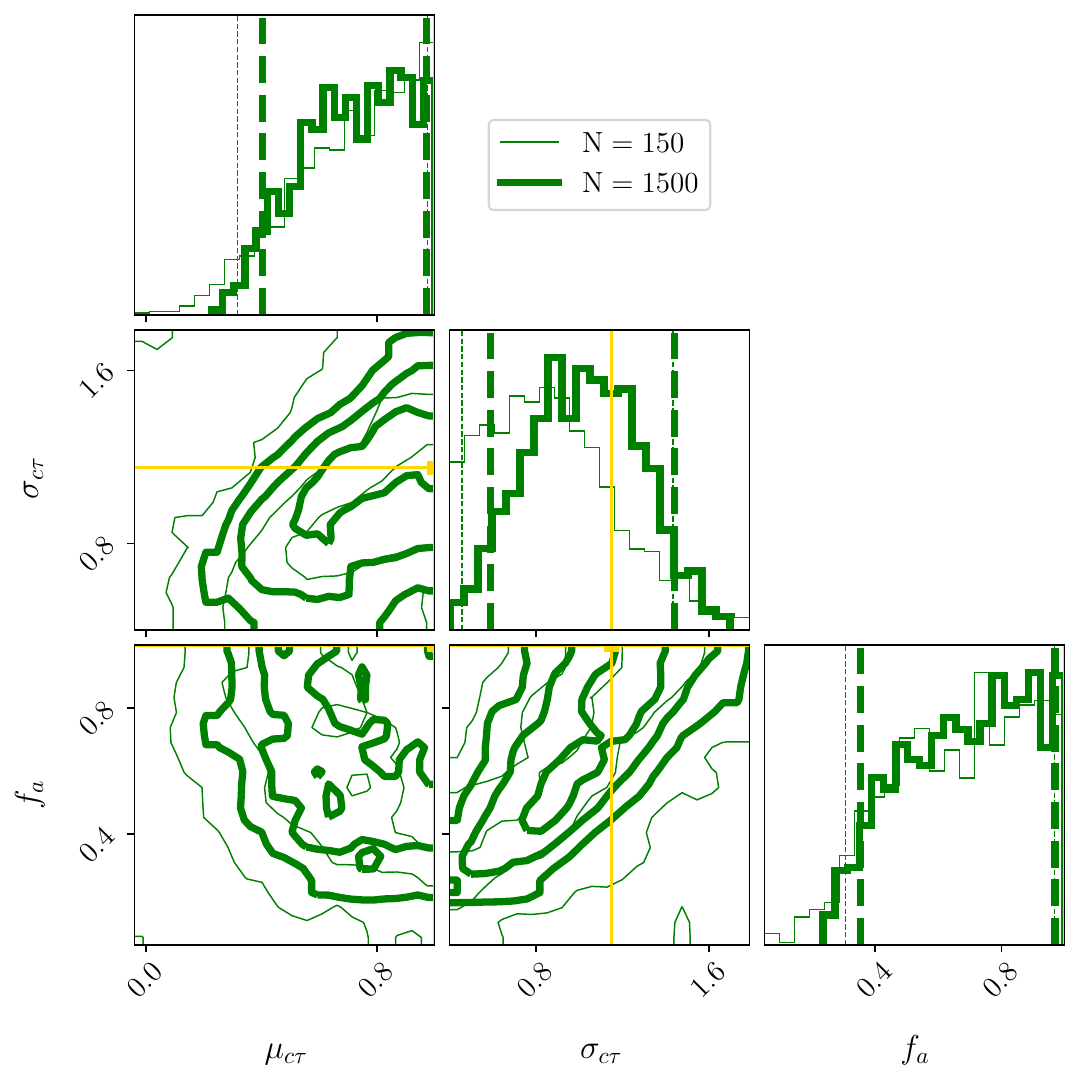}
    \caption{Parameters that control the \ct part of the \isogaus model. To keep the corner plot lighter we only plot results for $N=150$ and $N=1500$. The \truthcolor line shows the true values (note: there is no true value for $f_a$ since the true distribution has a $q$-dependent fraction of aligned events, Fig.~\ref{Fig.Truefa}). The contours in the 2D plots enclose 68.3\%, 95.4\%, and 99.7\% of the posterior volume. The vertical dashed lines in the diagonal panels enclose 90\% of the posterior. Priors are uniform for all parameters.}
    \label{Fig.cornerIsoGauss}
\end{figure}

It may be surprising that the uncertainty for \fa shrinks relatively slowly as the number of events increases. {In fact, for all the hyper parameters that control the spin tilt distributions it is \emph{not} the case that the uncertainties scale with the square root of the number of events}. The reason why this happens can be better understood by looking at a corner plot of all parameters that control the tilt distribution: \fa, \mct and \sct, which is shown in Fig.~\ref{Fig.cornerIsoGauss}. As more and more events are added to the catalog, the main effect is that the joint \fa-\sct projection shrinks along its semi-minor axis. This mainly helps exclude configurations with a large fraction of aligned sources (large \fa) and a narrow Gaussian peak (small \sct) centered close to $\mct=1$. Once that is done, it is harder to further exclude parts of the parameter space because---given the very uncertain measurement of \ct on an event by event basis---the analysis cannot easily differentiate between a universe with the true value of \sct and another universe with a smaller value of \sct which also produces slightly fewer sources with preferentially aligned spins\footnote{Similar patterns can be found in the \lvk model.}. We are left to contend with a correlated, high-dimensional {parameter space oddity}, where different combinations of the hyper parameters controlling the spin tilt distribution result in similar likelihoods. As we show below, quantities based on PPD should be preferred over marginalized hyper parameters. 

For the \isogaus model, the parameter \mct has a clear astrophysical interpretation as the location of the peak of the non-isotropic component of the \ct distribution. Because of this direct interpretation, we quote uncertainties for it,  the caveats about marginal posteriors notwithstanding. The top left panel of Fig.~\ref{Fig.cornerIsoGauss} shows the marginalized \mct posteriors for the catalogs with $150$ and $1500$ sources. We see that while in both cases the posterior peaks at 1, it is rather wide and smaller values are not excluded. We measure $\mct=0.74^{+0.24}_{-0.42}$, $\mct=0.72^{+0.25}_{-0.41}$ and $\mct=0.74^{+0.13}_{-0.30}$ for $N=150$, $N=500$ (not shown in the corner plot) and $N=1500$ sources, respectively. 

{\begin{center}{\small\textbf{PPD and derived quantities}}\end{center}}

\begin{figure}[t]
 \includegraphics[width=\columnwidth]{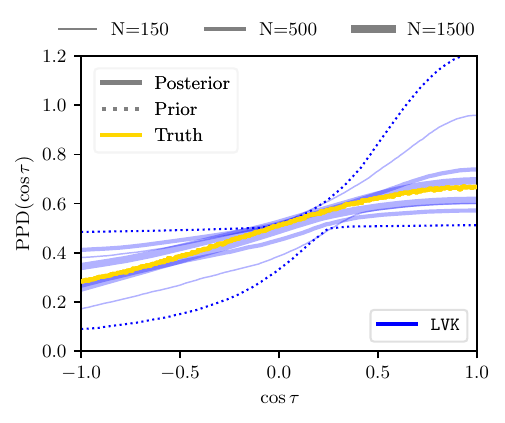}\\
     \includegraphics[width=\columnwidth]{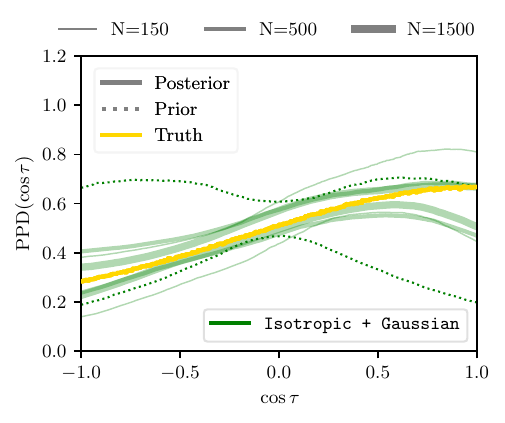}
    \caption{PPD for \ct for catalogs of $150$, $500$, $1500$ events using the uncorrelated \lvk model (top) and \isogaus model (bottom).The thickness of the curves indicates the catalog size and each set of curves spans the 90\% credible interval. The \truthcolor curve is the true distribution.}
    \label{Fig.PPDUncorrelated_a}
\end{figure}

In Fig.~\ref{Fig.PPDUncorrelated_a} we show the 90\% credible intervals of the PPDs for \ct (PPDs for the other parameters are shown in App.~\ref{App.PPD}) obtained using the \lvk and \isogaus models, with the \truthcolor curve indicating the true distribution and dotted curves enclosing 90\% of the prior. For each catalog size, the truth is included in the 90\% credible interval everywhere in the domain of \ct for both models. The main difference we observe is that for the two larger catalog sizes the \isogaus model yields a flatter distribution for positive \ct. 

\begin{figure}[t]
    \centering
    \includegraphics[width=\columnwidth]{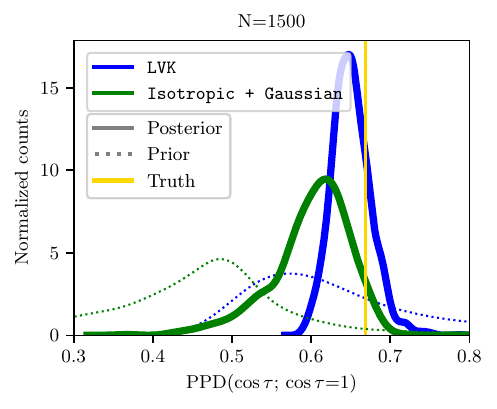}
    \caption{Slice of the PPD for \ct at $\ct=1$ for the catalog with 1500 sources. The truth is indicated with a \truthcolor line.}
    \label{Fig:LVKIsoGaussSliceAt1}
\end{figure}

We can more easily compare the width of the uncertainty regions by plotting slices of the PPD at fixed values of \ct. For example, we show a slice of the PPD $\ct=1$ for the catalog including $1500$ sources in Fig.~\ref{Fig:LVKIsoGaussSliceAt1}. Both models do well and, as one might have imagined, the simpler model with fewer parameters (and with $\mct=1$ by construction) yields a narrower PPD. The tail on the left of the \isogaus PPD corresponds to those flatter posteriors characteristic of the measurements obtained with that model. Smaller catalogs have correspondingly larger uncertainties: for the \isogaus model at $\ct=1$ the 90\% uncertainty is $0.16$ with $1500$ sources, which becomes $0.21$ ($0.36$) for catalogs of $500$ ($150$) sources. This is roughly a twofold reduction in the uncertainty at $\ct=1$ as the number of sources increases tenfold. Slicing at different values, we find shrinkage between 2 and 3 compared to the catalog with $150$ sources. Similar considerations can be made for the \lvk model, for which the shrinkage is larger, between $3$ and $4$ going from $150$ to $1500$ sources, depending on where one slices. We quote uncertainties on the PPD sliced at other values in Tabs.~\ref{tab:cosTau_slices_uncorr} and~\ref{tab:cosTau_slices_corr} in App.~\ref{App.Tables}.

\begin{figure}[t]
    \centering
    \includegraphics[width=1\columnwidth]{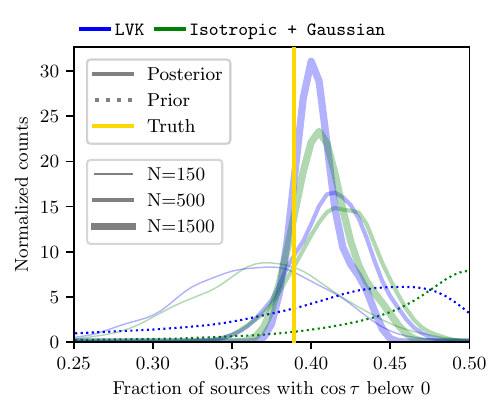}\\
    \includegraphics[width=1\columnwidth]{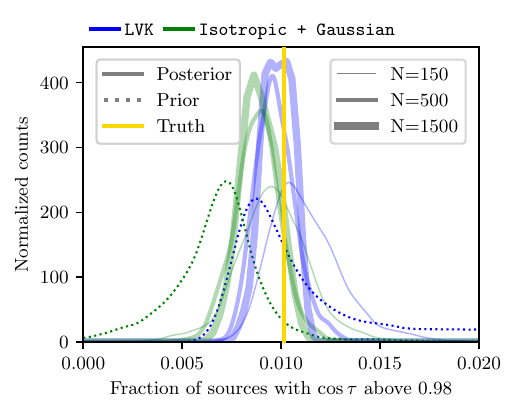}
    \caption{(Top) derived posterior on the fraction of sources with $\ct\leq 0$ for the \lvk and \isogaus models and three catalog sizes. The true value is indicated with a \truthcolor line. (Bottom) The same, but for the fraction of sources with $\ct\gtrsim0.98$ (i.e. within $10^\circ$ from perfect alignment). }
    \label{Fig.FractionCtBelow}
\end{figure}

We can recast our PPD into an integrated estimation of the fraction of tilts in the population that have \ct below (or above) any threshold. In Fig~\ref{Fig.FractionCtBelow} we show the fraction of BHs with $\ct\leq 0$ (top) and $\ct\gtrsim 0.98$ (bottom) -- that is, spin vectors closer than $10^\circ$ relative to perfect alignment -- for the \lvk and \isogaus models, with the \truthcolor lines marking the true values. We find that both models do well for both bounds. The true fraction of systems with negative tilts in the population is \truenegativectfrac; using the \lvk model we find \fractionBelowLvkOneFifty, \fractionBelowLvkFiveHundred and \fractionBelowLvkFifteenHundred with $N=150$, $500$, and $1500$ events, respectively. The \isogaus model yields \fractionBelowIsogaussOneFifty, \fractionBelowIsogaussFiveHundred, \fractionBelowIsogaussFifteenHundred.  The true fraction of sources with $\ct\gtrsim 0.98$ is \truepositivectfrac; using the \lvk model we find \fractionAboveLvkOneFifty, \fractionAboveLvkFiveHundred and \fractionAboveLvkFifteenHundred with $N=150$, $500$ and $1500$ events, respectively. The \isogaus model yields \fractionAboveIsogaussOneFifty, \fractionAboveIsogaussFiveHundred, \fractionAboveIsogaussFifteenHundred. 
For both bounds we find that uncertainty shrinks from by a factor of $\sim3-4$ as the number of events increases tenfold. Other values are tabulated in Tabs.~\ref{Tab.FractionsUncorr} and~\ref{Tab.FractionsCorr} in App.~\ref{App.Tables}.

\subsubsection{Uncorrelated models - \isobeta and \isotuk}\label{SubSec.UncorrModels_b}
\begin{figure}[t]
 \includegraphics[width=\columnwidth]{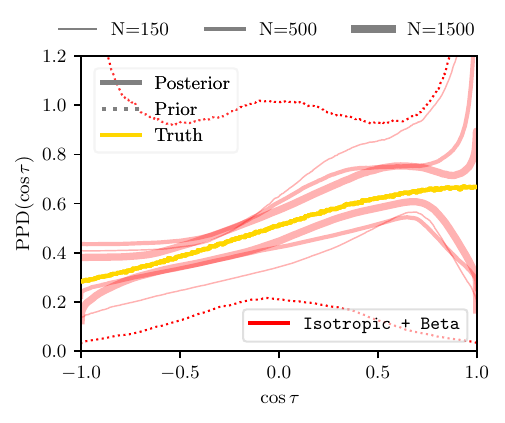}\\
 \includegraphics[width=\columnwidth]{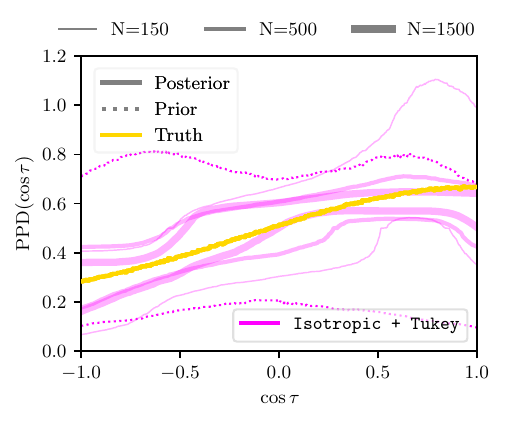}
    \caption{PPD for \ct for catalogs of $150$, $500$, $1500$ events using the uncorrelated \isobeta model (top) and \isotuk model (bottom).The thickness of the curves indicates the catalog size and each set of curves spans the 90\% credible interval. The \truthcolor curve is the true distribution.}
    \label{Fig.PPDUncorrelated_b}
\end{figure}

We now shift focus to the two other uncorrelated models included in this work, where the non-isotropic component is either a (possibly singular) beta distribution or a Tukey window. Those models have been introduced by Ref.~\cite{Vitale:2022dpa} and used to analyze \gwtcthree data. They are more generic than the \lvk model and its \isogaus extension in such that they can capture a more diverse set of morphologies, with non-Gaussian peaks or plateaus. We show the PPD for \ct obtained with both models in Fig.~\ref{Fig.PPDUncorrelated_b}. At $\ct\simeq 1$ the higher end of the 90\% credible interval for the \isobeta model is out of scale due to some hyper samples that result in a singular beta distribution. The number of these samples decreases as the catalog size increases. 
For the \isotuk model we find that the true is slightly outside of the 90\% credible interval for $\ct \gtrsim 0.7$, whilst it is included in the 90\% credible interval for all other cases. For $N=150$, the \isotuk model shows a peak in the upper 90\% credible interval, which is washed out as more events are added to the catalog.

\begin{figure}[t]
    \centering
    \includegraphics[width=\columnwidth]{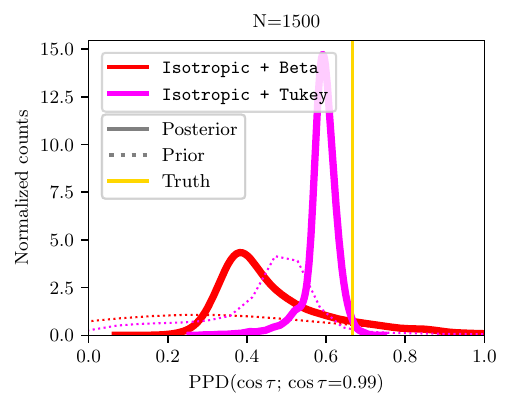}
    \caption{Slice of the PPD for \ct at $\ct=0.99$ for the catalog with 1500 sources. The truth is indicated with a \truthcolor line.}
    \label{Fig:IsoBetaIsoTukeySliceAt1}
\end{figure}

The fact that the \isotuk model slightly underestimates the value of the \ct PPD toward unity can be seen by plotting a slide of the PPD, as done for the \lvk and \isogaus models in Fig.~\ref{Fig:LVKIsoGaussSliceAt1}. We do this for the catalog comprising $1500$ sources and both \isogaus and \isotuk models in Fig.~\ref{Fig:IsoBetaIsoTukeySliceAt1}, where we actually slice at $\ct=0.99$ instead of $1.0$ to avoid numerical issues with the large values that singular beta posteriors can take in the last bin. The true value is in the right-hand side of the sliced PPD for both models.

\begin{figure}[t]
    \centering
    \includegraphics[width=1\columnwidth]{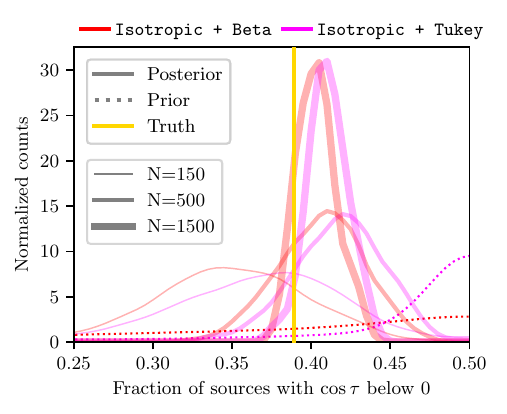}\\
    \includegraphics[width=1\columnwidth]{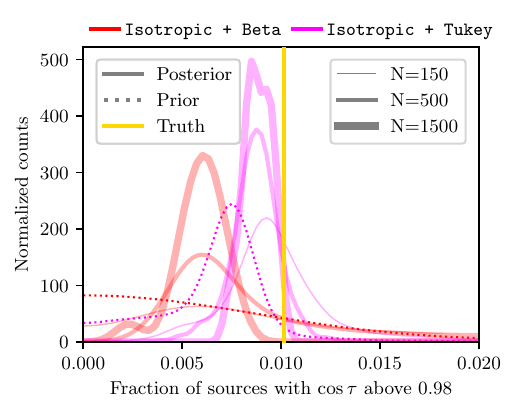}
    \caption{Same as Fig.~\ref{Fig.FractionCtBelow} but for the \isobeta and \isotuk models.}
    \label{Fig.FractionCtBelow_b}
\end{figure}

Relatedly, both models underestimate \fa, the fraction of tilts in the non-isotropic component. However, it's worth stressing again that because these two models are more elastic and do \emph{not} enforce a Gaussian peak, the possible morphologies of the non-isotropic component are much richer, and include anything from a cosine peak, to a broad plateau or a skewed peak. This implies that the same value of \fa across models can result in rather different PPDs for \ct.  This is another reason why is useful to work directly with quantities that are born off the PPD: they can be readily compared across models, irrespective of how they are parametrized (or even if they are non-parametric). We do this by looking again at the fraction of events with negative \ct and with \ct above $\sim0.98$ for both models, Fig.~\ref{Fig.FractionCtBelow_b}. In most cases, the true value are included within the 90\% credible intervals. Specifically, for the fraction of sources with $\ct\leq 0$ the \isobeta (\isotuk) model yields \fractionBelowIsoBetaOneFifty, \fractionBelowIsoBetaFiveHundred and \fractionBelowIsoBetaFifteenHundred (\fractionBelowIsoTukeyOneFifty, \fractionBelowIsoTukeyFiveHundred and \fractionBelowIsoTukeyFifteenHundred) for $N=150$, $500$ and $1500$ respectively, where the true value was \truenegativectfrac.
The fraction of systems with $\ct\gtrsim 0.98$ for the \isobeta (\isotuk) model is measured to be \fractionAboveIsoBetaOneFifty, \fractionAboveIsoBetaFiveHundred, \fractionAboveIsoBetaFifteenHundred (\fractionAboveIsoTukeyOneFifty, \fractionAboveIsoTukeyFiveHundred, \fractionAboveIsoTukeyFifteenHundred) for $N=150,500,1500$, respectively, where the true value was \truepositivectfrac.

\subsubsection{Correlated models}\label{SubSec.CorrModels}

In this section we focus on the results obtained when the branching fraction between the two components of the tilt models is allowed to vary with the mass ratio. This introduces another hyper parameter that we call $n$. The mass ratio-dependent branching fraction is thus parametrized by two quantities that are measured from the data, \fatone and \enn, which are combined as in Eq.~\ref{Eq.Correlatedfa} to give $\fa(q)$. Qualitatively speaking, \fatone is the fraction of sources in the non-isotropic component at $q=1$ and $n$ controls how quickly the branching ratio evolves as $q$ increases, with $n=0$ yielding a uniform branching ratio, and larger $n$ corresponding to a steeper increase of $f(q)$ as $q$ approaches 1.

\begin{figure}[t]
    \centering
    \includegraphics[width=\columnwidth]{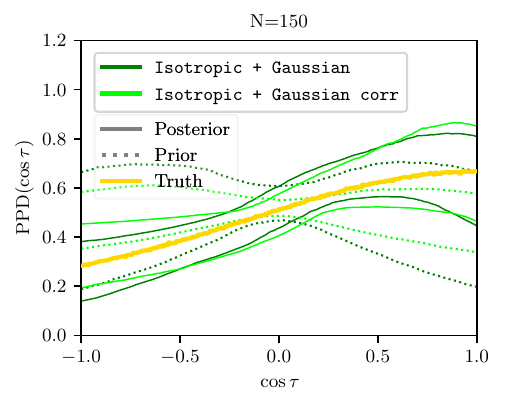}
    \caption{PPD for \ct for catalogs of $150$ events using the \isogaus model (green solid lines) and \isocgaus model (lime solid lines). The \truthcolor curve is the true distribution and the dotted lines are the priors.}
    \label{Fig.PPDCorrelated_a}
\end{figure}

We begin by anticipating that the results in this section are to a large extent the same as what presented in Secs.~\ref{SubSec.UncorrModels_a} and~\ref{SubSec.UncorrModels_b} because even for our largest catalogs it proves to be extremely challenging to measure this correlation. The main difference we observe is that the PPD for \ct obtained with the correlated models can be slightly more uncertain than what obtained with the corresponding non-correlated model even when the prior 90\% region is smaller, i.e., favoring more uniform distributions. For example, in Fig.~\ref{Fig.PPDCorrelated_a} we compare the PPD obtained with the \isocgaus model with that of the \isogaus, for the catalog with $150$ sources. 

\begin{figure}[t]
    \centering
    \includegraphics[width=\columnwidth]{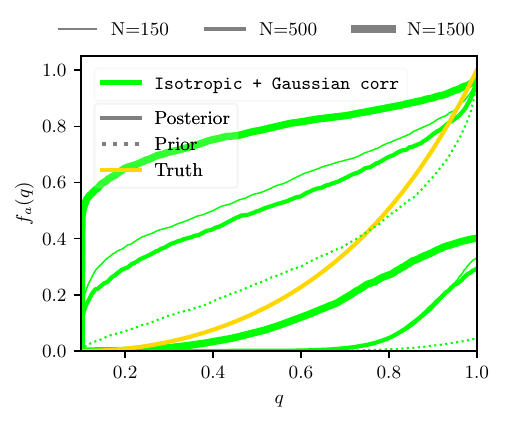}
    \caption{Posterior on the mass ratio-dependent branching ratio \faq obtained with the \isocgaus model. The green curves enclose the 90\% credible regions  while the \truthcolor line is the true branching ratio and the black dotted lines enclose the 90\% credible region of hyper prior draws.}
    \label{Fig:f_of_a_IsoGaussian_c}
\end{figure}

Therefore, in this section we will not show updated versions of the plots of Secs.~\ref{SubSec.UncorrModels_a} and~\ref{SubSec.UncorrModels_b} for the correlated models, as they will too look similar to justify the space they take. Instead, we will focus specifically on how well the mass ratio dependent branching fraction can be measured and whether the analysis can prove that there indeed is correlation in the data. In Fig.~\ref{Fig:f_of_a_IsoGaussian_c} we show the 90\% credible interval for the measurement of \faq obtained using the \isocgaus model (the \lvkc model yields virtually the same set of curves). The truth (\truthcolor solid curve) is included in the 90\% credible intervals except at the right edge, which is expected given that the $95^\mathrm{th}$ percentile must be smaller than the maximum, and the maximum cannot be larger than 1 for the branching ratio. 

Three main things are worth stressing. First, the measurement is somewhat informative, meaning that we do not simply get the hyper prior back. That is shown in the figure as two dotted black lines enclosing 90\% of the hyper prior. Second, the uncertainty bands are extremely wide, even for the largest catalog, highlighting the difficulty of measuring \faq even with a model that can perfectly match the truth. Third, the width of the uncertainty bands does not trivially shrink as the catalog size increases, especially at small $q$. We interpret this as a sign that not only the catalog size matters, but also which specific sources are included in the catalog. This can be understood because a good measurement of \faq at small $q$ requires the detection of sources with small $q$. Given that we are using a true mass-ratio distribution that favors equal-mass sources (compatibly with the LVK's measurement in \gwtcthree) there are not that many small mass ratio sources in our catalogs to start with. This, together with the fact that \fatone and \enn are heavily correlated with one another and partially correlated with the other spin magnitude and spin tilt parameters implies that the resulting posterior on \faq can appear to have this non-obvious progression with the number of sources. 

To verify that nothing is wrong with the model itself (meaning that we are not affected by a bug), we have verified that running the population code with the \isocgaus model fixing all parameters to their true values but either of \fatone or \enn in turn, returns a 1D posterior that includes the corresponding true value. Even in the unrealistic scenario captured by this test (all hyper parameters known but one) the measurements are noisy. For example, when only \enn is assumed unknown we get $\enn=1.30^{+5.96}_{-1.17}$, $\enn=5.00^{+6.10}_{-3.21}$ and $\enn=2.15^{+1.73}_{-1.04}$ for $N=150,500$ and $1500$ respectively. Only the largest catalog yields a 1D posterior that peaks at the true value, but still has a relative uncertainty of $\gtrsim 100\%$.
The two remaining models yield the same qualitative results, with the \isocbeta underestimating \faq for large $q$ in the smaller catalogs. 

We end this section by focusing on the remaining question we wanted to address: can the analyst even claim that this $q-\ct$ correlation exists in the detected population? Based on the fact that the PPDs of the correlated and uncorrelated models are so similar, it should not be surprising that the answer is not positive. Indeed, a constant, horizontal, \faq line can be drawn within the 90\% credible intervals of Fig.~\ref{Fig:f_of_a_IsoGaussian_c} for any catalog size (the sharp decline on the left is due to the prior).  In Tab.~\ref{Tab:model_selection} we show for all our runs the natural log Bayes factor between each model and the "correct" model, i.e. the \lvkc. We find that for all catalog sizes the simpler \lvk model that does not allow for the correlation is slightly preferred (though it should be remembered that the uncertainty on the Bayes factors we calculate with \texttt{Dynesty} with our settings is of the order of $\sim 0.4$).
The model with the largest number of hyper parameters, \isoctuk is consistently ranked last. The table highlights that none of the models is definitively ruled out, and that it is not the case that the models allowing for the correlation are ranked higher.

\begin{table}[h]
\centering
\begin{tabular}{lr}
\hline\hline
Model & ln Bayes Factor \\
\hline
\multicolumn{2}{c}{\textbf{150 events}} \\
\hline
\lvk                & $+0.5$ \\
\isogaus     & $-0.1$ \\
\isobeta        & $-0.1$ \\
\isocbeta    & $-0.4$ \\
\isotuk        & $-0.5$ \\
\isocgaus & $-0.7$ \\
\isoctuk   & $-0.7$ \\
\hline
\multicolumn{2}{c}{\textbf{500 events}} \\
\hline
\lvk                & $+0.6$ \\
\isogaus     & $0.0$ \\
\isobeta        & $-0.4$ \\
\isocbeta   & $-0.4$ \\
\isotuk        & $-0.4$ \\
\isocgaus & $-0.5$ \\
\isoctuk   & $-0.8$ \\
\hline
\multicolumn{2}{c}{\textbf{1500 events}} \\
\hline
\lvk                & $+1.1$ \\
\isogaus     & $+0.5$ \\
\isotuk        & $0.0$ \\
\isobeta         & $-0.1$ \\
\isocgaus & $-0.7$ \\
\isoctuk   & $-1.0$ \\
\isocbeta   & $-1.2$ \\
\hline\hline
\end{tabular}
\caption{For each group of runs with the same number of sources, the second column reports the natural log of the Bayes factor between the model given in the first column and the \lvkc model run with the corresponding number of events. A positive value favors the model in the row over the \lvkc. In each group, models are sorted by the Bayes factor.}
\label{Tab:model_selection}
\end{table}

\section{Conclusions}\label{Sec.Conclusions}

In this paper, we looked into the prospects for measuring the astrophysical distribution of spin tilts (i.e., the angle between the spin vectors and the binary orbital angular momentum) for the stellar-mass black holes in the binaries detected by LIGO, Virgo and KAGRA. This is known to be a challenging problem, owing to the uncertain event-by-event measurements of this quantity. We have generated a synthetic population of 1599 BBH whose masses, redshifts and spins are drawn from a true underlying population that is consistent with what the LVK has measured in \gwtcthree data. Specifically, the spin tilts for both black holes are drawn from a distribution that has two components: one that produces isotropically distributed spin vectors and one that produces spins vectors that are drawn from a Gaussian distribution centered at $\ct=1$ with a width of $1.15$. The branching ratio between these two components is mass-ratio dependent -- Fig.~\ref{Fig.Truefa} --- representing an hypothetical scenario in which isolated binaries form with more equal masses while dynamically-formed binaries tend to have unequal masses. 
We used this large dataset to tackle four main questions.

\textbf{Is a peak at $\ct\neq 1$ surprising?} - In Ref.~\cite{Vitale:2022dpa} we showed that \gwtcthree is consistent with the astrophysical \ct distribution having a peak \emph{away} from $\ct=1$, a result later corroborated by others. This is surprising since such a peak away from $1$ does not have any known astrophysical interpretation. We have formed 20 catalogs with a random collection of $69$ BBHs each, i.e., having the same size of \gwtcthree and analyzed them with the \isogaus model, where the mean of preferentially aligned \ct component is a free hyper parameter. We found that spurious peaks away from $\ct=1$ are not uncommon. We repeated this experiment with 20 catalogs of $150$ or $300$ sources, which are realistic catalog sizes at the end of O4a and O4b\footnote{When we performed the bulk of these analysis, the LVK had not yet announced that there would be an O4c.~\cite{LVK:ObservingPlan}}. We find that, even for the largest catalog sizes, peaks away from $1$ are not impossible, even though they become less common. Broad plateaus that span most of the positive \ct domain remain somewhat common. Indeed, it becomes much easier to at least constrain the mean of the preferentially aligned \ct component -- \mct -- to be positive. That is possible for only $10\%$ of the catalogs with $69$ sources but with $100\%$ for the catalogs with $300$ sources. 
We conclude that the results based on \gwtcthree alone do not necessarily imply that the astrophysical \ct distribution does not have a peak at $\ct=1$.

\emph{To the extent that our assumed population is representative of nature's, these results suggest that several hundred sources will be required before claims about the location of such peak, if one exists, can be made. }

\textbf{Can we measure the location of a peak?} - Next, we used our simulated sources to create even larger catalogs with $N=150$, $500$ or $1500$ sources. Each of these was analyzed using 4 different spin tilt models (from Ref.~\cite{Vitale:2022dpa}) that do \emph{not} allow for any correlations. We found that measuring individual fully marginalized hyper parameters is challenging, due to the non-trivial correlations between them, and can depict a landscape grimmer than what we are actually facing. These includes the hyper parameters that control the fraction of non-isotropic spins, and the location of the non-isotropic component. In this sense, revealing the presence of a peak at $\ct=1$ like the one we simulated remains extremely challenging. We also looked at the PPD of the spin tilts. The \isogaus and \isotuk models find very broad plateaus for the \ct distribution, spanning most of the positive \ct range even for the catalog with $1500$ sources. The \isobeta model does find a peak at $\ct\simeq 0.6$. However, all models yield PPDs consistent with the truth within 90\% credible intervals for all catalog sizes with the exception of the \isotuk model, that for $N=1500$ slightly underestimates the true \ct distribution at $\ct \gtrsim 0.6$. These results are partially driven by the fact that the true shape of the preferentially-aligned tilt component is quite broad (consistently with what found in \gwtcthree).

\emph{These findings highlight the inherent difficulty in conclusively identifying a peak at perfect alignment, unless the underlying astrophysical distribution is significantly narrower than the one we have modeled.}

\textbf{Can we measure precisely the fraction of events with negative tilt or nearly aligned spins?} - 
While finding the evidence of a peak might be elusive, all of the models we used did well in quantifying the fraction of systems with either negative tilts or with nearly aligned tilts. For example, using the \isogaus we measured the fraction of sources with negative \ct to be \fractionBelowIsogaussOneFifty with the catalog comprising $150$ sources. The true value for our population is \truenegativectfrac and calculating this quantity with prior draws\footnote{For the simple \isogaus model, this number can actually be calculated analytically as $50^{+13}_{-13}\%$.} one gets for the \isogaus model $50.0^{+15.1}_{-15.8}$, thus this is truly a data-driven measurement. For our largest catalog, one gets \fractionBelowIsogaussFifteenHundred, that is a reduction of the uncertainties by a factor of $\sim 3$ with a tenfold increase in the number of sources. All models measure this parameter well, with the true value contained inside the 90\% credible interval. 

Perhaps even more interestingly, given the apparent difficulty in revealing peaks in the PPD for \ct, is the measurement of the fraction of sources with tilts within a $10^\circ$ angle from alignment (i.e., with $\ct\gtrsim 0.98$). For our population, this is only 1\% of black holes. For the \lvk and \isogaus models the effective prior on this fraction peaks slightly below 1\% (Fig.~\ref{Fig.FractionCtBelow} bottom panel). For all catalog sizes the posterior shifts toward the truth, compared to the prior and shrinks as more sources are added. For the catalog with $1500$ sources the \isogaus model measures this fraction to be \fractionAboveIsogaussFifteenHundred. The more general \isobeta and \isotuk models do a similarly well for the fraction of negative tilts, with posteriors that include the truth and are significantly different from the respective priors. While the \isotuk model does well even in measuring the fraction of sources with nearly aligned spin, the \isobeta model underestimate that fraction. The catalog with $150$ sources yields a measurement that is barely different from the prior. The precision increases with the number of sources, at the expense of the accuracy: for the largest catalog the truth is just outside the 90\% credible interval. 

\emph{Overall, these results suggest that integral quantities---such as the fraction of sources within a particular parameter range---are both less model dependent and have lower uncertainties than the marginal posteriors of model parameters, even when the two seem to have similar meanings.}

\textbf{Can we reveal a correlation between tilt angles and mass ratio?} - Given the large number of BBH sources simulated for this work, we thought it would be worth to not only assess the measurability of the spin tilt distribution but also whether an underlying astrophysical correlation can be revealed in the data. Given the \gwtcthree median merger rate estimate, even at O5 sensitivity it will take years to collect $1500$ sources~\cite{Kiendrebeogo:2023hzf}. We therefore are probing what could be done toward the end of this decade. Sadly, the answer is disappointing. At least with the correlation and population we have assumed, we find that even the largest catalogs cannot measure a mass ratio-dependent branching fraction \faq that is definitively non-constant, i.e. correlated. The PPD on \ct obtained with models that allow for this correlation are not very different from the corresponding measurements performed with models that do not include such correlation. The fact that the data is not very informative about the correlation in our population was confirmed calculating Bayes factors between each model (with or without correlation) and the \lvkc model, which is the ``true'' one that was used to simulate the sources. We find that for all catalog sizes, the simpler \lvk model that does not include correlations is favored, with natural log of the Bayes factor between $0.5$ and $1.0$ relative to the \lvkc model. All other models have either Bayes factors close to zero or are slightly disfavored relative to the \lvkc model. We notice that given our sampler settings, Bayes factors whose absolute values is smaller than $\sim 0.4$ should not be considered significant.

\emph{These results indicate that confidently revealing subtle correlations between spin tilt angles and mass ratios---like the one we simulated---from GW observations will likely require substantially larger datasets, beyond what advanced detectors alone will provide this decade (but see caveats below).}

\vskip 0.5cm
\textbf{Outlook} - Is it possible our choices for the simulated BBH population led to results there are too pessimistic? Here we lay down a series of caveats that might affect how the informed reader will think of this question.  
A peak at $\ct=1$ would likely be easier to reveal than what we reported here under a few scenarios: a) if there is a preferentially aligned component with a peak significantly narrower (e.g., a small \sct in our \isogaus and \lvk models) than what we simulated; b) if the true distribution of spin \emph{magnitude} were to produce more high-spin sources than what simulated here, since the tilt of large spins is usually easier to measure than that of small spins~\cite{Vitale:2016avz}; c) if the true distribution of mass ratios were to produce more often unequal mass black holes than what assumed here, since the primary spins of black holes in large mass ratio systems are usually easier to measure~\footnote{This happens because for large mass ratios, the primary spin is more and more similar to $\chi_{\mathrm{eff}}$~\cite{Damour:2001tu,Racine:2008qv,Santamaria:2010yb} which is the best measured spin parameter~\cite{LIGOScientific:2021djp}}. Any combination of these three factors would likely improve the measurement of the \ct distribution, and in particular the existence of peak at nearly aligned spins. 

However, it is important to admit that is possible that our simulated population is in fact more generous than nature's: we have assumed that systems with comparable masses (i.e. the majority) $\fatone=1$, while in reality the fraction of systems in the nearly-aligned component might be much smaller, making the measurement of the properties of that component more challenging. 
It is also entirely possible that the true spin magnitude distribution produces fewer systems with medium or large spin magnitudes, leaving the tilt measurements {comfortably numb} to subtle features.

It is also worth discussing caveats associated with inference models and waveforms. 
We included in our suites of models two that can perfectly well match the truth (\lvkc and \isocgaus). We have done that to assess possible best-case scenarios that set a lower limit for the statistical and systematic uncertainties arising from the modeling of the populations. In reality, it is very unlikely any one of the parametric models currently used is a good representation of nature. Non-parametric models can be used, in either one or multiple dimensions, but the mitigated risk of systematics comes at the expense of increased statistical uncertainties, implying that even more sources than what we have considered would be required to reach a comparable level of statistical uncertainty. Regardless, it is noteworthy that, despite potential challenges from likelihood approximations~\cite{2019RNAAS...3...66F,Golomb:2022bon} and models' systematics \cite{Miller:2024sui} we are always able to correctly infer the underlying BBH distributions within uncertainties for all our models and parameters (App.~\ref{App.PPD}).

Everywhere in our analysis we used the same waveform family - to simulate the BBH added into synthetic noise, to analyze their properties and to evaluate the sensitivity of the detectors. 
This has removed the risk of systematic errors in a way that is probably not representative of what would happen when analyzing real data~\cite{Dhani:2024jja}. Additionally, we have not used waveforms with higher order modes, which could improve the measurements of some parameters, for heavy systems~\cite{Varma:2019csw}. Furthermore, there is some evidence that numerical relativity surrogates~\cite{Varma:2018mmi} might yield more precise spin posteriors - at least for certain sources~\cite{Islam:2023zzj}. Should that be the case for enough sources, it might lead to more stringent constraints of the spin population parameters. 

As advanced ground based detectors get more sensitive in the next few years, hundreds and then thousands of binary black holes will be detected. Next-generation detectors will reveal hundreds \emph{of} thousands of these sources annually. We emphasize that accurate measurements of integrated properties of the astrophysical distribution of spin tilts are at reach already in the advanced detector era, though uncertainties are likely to remain larger than $10\%$. We encourage the population synthesis community to produce and quote these integrated numbers (e.g., the fraction of tilts above or below certain thresholds), 
wherever possible.
Obtaining percent-level measurements or revealing more subtle features such as correlation between tilts and other parameters might have to wait for next-generation detectors.
Eventually, {under pressure} from larger datasets, the balance will tilt toward a more detailed understanding of the origins of merging black hole binaries. {One fine day we shall see} clearly the details, as well as {the big picture}. 
\vspace{1em}
    \acknowledgments
    
S.V. is partially supported by NSF through the grant PHY-2045740.  
M.M. is supported by the LIGO Laboratory through the National Science Foundation awards PHY-1764464 and PHY-2309200.
The authors are grateful for computational resources provided by subMIT at MIT Physics and the LIGO Laboratory supported by National Science Foundation Grants PHY-0757058 and PHY-0823459. This material is based upon work supported by NSF's LIGO Laboratory which is a major facility fully funded by the National Science Foundation. We thank the Institute for Nuclear Theory at the University of Washington for its kind hospitality and stimulating research environment. This research was supported in part by the INT's U.S. Department of Energy grant No. DE-FG02- 00ER41132.
We thank Sofia Alvarez Lopez, Cristopher Berry, Sylvia Biscoveanu, Tom Callister, Tom Dent, Jack Heinzel, Simona Miller, Cailin Plunkett, Colm Talbot and Noah Wolfe for useful comments and feedback on this work. We thanks Davide Gerosa and our other colleagues in the SPINS: our spins might be non-aligned but our interests are.

The hyper posteriors samples produced for this work will be available in Zenodo~\cite{vitale_data_2025} at the time of publication.
\vskip 0.3cm
S.V. would like to dedicate this work to the memory of Aldo Francesco Vitale.

\bibliography{tilt}

\begin{thebibliography}{102}%
\makeatletter
\providecommand \@ifxundefined [1]{%
 \@ifx{#1\undefined}
}%
\providecommand \@ifnum [1]{%
 \ifnum #1\expandafter \@firstoftwo
 \else \expandafter \@secondoftwo
 \fi
}%
\providecommand \@ifx [1]{%
 \ifx #1\expandafter \@firstoftwo
 \else \expandafter \@secondoftwo
 \fi
}%
\providecommand \natexlab [1]{#1}%
\providecommand \enquote  [1]{``#1''}%
\providecommand \bibnamefont  [1]{#1}%
\providecommand \bibfnamefont [1]{#1}%
\providecommand \citenamefont [1]{#1}%
\providecommand \href@noop [0]{\@secondoftwo}%
\providecommand \href [0]{\begingroup \@sanitize@url \@href}%
\providecommand \@href[1]{\@@startlink{#1}\@@href}%
\providecommand \@@href[1]{\endgroup#1\@@endlink}%
\providecommand \@sanitize@url [0]{\catcode `\\12\catcode `\$12\catcode
  `\&12\catcode `\#12\catcode `\^12\catcode `\_12\catcode `\%12\relax}%
\providecommand \@@startlink[1]{}%
\providecommand \@@endlink[0]{}%
\providecommand \url  [0]{\begingroup\@sanitize@url \@url }%
\providecommand \@url [1]{\endgroup\@href {#1}{\urlprefix }}%
\providecommand \urlprefix  [0]{URL }%
\providecommand \Eprint [0]{\href }%
\providecommand \doibase [0]{https://doi.org/}%
\providecommand \selectlanguage [0]{\@gobble}%
\providecommand \bibinfo  [0]{\@secondoftwo}%
\providecommand \bibfield  [0]{\@secondoftwo}%
\providecommand \translation [1]{[#1]}%
\providecommand \BibitemOpen [0]{}%
\providecommand \bibitemStop [0]{}%
\providecommand \bibitemNoStop [0]{.\EOS\space}%
\providecommand \EOS [0]{\spacefactor3000\relax}%
\providecommand \BibitemShut  [1]{\csname bibitem#1\endcsname}%
\let\auto@bib@innerbib\@empty
\bibitem [{\citenamefont {Abbott}\ \emph {et~al.}(2016)\citenamefont {Abbott}
  \emph {et~al.}}]{LIGOScientific:2016aoc}%
  \BibitemOpen
  \bibfield  {author} {\bibinfo {author} {\bibfnamefont {B.~P.}\ \bibnamefont
  {Abbott}} \emph {et~al.} (\bibinfo {collaboration} {LIGO Scientific,
  Virgo}),\ }\href {https://doi.org/10.1103/PhysRevLett.116.061102} {\bibfield
  {journal} {\bibinfo  {journal} {Phys. Rev. Lett.}\ }\textbf {\bibinfo
  {volume} {116}},\ \bibinfo {pages} {061102} (\bibinfo {year} {2016})},\
  \Eprint {https://arxiv.org/abs/1602.03837} {arXiv:1602.03837 [gr-qc]}
  \BibitemShut {NoStop}%
\bibitem [{\citenamefont {Aasi}\ \emph {et~al.}(2015)\citenamefont {Aasi} \emph
  {et~al.}}]{TheLIGOScientific:2014jea}%
  \BibitemOpen
  \bibfield  {author} {\bibinfo {author} {\bibfnamefont {J.}~\bibnamefont
  {Aasi}} \emph {et~al.} (\bibinfo {collaboration} {LIGO Scientific}),\ }\href
  {https://doi.org/10.1088/0264-9381/32/7/074001} {\bibfield  {journal}
  {\bibinfo  {journal} {Class. Quant. Grav.}\ }\textbf {\bibinfo {volume}
  {32}},\ \bibinfo {pages} {074001} (\bibinfo {year} {2015})},\ \Eprint
  {https://arxiv.org/abs/1411.4547} {arXiv:1411.4547 [gr-qc]} \BibitemShut
  {NoStop}%
\bibitem [{\citenamefont {Acernese}\ \emph {et~al.}(2015)\citenamefont
  {Acernese} \emph {et~al.}}]{TheVirgo:2014hva}%
  \BibitemOpen
  \bibfield  {author} {\bibinfo {author} {\bibfnamefont {F.}~\bibnamefont
  {Acernese}} \emph {et~al.} (\bibinfo {collaboration} {VIRGO}),\ }\href
  {https://doi.org/10.1088/0264-9381/32/2/024001} {\bibfield  {journal}
  {\bibinfo  {journal} {Class. Quant. Grav.}\ }\textbf {\bibinfo {volume}
  {32}},\ \bibinfo {pages} {024001} (\bibinfo {year} {2015})},\ \Eprint
  {https://arxiv.org/abs/1408.3978} {arXiv:1408.3978 [gr-qc]} \BibitemShut
  {NoStop}%
\bibitem [{\citenamefont {Aso}\ \emph {et~al.}(2013)\citenamefont {Aso},
  \citenamefont {Michimura}, \citenamefont {Somiya}, \citenamefont {Ando},
  \citenamefont {Miyakawa}, \citenamefont {Sekiguchi}, \citenamefont
  {Tatsumi},\ and\ \citenamefont {Yamamoto}}]{Aso:2013eba}%
  \BibitemOpen
  \bibfield  {author} {\bibinfo {author} {\bibfnamefont {Y.}~\bibnamefont
  {Aso}}, \bibinfo {author} {\bibfnamefont {Y.}~\bibnamefont {Michimura}},
  \bibinfo {author} {\bibfnamefont {K.}~\bibnamefont {Somiya}}, \bibinfo
  {author} {\bibfnamefont {M.}~\bibnamefont {Ando}}, \bibinfo {author}
  {\bibfnamefont {O.}~\bibnamefont {Miyakawa}}, \bibinfo {author}
  {\bibfnamefont {T.}~\bibnamefont {Sekiguchi}}, \bibinfo {author}
  {\bibfnamefont {D.}~\bibnamefont {Tatsumi}},\ and\ \bibinfo {author}
  {\bibfnamefont {H.}~\bibnamefont {Yamamoto}} (\bibinfo {collaboration}
  {KAGRA}),\ }\href {https://doi.org/10.1103/PhysRevD.88.043007} {\bibfield
  {journal} {\bibinfo  {journal} {Phys. Rev. D}\ }\textbf {\bibinfo {volume}
  {88}},\ \bibinfo {pages} {043007} (\bibinfo {year} {2013})},\ \Eprint
  {https://arxiv.org/abs/1306.6747} {arXiv:1306.6747 [gr-qc]} \BibitemShut
  {NoStop}%
\bibitem [{\citenamefont {Abbott}\ \emph {et~al.}(2023)\citenamefont {Abbott}
  \emph {et~al.}}]{KAGRA:2021vkt}%
  \BibitemOpen
  \bibfield  {author} {\bibinfo {author} {\bibfnamefont {R.}~\bibnamefont
  {Abbott}} \emph {et~al.} (\bibinfo {collaboration} {KAGRA, VIRGO, LIGO
  Scientific}),\ }\href {https://doi.org/10.1103/PhysRevX.13.041039} {\bibfield
   {journal} {\bibinfo  {journal} {Phys. Rev. X}\ }\textbf {\bibinfo {volume}
  {13}},\ \bibinfo {pages} {041039} (\bibinfo {year} {2023})},\ \Eprint
  {https://arxiv.org/abs/2111.03606} {arXiv:2111.03606 [gr-qc]} \BibitemShut
  {NoStop}%
\bibitem [{\citenamefont {Nitz}\ \emph {et~al.}(2021)\citenamefont {Nitz},
  \citenamefont {Capano}, \citenamefont {Kumar}, \citenamefont {Wang},
  \citenamefont {Kastha}, \citenamefont {Sch\"afer}, \citenamefont
  {Dhurkunde},\ and\ \citenamefont {Cabero}}]{Nitz:2021uxj}%
  \BibitemOpen
  \bibfield  {author} {\bibinfo {author} {\bibfnamefont {A.~H.}\ \bibnamefont
  {Nitz}}, \bibinfo {author} {\bibfnamefont {C.~D.}\ \bibnamefont {Capano}},
  \bibinfo {author} {\bibfnamefont {S.}~\bibnamefont {Kumar}}, \bibinfo
  {author} {\bibfnamefont {Y.-F.}\ \bibnamefont {Wang}}, \bibinfo {author}
  {\bibfnamefont {S.}~\bibnamefont {Kastha}}, \bibinfo {author} {\bibfnamefont
  {M.}~\bibnamefont {Sch\"afer}}, \bibinfo {author} {\bibfnamefont
  {R.}~\bibnamefont {Dhurkunde}},\ and\ \bibinfo {author} {\bibfnamefont
  {M.}~\bibnamefont {Cabero}},\ }\href
  {https://doi.org/10.3847/1538-4357/ac1c03} {\bibfield  {journal} {\bibinfo
  {journal} {Astrophys. J.}\ }\textbf {\bibinfo {volume} {922}},\ \bibinfo
  {pages} {76} (\bibinfo {year} {2021})},\ \Eprint
  {https://arxiv.org/abs/2105.09151} {arXiv:2105.09151 [astro-ph.HE]}
  \BibitemShut {NoStop}%
\bibitem [{\citenamefont {{Olsen}}\ \emph {et~al.}(2022)\citenamefont
  {{Olsen}}, \citenamefont {{Venumadhav}}, \citenamefont {{Mushkin}},
  \citenamefont {{Roulet}}, \citenamefont {{Zackay}},\ and\ \citenamefont
  {{Zaldarriaga}}}]{Olsen:2022pin}%
  \BibitemOpen
  \bibfield  {author} {\bibinfo {author} {\bibfnamefont {S.}~\bibnamefont
  {{Olsen}}}, \bibinfo {author} {\bibfnamefont {T.}~\bibnamefont
  {{Venumadhav}}}, \bibinfo {author} {\bibfnamefont {J.}~\bibnamefont
  {{Mushkin}}}, \bibinfo {author} {\bibfnamefont {J.}~\bibnamefont {{Roulet}}},
  \bibinfo {author} {\bibfnamefont {B.}~\bibnamefont {{Zackay}}},\ and\
  \bibinfo {author} {\bibfnamefont {M.}~\bibnamefont {{Zaldarriaga}}},\ }\href
  {https://doi.org/10.1103/PhysRevD.106.043009} {\bibfield  {journal} {\bibinfo
   {journal} {\prd}\ }\textbf {\bibinfo {volume} {106}},\ \bibinfo {eid}
  {043009} (\bibinfo {year} {2022})},\ \Eprint
  {https://arxiv.org/abs/2201.02252} {arXiv:2201.02252 [astro-ph.HE]}
  \BibitemShut {NoStop}%
\bibitem [{\citenamefont {{LIGO, Virgo, and KAGRA
  Collaborations}}(2024)}]{LVK:gracedbO4}%
  \BibitemOpen
  \bibfield  {author} {\bibinfo {author} {\bibnamefont {{LIGO, Virgo, and KAGRA
  Collaborations}}},\ }\href@noop {} {\bibinfo {title} {{Public alerts during
  the O4 observing run}}},\ \bibinfo {howpublished}
  {\url{https://gracedb.ligo.org/superevents/public/O4/}} (\bibinfo {year}
  {2024})\BibitemShut {NoStop}%
\bibitem [{\citenamefont {{LIGO, Virgo, and KAGRA
  Collaborations}}(2025)}]{LVK:ObservingPlan}%
  \BibitemOpen
  \bibfield  {author} {\bibinfo {author} {\bibnamefont {{LIGO, Virgo, and KAGRA
  Collaborations}}},\ }\href@noop {} {\bibinfo {title} {{LIGO-Virgo-KAGRA
  Observing Plans}}},\ \bibinfo {howpublished}
  {\url{https://observing.docs.ligo.org/plan/}} (\bibinfo {year}
  {2025})\BibitemShut {NoStop}%
\bibitem [{\citenamefont {Mapelli}(2021)}]{Mapelli:2021taw}%
  \BibitemOpen
  \bibfield  {author} {\bibinfo {author} {\bibfnamefont {M.}~\bibnamefont
  {Mapelli}},\ }\bibinfo {title} {{Formation Channels of Single and Binary
  Stellar-Mass Black Holes}}\ (\bibinfo {year} {2021})\ \Eprint
  {https://arxiv.org/abs/2106.00699} {arXiv:2106.00699 [astro-ph.HE]}
  \BibitemShut {NoStop}%
\bibitem [{\citenamefont {Callister}(2024)}]{Callister:2024cdx}%
  \BibitemOpen
  \bibfield  {author} {\bibinfo {author} {\bibfnamefont {T.~A.}\ \bibnamefont
  {Callister}},\ }\href@noop {} {\bibfield  {journal} {\bibinfo  {journal}
  {Arxiv}\ } (\bibinfo {year} {2024})},\ \Eprint
  {https://arxiv.org/abs/2410.19145} {arXiv:2410.19145 [astro-ph.HE]}
  \BibitemShut {NoStop}%
\bibitem [{\citenamefont {Favata}\ \emph {et~al.}(2022)\citenamefont {Favata},
  \citenamefont {Kim}, \citenamefont {Arun}, \citenamefont {Kim},\ and\
  \citenamefont {Lee}}]{Favata:2021vhw}%
  \BibitemOpen
  \bibfield  {author} {\bibinfo {author} {\bibfnamefont {M.}~\bibnamefont
  {Favata}}, \bibinfo {author} {\bibfnamefont {C.}~\bibnamefont {Kim}},
  \bibinfo {author} {\bibfnamefont {K.~G.}\ \bibnamefont {Arun}}, \bibinfo
  {author} {\bibfnamefont {J.}~\bibnamefont {Kim}},\ and\ \bibinfo {author}
  {\bibfnamefont {H.~W.}\ \bibnamefont {Lee}},\ }\href
  {https://doi.org/10.1103/PhysRevD.105.023003} {\bibfield  {journal} {\bibinfo
   {journal} {Phys. Rev. D}\ }\textbf {\bibinfo {volume} {105}},\ \bibinfo
  {pages} {023003} (\bibinfo {year} {2022})},\ \Eprint
  {https://arxiv.org/abs/2108.05861} {arXiv:2108.05861 [gr-qc]} \BibitemShut
  {NoStop}%
\bibitem [{\citenamefont {Belczynski}\ \emph {et~al.}(2008)\citenamefont
  {Belczynski}, \citenamefont {Kalogera}, \citenamefont {Rasio}, \citenamefont
  {Taam}, \citenamefont {Zezas}, \citenamefont {Bulik}, \citenamefont
  {Maccarone},\ and\ \citenamefont {Ivanova}}]{Belczynski:2005mr}%
  \BibitemOpen
  \bibfield  {author} {\bibinfo {author} {\bibfnamefont {K.}~\bibnamefont
  {Belczynski}}, \bibinfo {author} {\bibfnamefont {V.}~\bibnamefont
  {Kalogera}}, \bibinfo {author} {\bibfnamefont {F.~A.}\ \bibnamefont {Rasio}},
  \bibinfo {author} {\bibfnamefont {R.~E.}\ \bibnamefont {Taam}}, \bibinfo
  {author} {\bibfnamefont {A.}~\bibnamefont {Zezas}}, \bibinfo {author}
  {\bibfnamefont {T.}~\bibnamefont {Bulik}}, \bibinfo {author} {\bibfnamefont
  {T.~J.}\ \bibnamefont {Maccarone}},\ and\ \bibinfo {author} {\bibfnamefont
  {N.}~\bibnamefont {Ivanova}},\ }\href {https://doi.org/10.1086/521026}
  {\bibfield  {journal} {\bibinfo  {journal} {Astrophys. J. Suppl.}\ }\textbf
  {\bibinfo {volume} {174}},\ \bibinfo {pages} {223} (\bibinfo {year}
  {2008})},\ \Eprint {https://arxiv.org/abs/astro-ph/0511811}
  {arXiv:astro-ph/0511811} \BibitemShut {NoStop}%
\bibitem [{\citenamefont {{Eldridge}}\ \emph {et~al.}(2017)\citenamefont
  {{Eldridge}}, \citenamefont {{Stanway}}, \citenamefont {{Xiao}},
  \citenamefont {{McClelland}}, \citenamefont {{Taylor}}, \citenamefont {{Ng}},
  \citenamefont {{Greis}},\ and\ \citenamefont {{Bray}}}]{2017PASA...34...58E}%
  \BibitemOpen
  \bibfield  {author} {\bibinfo {author} {\bibfnamefont {J.~J.}\ \bibnamefont
  {{Eldridge}}}, \bibinfo {author} {\bibfnamefont {E.~R.}\ \bibnamefont
  {{Stanway}}}, \bibinfo {author} {\bibfnamefont {L.}~\bibnamefont {{Xiao}}},
  \bibinfo {author} {\bibfnamefont {L.~A.~S.}\ \bibnamefont {{McClelland}}},
  \bibinfo {author} {\bibfnamefont {G.}~\bibnamefont {{Taylor}}}, \bibinfo
  {author} {\bibfnamefont {M.}~\bibnamefont {{Ng}}}, \bibinfo {author}
  {\bibfnamefont {S.~M.~L.}\ \bibnamefont {{Greis}}},\ and\ \bibinfo {author}
  {\bibfnamefont {J.~C.}\ \bibnamefont {{Bray}}},\ }\href
  {https://doi.org/10.1017/pasa.2017.51} {\bibfield  {journal} {\bibinfo
  {journal} {Publications of the Astronomical Society of Australia}\ }\textbf
  {\bibinfo {volume} {34}},\ \bibinfo {eid} {e058} (\bibinfo {year} {2017})},\
  \Eprint {https://arxiv.org/abs/1710.02154} {arXiv:1710.02154 [astro-ph.SR]}
  \BibitemShut {NoStop}%
\bibitem [{\citenamefont {{Fragos}}\ \emph {et~al.}(2023)\citenamefont
  {{Fragos}}, \citenamefont {{Andrews}}, \citenamefont {{Bavera}},
  \citenamefont {{Berry}}, \citenamefont {{Coughlin}}, \citenamefont
  {{Dotter}}, \citenamefont {{Giri}}, \citenamefont {{Kalogera}}, \citenamefont
  {{Katsaggelos}}, \citenamefont {{Kovlakas}}, \citenamefont {{Lalvani}},
  \citenamefont {{Misra}}, \citenamefont {{Srivastava}}, \citenamefont {{Qin}},
  \citenamefont {{Rocha}}, \citenamefont {{Rom{\'a}n-Garza}}, \citenamefont
  {{Serra}}, \citenamefont {{Stahle}}, \citenamefont {{Sun}}, \citenamefont
  {{Teng}}, \citenamefont {{Trajcevski}}, \citenamefont {{Tran}}, \citenamefont
  {{Xing}}, \citenamefont {{Zapartas}},\ and\ \citenamefont
  {{Zevin}}}]{2023ApJS..264...45F}%
  \BibitemOpen
  \bibfield  {author} {\bibinfo {author} {\bibfnamefont {T.}~\bibnamefont
  {{Fragos}}}, \bibinfo {author} {\bibfnamefont {J.~J.}\ \bibnamefont
  {{Andrews}}}, \bibinfo {author} {\bibfnamefont {S.~S.}\ \bibnamefont
  {{Bavera}}}, \bibinfo {author} {\bibfnamefont {C.~P.~L.}\ \bibnamefont
  {{Berry}}}, \bibinfo {author} {\bibfnamefont {S.}~\bibnamefont {{Coughlin}}},
  \bibinfo {author} {\bibfnamefont {A.}~\bibnamefont {{Dotter}}}, \bibinfo
  {author} {\bibfnamefont {P.}~\bibnamefont {{Giri}}}, \bibinfo {author}
  {\bibfnamefont {V.}~\bibnamefont {{Kalogera}}}, \bibinfo {author}
  {\bibfnamefont {A.}~\bibnamefont {{Katsaggelos}}}, \bibinfo {author}
  {\bibfnamefont {K.}~\bibnamefont {{Kovlakas}}}, \bibinfo {author}
  {\bibfnamefont {S.}~\bibnamefont {{Lalvani}}}, \bibinfo {author}
  {\bibfnamefont {D.}~\bibnamefont {{Misra}}}, \bibinfo {author} {\bibfnamefont
  {P.~M.}\ \bibnamefont {{Srivastava}}}, \bibinfo {author} {\bibfnamefont
  {Y.}~\bibnamefont {{Qin}}}, \bibinfo {author} {\bibfnamefont {K.~A.}\
  \bibnamefont {{Rocha}}}, \bibinfo {author} {\bibfnamefont {J.}~\bibnamefont
  {{Rom{\'a}n-Garza}}}, \bibinfo {author} {\bibfnamefont {J.~G.}\ \bibnamefont
  {{Serra}}}, \bibinfo {author} {\bibfnamefont {P.}~\bibnamefont {{Stahle}}},
  \bibinfo {author} {\bibfnamefont {M.}~\bibnamefont {{Sun}}}, \bibinfo
  {author} {\bibfnamefont {X.}~\bibnamefont {{Teng}}}, \bibinfo {author}
  {\bibfnamefont {G.}~\bibnamefont {{Trajcevski}}}, \bibinfo {author}
  {\bibfnamefont {N.~H.}\ \bibnamefont {{Tran}}}, \bibinfo {author}
  {\bibfnamefont {Z.}~\bibnamefont {{Xing}}}, \bibinfo {author} {\bibfnamefont
  {E.}~\bibnamefont {{Zapartas}}},\ and\ \bibinfo {author} {\bibfnamefont
  {M.}~\bibnamefont {{Zevin}}},\ }\href
  {https://doi.org/10.3847/1538-4365/ac90c1} {\bibfield  {journal} {\bibinfo
  {journal} {Astrophysical Journal Supplement Series}\ }\textbf {\bibinfo
  {volume} {264}},\ \bibinfo {eid} {45} (\bibinfo {year} {2023})},\ \Eprint
  {https://arxiv.org/abs/2202.05892} {arXiv:2202.05892 [astro-ph.SR]}
  \BibitemShut {NoStop}%
\bibitem [{\citenamefont {Riley}\ \emph {et~al.}(2022)\citenamefont {Riley}
  \emph {et~al.}}]{COMPASTeam:2021tbl}%
  \BibitemOpen
  \bibfield  {author} {\bibinfo {author} {\bibfnamefont {J.}~\bibnamefont
  {Riley}} \emph {et~al.} (\bibinfo {collaboration} {COMPAS Team, Team
  COMPAS}),\ }\href {https://doi.org/10.3847/1538-4365/ac416c} {\bibfield
  {journal} {\bibinfo  {journal} {Astrophys. J. Supp.}\ }\textbf {\bibinfo
  {volume} {258}},\ \bibinfo {pages} {34} (\bibinfo {year} {2022})},\ \Eprint
  {https://arxiv.org/abs/2109.10352} {arXiv:2109.10352 [astro-ph.IM]}
  \BibitemShut {NoStop}%
\bibitem [{\citenamefont {Spera}\ \emph {et~al.}(2019)\citenamefont {Spera},
  \citenamefont {Mapelli}, \citenamefont {Giacobbo}, \citenamefont {Trani},
  \citenamefont {Bressan},\ and\ \citenamefont {Costa}}]{Spera:2018wnw}%
  \BibitemOpen
  \bibfield  {author} {\bibinfo {author} {\bibfnamefont {M.}~\bibnamefont
  {Spera}}, \bibinfo {author} {\bibfnamefont {M.}~\bibnamefont {Mapelli}},
  \bibinfo {author} {\bibfnamefont {N.}~\bibnamefont {Giacobbo}}, \bibinfo
  {author} {\bibfnamefont {A.~A.}\ \bibnamefont {Trani}}, \bibinfo {author}
  {\bibfnamefont {A.}~\bibnamefont {Bressan}},\ and\ \bibinfo {author}
  {\bibfnamefont {G.}~\bibnamefont {Costa}},\ }\href
  {https://doi.org/10.1093/mnras/stz359} {\bibfield  {journal} {\bibinfo
  {journal} {Mon. Not. Roy. Astron. Soc.}\ }\textbf {\bibinfo {volume} {485}},\
  \bibinfo {pages} {889} (\bibinfo {year} {2019})},\ \Eprint
  {https://arxiv.org/abs/1809.04605} {arXiv:1809.04605 [astro-ph.HE]}
  \BibitemShut {NoStop}%
\bibitem [{\citenamefont {{Rodriguez}}\ \emph {et~al.}(2022)\citenamefont
  {{Rodriguez}}, \citenamefont {{Weatherford}}, \citenamefont {{Coughlin}},
  \citenamefont {{Amaro-Seoane}}, \citenamefont {{Breivik}}, \citenamefont
  {{Chatterjee}}, \citenamefont {{Fragione}}, \citenamefont
  {{K{\i}ro{\u{g}}lu}}, \citenamefont {{Kremer}}, \citenamefont {{Rui}},
  \citenamefont {{Ye}}, \citenamefont {{Zevin}},\ and\ \citenamefont
  {{Rasio}}}]{2022ApJS..258...22R}%
  \BibitemOpen
  \bibfield  {author} {\bibinfo {author} {\bibfnamefont {C.~L.}\ \bibnamefont
  {{Rodriguez}}}, \bibinfo {author} {\bibfnamefont {N.~C.}\ \bibnamefont
  {{Weatherford}}}, \bibinfo {author} {\bibfnamefont {S.~C.}\ \bibnamefont
  {{Coughlin}}}, \bibinfo {author} {\bibfnamefont {P.}~\bibnamefont
  {{Amaro-Seoane}}}, \bibinfo {author} {\bibfnamefont {K.}~\bibnamefont
  {{Breivik}}}, \bibinfo {author} {\bibfnamefont {S.}~\bibnamefont
  {{Chatterjee}}}, \bibinfo {author} {\bibfnamefont {G.}~\bibnamefont
  {{Fragione}}}, \bibinfo {author} {\bibfnamefont {F.}~\bibnamefont
  {{K{\i}ro{\u{g}}lu}}}, \bibinfo {author} {\bibfnamefont {K.}~\bibnamefont
  {{Kremer}}}, \bibinfo {author} {\bibfnamefont {N.~Z.}\ \bibnamefont {{Rui}}},
  \bibinfo {author} {\bibfnamefont {C.~S.}\ \bibnamefont {{Ye}}}, \bibinfo
  {author} {\bibfnamefont {M.}~\bibnamefont {{Zevin}}},\ and\ \bibinfo {author}
  {\bibfnamefont {F.~A.}\ \bibnamefont {{Rasio}}},\ }\href
  {https://doi.org/10.3847/1538-4365/ac2edf} {\bibfield  {journal} {\bibinfo
  {journal} {Astrophysical Journal Supplement Series}\ }\textbf {\bibinfo
  {volume} {258}},\ \bibinfo {eid} {22} (\bibinfo {year} {2022})},\ \Eprint
  {https://arxiv.org/abs/2106.02643} {arXiv:2106.02643 [astro-ph.GA]}
  \BibitemShut {NoStop}%
\bibitem [{\citenamefont {Vitale}\ \emph {et~al.}(2019)\citenamefont {Vitale},
  \citenamefont {Farr}, \citenamefont {Ng},\ and\ \citenamefont
  {Rodriguez}}]{Vitale:2018yhm}%
  \BibitemOpen
  \bibfield  {author} {\bibinfo {author} {\bibfnamefont {S.}~\bibnamefont
  {Vitale}}, \bibinfo {author} {\bibfnamefont {W.~M.}\ \bibnamefont {Farr}},
  \bibinfo {author} {\bibfnamefont {K.}~\bibnamefont {Ng}},\ and\ \bibinfo
  {author} {\bibfnamefont {C.~L.}\ \bibnamefont {Rodriguez}},\ }\href
  {https://doi.org/10.3847/2041-8213/ab50c0} {\bibfield  {journal} {\bibinfo
  {journal} {Astrophys. J. Lett.}\ }\textbf {\bibinfo {volume} {886}},\
  \bibinfo {pages} {L1} (\bibinfo {year} {2019})},\ \Eprint
  {https://arxiv.org/abs/1808.00901} {arXiv:1808.00901 [astro-ph.HE]}
  \BibitemShut {NoStop}%
\bibitem [{\citenamefont {Ray}\ \emph {et~al.}(2023)\citenamefont {Ray},
  \citenamefont {Maga\~na Hernandez}, \citenamefont {Mohite}, \citenamefont
  {Creighton},\ and\ \citenamefont {Kapadia}}]{Ray:2023upk}%
  \BibitemOpen
  \bibfield  {author} {\bibinfo {author} {\bibfnamefont {A.}~\bibnamefont
  {Ray}}, \bibinfo {author} {\bibfnamefont {I.}~\bibnamefont {Maga\~na
  Hernandez}}, \bibinfo {author} {\bibfnamefont {S.}~\bibnamefont {Mohite}},
  \bibinfo {author} {\bibfnamefont {J.}~\bibnamefont {Creighton}},\ and\
  \bibinfo {author} {\bibfnamefont {S.}~\bibnamefont {Kapadia}},\ }\href
  {https://doi.org/10.3847/1538-4357/acf452} {\bibfield  {journal} {\bibinfo
  {journal} {Astrophys. J.}\ }\textbf {\bibinfo {volume} {957}},\ \bibinfo
  {pages} {37} (\bibinfo {year} {2023})},\ \Eprint
  {https://arxiv.org/abs/2304.08046} {arXiv:2304.08046 [gr-qc]} \BibitemShut
  {NoStop}%
\bibitem [{\citenamefont {Callister}\ and\ \citenamefont
  {Farr}(2024)}]{Callister:2023tgi}%
  \BibitemOpen
  \bibfield  {author} {\bibinfo {author} {\bibfnamefont {T.~A.}\ \bibnamefont
  {Callister}}\ and\ \bibinfo {author} {\bibfnamefont {W.~M.}\ \bibnamefont
  {Farr}},\ }\href {https://doi.org/10.1103/PhysRevX.14.021005} {\bibfield
  {journal} {\bibinfo  {journal} {Phys. Rev. X}\ }\textbf {\bibinfo {volume}
  {14}},\ \bibinfo {pages} {021005} (\bibinfo {year} {2024})},\ \Eprint
  {https://arxiv.org/abs/2302.07289} {arXiv:2302.07289 [astro-ph.HE]}
  \BibitemShut {NoStop}%
\bibitem [{\citenamefont {Farah}\ \emph {et~al.}(2025)\citenamefont {Farah},
  \citenamefont {Callister}, \citenamefont {Ezquiaga}, \citenamefont {Zevin},\
  and\ \citenamefont {Holz}}]{Farah:2024xub}%
  \BibitemOpen
  \bibfield  {author} {\bibinfo {author} {\bibfnamefont {A.~M.}\ \bibnamefont
  {Farah}}, \bibinfo {author} {\bibfnamefont {T.~A.}\ \bibnamefont
  {Callister}}, \bibinfo {author} {\bibfnamefont {J.~M.}\ \bibnamefont
  {Ezquiaga}}, \bibinfo {author} {\bibfnamefont {M.}~\bibnamefont {Zevin}},\
  and\ \bibinfo {author} {\bibfnamefont {D.~E.}\ \bibnamefont {Holz}},\ }\href
  {https://doi.org/10.3847/1538-4357/ad9253} {\bibfield  {journal} {\bibinfo
  {journal} {Astrophys. J.}\ }\textbf {\bibinfo {volume} {978}},\ \bibinfo
  {pages} {153} (\bibinfo {year} {2025})},\ \Eprint
  {https://arxiv.org/abs/2404.02210} {arXiv:2404.02210 [astro-ph.CO]}
  \BibitemShut {NoStop}%
\bibitem [{\citenamefont {Edelman}\ \emph {et~al.}(2022)\citenamefont
  {Edelman}, \citenamefont {Doctor}, \citenamefont {Godfrey},\ and\
  \citenamefont {Farr}}]{Edelman:2021zkw}%
  \BibitemOpen
  \bibfield  {author} {\bibinfo {author} {\bibfnamefont {B.}~\bibnamefont
  {Edelman}}, \bibinfo {author} {\bibfnamefont {Z.}~\bibnamefont {Doctor}},
  \bibinfo {author} {\bibfnamefont {J.}~\bibnamefont {Godfrey}},\ and\ \bibinfo
  {author} {\bibfnamefont {B.}~\bibnamefont {Farr}},\ }\href
  {https://doi.org/10.3847/1538-4357/ac3667} {\bibfield  {journal} {\bibinfo
  {journal} {Astrophys. J.}\ }\textbf {\bibinfo {volume} {924}},\ \bibinfo
  {pages} {101} (\bibinfo {year} {2022})},\ \Eprint
  {https://arxiv.org/abs/2109.06137} {arXiv:2109.06137 [astro-ph.HE]}
  \BibitemShut {NoStop}%
\bibitem [{\citenamefont {Edelman}\ \emph {et~al.}(2023)\citenamefont
  {Edelman}, \citenamefont {Farr},\ and\ \citenamefont
  {Doctor}}]{Edelman:2022ydv}%
  \BibitemOpen
  \bibfield  {author} {\bibinfo {author} {\bibfnamefont {B.}~\bibnamefont
  {Edelman}}, \bibinfo {author} {\bibfnamefont {B.}~\bibnamefont {Farr}},\ and\
  \bibinfo {author} {\bibfnamefont {Z.}~\bibnamefont {Doctor}},\ }\href
  {https://doi.org/10.3847/1538-4357/acb5ed} {\bibfield  {journal} {\bibinfo
  {journal} {Astrophys. J.}\ }\textbf {\bibinfo {volume} {946}},\ \bibinfo
  {pages} {16} (\bibinfo {year} {2023})},\ \Eprint
  {https://arxiv.org/abs/2210.12834} {arXiv:2210.12834 [astro-ph.HE]}
  \BibitemShut {NoStop}%
\bibitem [{\citenamefont {Golomb}\ and\ \citenamefont
  {Talbot}(2023)}]{Golomb:2022bon}%
  \BibitemOpen
  \bibfield  {author} {\bibinfo {author} {\bibfnamefont {J.}~\bibnamefont
  {Golomb}}\ and\ \bibinfo {author} {\bibfnamefont {C.}~\bibnamefont
  {Talbot}},\ }\href {https://doi.org/10.1103/PhysRevD.108.103009} {\bibfield
  {journal} {\bibinfo  {journal} {Phys. Rev. D}\ }\textbf {\bibinfo {volume}
  {108}},\ \bibinfo {pages} {103009} (\bibinfo {year} {2023})},\ \Eprint
  {https://arxiv.org/abs/2210.12287} {arXiv:2210.12287 [astro-ph.HE]}
  \BibitemShut {NoStop}%
\bibitem [{\citenamefont {Heinzel}\ \emph
  {et~al.}(2025{\natexlab{a}})\citenamefont {Heinzel}, \citenamefont {Mould},\
  and\ \citenamefont {Vitale}}]{Heinzel:2024hva}%
  \BibitemOpen
  \bibfield  {author} {\bibinfo {author} {\bibfnamefont {J.}~\bibnamefont
  {Heinzel}}, \bibinfo {author} {\bibfnamefont {M.}~\bibnamefont {Mould}},\
  and\ \bibinfo {author} {\bibfnamefont {S.}~\bibnamefont {Vitale}},\ }\href
  {https://doi.org/10.1103/PhysRevD.111.L061305} {\bibfield  {journal}
  {\bibinfo  {journal} {Phys. Rev. D}\ }\textbf {\bibinfo {volume} {111}},\
  \bibinfo {pages} {L061305} (\bibinfo {year} {2025}{\natexlab{a}})},\ \Eprint
  {https://arxiv.org/abs/2406.16844} {arXiv:2406.16844 [astro-ph.HE]}
  \BibitemShut {NoStop}%
\bibitem [{\citenamefont {Mandel}\ \emph {et~al.}(2017)\citenamefont {Mandel},
  \citenamefont {Farr}, \citenamefont {Colonna}, \citenamefont {Stevenson},
  \citenamefont {Ti\v{n}o},\ and\ \citenamefont {Veitch}}]{Mandel:2016prl}%
  \BibitemOpen
  \bibfield  {author} {\bibinfo {author} {\bibfnamefont {I.}~\bibnamefont
  {Mandel}}, \bibinfo {author} {\bibfnamefont {W.~M.}\ \bibnamefont {Farr}},
  \bibinfo {author} {\bibfnamefont {A.}~\bibnamefont {Colonna}}, \bibinfo
  {author} {\bibfnamefont {S.}~\bibnamefont {Stevenson}}, \bibinfo {author}
  {\bibfnamefont {P.}~\bibnamefont {Ti\v{n}o}},\ and\ \bibinfo {author}
  {\bibfnamefont {J.}~\bibnamefont {Veitch}},\ }\href
  {https://doi.org/10.1093/mnras/stw2883} {\bibfield  {journal} {\bibinfo
  {journal} {Mon. Not. Roy. Astron. Soc.}\ }\textbf {\bibinfo {volume} {465}},\
  \bibinfo {pages} {3254} (\bibinfo {year} {2017})},\ \Eprint
  {https://arxiv.org/abs/1608.08223} {arXiv:1608.08223 [astro-ph.HE]}
  \BibitemShut {NoStop}%
\bibitem [{\citenamefont {Heinzel}\ \emph
  {et~al.}(2025{\natexlab{b}})\citenamefont {Heinzel}, \citenamefont {Mould},
  \citenamefont {\'Alvarez-L\'opez},\ and\ \citenamefont
  {Vitale}}]{Heinzel:2024jlc}%
  \BibitemOpen
  \bibfield  {author} {\bibinfo {author} {\bibfnamefont {J.}~\bibnamefont
  {Heinzel}}, \bibinfo {author} {\bibfnamefont {M.}~\bibnamefont {Mould}},
  \bibinfo {author} {\bibfnamefont {S.}~\bibnamefont {\'Alvarez-L\'opez}},\
  and\ \bibinfo {author} {\bibfnamefont {S.}~\bibnamefont {Vitale}},\ }\href
  {https://doi.org/10.1103/PhysRevD.111.063043} {\bibfield  {journal} {\bibinfo
   {journal} {Phys. Rev. D}\ }\textbf {\bibinfo {volume} {111}},\ \bibinfo
  {pages} {063043} (\bibinfo {year} {2025}{\natexlab{b}})},\ \Eprint
  {https://arxiv.org/abs/2406.16813} {arXiv:2406.16813 [astro-ph.HE]}
  \BibitemShut {NoStop}%
\bibitem [{\citenamefont {Toubiana}\ \emph {et~al.}(2023)\citenamefont
  {Toubiana}, \citenamefont {Katz},\ and\ \citenamefont
  {Gair}}]{Toubiana:2023egi}%
  \BibitemOpen
  \bibfield  {author} {\bibinfo {author} {\bibfnamefont {A.}~\bibnamefont
  {Toubiana}}, \bibinfo {author} {\bibfnamefont {M.~L.}\ \bibnamefont {Katz}},\
  and\ \bibinfo {author} {\bibfnamefont {J.~R.}\ \bibnamefont {Gair}},\ }\href
  {https://doi.org/10.1093/mnras/stad2215} {\bibfield  {journal} {\bibinfo
  {journal} {Mon. Not. Roy. Astron. Soc.}\ }\textbf {\bibinfo {volume} {524}},\
  \bibinfo {pages} {5844} (\bibinfo {year} {2023})},\ \Eprint
  {https://arxiv.org/abs/2305.08909} {arXiv:2305.08909 [gr-qc]} \BibitemShut
  {NoStop}%
\bibitem [{\citenamefont {Tiwari}\ and\ \citenamefont
  {Fairhurst}(2021)}]{Tiwari:2020otp}%
  \BibitemOpen
  \bibfield  {author} {\bibinfo {author} {\bibfnamefont {V.}~\bibnamefont
  {Tiwari}}\ and\ \bibinfo {author} {\bibfnamefont {S.}~\bibnamefont
  {Fairhurst}},\ }\href {https://doi.org/10.3847/2041-8213/abfbe7} {\bibfield
  {journal} {\bibinfo  {journal} {Astrophys. J. Lett.}\ }\textbf {\bibinfo
  {volume} {913}},\ \bibinfo {pages} {L19} (\bibinfo {year} {2021})},\ \Eprint
  {https://arxiv.org/abs/2011.04502} {arXiv:2011.04502 [astro-ph.HE]}
  \BibitemShut {NoStop}%
\bibitem [{\citenamefont {Godfrey}\ \emph {et~al.}(2023)\citenamefont
  {Godfrey}, \citenamefont {Edelman},\ and\ \citenamefont
  {Farr}}]{Godfrey:2023oxb}%
  \BibitemOpen
  \bibfield  {author} {\bibinfo {author} {\bibfnamefont {J.}~\bibnamefont
  {Godfrey}}, \bibinfo {author} {\bibfnamefont {B.}~\bibnamefont {Edelman}},\
  and\ \bibinfo {author} {\bibfnamefont {B.}~\bibnamefont {Farr}},\ }\href@noop
  {} {\bibfield  {journal} {\bibinfo  {journal} {Arxiv}\ } (\bibinfo {year}
  {2023})},\ \Eprint {https://arxiv.org/abs/2304.01288} {arXiv:2304.01288
  [astro-ph.HE]} \BibitemShut {NoStop}%
\bibitem [{\citenamefont {Sadiq}\ \emph {et~al.}(2024)\citenamefont {Sadiq},
  \citenamefont {Dent},\ and\ \citenamefont {Gieles}}]{Sadiq:2023zee}%
  \BibitemOpen
  \bibfield  {author} {\bibinfo {author} {\bibfnamefont {J.}~\bibnamefont
  {Sadiq}}, \bibinfo {author} {\bibfnamefont {T.}~\bibnamefont {Dent}},\ and\
  \bibinfo {author} {\bibfnamefont {M.}~\bibnamefont {Gieles}},\ }\href
  {https://doi.org/10.3847/1538-4357/ad0ce6} {\bibfield  {journal} {\bibinfo
  {journal} {Astrophys. J.}\ }\textbf {\bibinfo {volume} {960}},\ \bibinfo
  {pages} {65} (\bibinfo {year} {2024})},\ \Eprint
  {https://arxiv.org/abs/2307.12092} {arXiv:2307.12092 [astro-ph.HE]}
  \BibitemShut {NoStop}%
\bibitem [{\citenamefont {Vitale}\ \emph
  {et~al.}(2017{\natexlab{a}})\citenamefont {Vitale}, \citenamefont {Lynch},
  \citenamefont {Sturani},\ and\ \citenamefont {Graff}}]{Vitale:2015tea}%
  \BibitemOpen
  \bibfield  {author} {\bibinfo {author} {\bibfnamefont {S.}~\bibnamefont
  {Vitale}}, \bibinfo {author} {\bibfnamefont {R.}~\bibnamefont {Lynch}},
  \bibinfo {author} {\bibfnamefont {R.}~\bibnamefont {Sturani}},\ and\ \bibinfo
  {author} {\bibfnamefont {P.}~\bibnamefont {Graff}},\ }\href
  {https://doi.org/10.1088/1361-6382/aa552e} {\bibfield  {journal} {\bibinfo
  {journal} {Class. Quant. Grav.}\ }\textbf {\bibinfo {volume} {34}},\ \bibinfo
  {pages} {03LT01} (\bibinfo {year} {2017}{\natexlab{a}})},\ \Eprint
  {https://arxiv.org/abs/1503.04307} {arXiv:1503.04307 [gr-qc]} \BibitemShut
  {NoStop}%
\bibitem [{\citenamefont {Talbot}\ and\ \citenamefont
  {Thrane}(2018)}]{Talbot:2018cva}%
  \BibitemOpen
  \bibfield  {author} {\bibinfo {author} {\bibfnamefont {C.}~\bibnamefont
  {Talbot}}\ and\ \bibinfo {author} {\bibfnamefont {E.}~\bibnamefont
  {Thrane}},\ }\href {https://doi.org/10.3847/1538-4357/aab34c} {\bibfield
  {journal} {\bibinfo  {journal} {Astrophys. J.}\ }\textbf {\bibinfo {volume}
  {856}},\ \bibinfo {pages} {173} (\bibinfo {year} {2018})},\ \Eprint
  {https://arxiv.org/abs/1801.02699} {arXiv:1801.02699 [astro-ph.HE]}
  \BibitemShut {NoStop}%
\bibitem [{\citenamefont {Talbot}\ and\ \citenamefont
  {Thrane}(2017)}]{Talbot:2017yur}%
  \BibitemOpen
  \bibfield  {author} {\bibinfo {author} {\bibfnamefont {C.}~\bibnamefont
  {Talbot}}\ and\ \bibinfo {author} {\bibfnamefont {E.}~\bibnamefont
  {Thrane}},\ }\href {https://doi.org/10.1103/PhysRevD.96.023012} {\bibfield
  {journal} {\bibinfo  {journal} {Phys. Rev. D}\ }\textbf {\bibinfo {volume}
  {96}},\ \bibinfo {pages} {023012} (\bibinfo {year} {2017})},\ \Eprint
  {https://arxiv.org/abs/1704.08370} {arXiv:1704.08370 [astro-ph.HE]}
  \BibitemShut {NoStop}%
\bibitem [{\citenamefont {Fishbach}\ \emph {et~al.}(2018)\citenamefont
  {Fishbach}, \citenamefont {Holz},\ and\ \citenamefont
  {Farr}}]{Fishbach:2018edt}%
  \BibitemOpen
  \bibfield  {author} {\bibinfo {author} {\bibfnamefont {M.}~\bibnamefont
  {Fishbach}}, \bibinfo {author} {\bibfnamefont {D.~E.}\ \bibnamefont {Holz}},\
  and\ \bibinfo {author} {\bibfnamefont {W.~M.}\ \bibnamefont {Farr}},\ }\href
  {https://doi.org/10.3847/2041-8213/aad800} {\bibfield  {journal} {\bibinfo
  {journal} {Astrophys. J. Lett.}\ }\textbf {\bibinfo {volume} {863}},\
  \bibinfo {pages} {L41} (\bibinfo {year} {2018})},\ \Eprint
  {https://arxiv.org/abs/1805.10270} {arXiv:1805.10270 [astro-ph.HE]}
  \BibitemShut {NoStop}%
\bibitem [{\citenamefont {Callister}\ \emph {et~al.}(2021)\citenamefont
  {Callister}, \citenamefont {Haster}, \citenamefont {Ng}, \citenamefont
  {Vitale},\ and\ \citenamefont {Farr}}]{Callister:2021fpo}%
  \BibitemOpen
  \bibfield  {author} {\bibinfo {author} {\bibfnamefont {T.~A.}\ \bibnamefont
  {Callister}}, \bibinfo {author} {\bibfnamefont {C.-J.}\ \bibnamefont
  {Haster}}, \bibinfo {author} {\bibfnamefont {K.~K.~Y.}\ \bibnamefont {Ng}},
  \bibinfo {author} {\bibfnamefont {S.}~\bibnamefont {Vitale}},\ and\ \bibinfo
  {author} {\bibfnamefont {W.~M.}\ \bibnamefont {Farr}},\ }\href
  {https://doi.org/10.3847/2041-8213/ac2ccc} {\bibfield  {journal} {\bibinfo
  {journal} {Astrophys. J. Lett.}\ }\textbf {\bibinfo {volume} {922}},\
  \bibinfo {pages} {L5} (\bibinfo {year} {2021})},\ \Eprint
  {https://arxiv.org/abs/2106.00521} {arXiv:2106.00521 [astro-ph.HE]}
  \BibitemShut {NoStop}%
\bibitem [{\citenamefont {Biscoveanu}\ \emph {et~al.}(2022)\citenamefont
  {Biscoveanu}, \citenamefont {Callister}, \citenamefont {Haster},
  \citenamefont {Ng}, \citenamefont {Vitale},\ and\ \citenamefont
  {Farr}}]{Biscoveanu:2022qac}%
  \BibitemOpen
  \bibfield  {author} {\bibinfo {author} {\bibfnamefont {S.}~\bibnamefont
  {Biscoveanu}}, \bibinfo {author} {\bibfnamefont {T.~A.}\ \bibnamefont
  {Callister}}, \bibinfo {author} {\bibfnamefont {C.-J.}\ \bibnamefont
  {Haster}}, \bibinfo {author} {\bibfnamefont {K.~K.~Y.}\ \bibnamefont {Ng}},
  \bibinfo {author} {\bibfnamefont {S.}~\bibnamefont {Vitale}},\ and\ \bibinfo
  {author} {\bibfnamefont {W.~M.}\ \bibnamefont {Farr}},\ }\href
  {https://doi.org/10.3847/2041-8213/ac71a8} {\bibfield  {journal} {\bibinfo
  {journal} {Astrophys. J. Lett.}\ }\textbf {\bibinfo {volume} {932}},\
  \bibinfo {pages} {L19} (\bibinfo {year} {2022})},\ \Eprint
  {https://arxiv.org/abs/2204.01578} {arXiv:2204.01578 [astro-ph.HE]}
  \BibitemShut {NoStop}%
\bibitem [{\citenamefont {Abbott}\ \emph
  {et~al.}(2021{\natexlab{a}})\citenamefont {Abbott} \emph
  {et~al.}}]{LIGOScientific:2020kqk}%
  \BibitemOpen
  \bibfield  {author} {\bibinfo {author} {\bibfnamefont {R.}~\bibnamefont
  {Abbott}} \emph {et~al.} (\bibinfo {collaboration} {LIGO Scientific,
  Virgo}),\ }\href {https://doi.org/10.3847/2041-8213/abe949} {\bibfield
  {journal} {\bibinfo  {journal} {Astrophys. J. Lett.}\ }\textbf {\bibinfo
  {volume} {913}},\ \bibinfo {pages} {L7} (\bibinfo {year}
  {2021}{\natexlab{a}})},\ \Eprint {https://arxiv.org/abs/2010.14533}
  {arXiv:2010.14533 [astro-ph.HE]} \BibitemShut {NoStop}%
\bibitem [{\citenamefont {Abbott}\ \emph
  {et~al.}(2021{\natexlab{b}})\citenamefont {Abbott} \emph
  {et~al.}}]{LIGOScientific:2021psn}%
  \BibitemOpen
  \bibfield  {author} {\bibinfo {author} {\bibfnamefont {R.}~\bibnamefont
  {Abbott}} \emph {et~al.} (\bibinfo {collaboration} {LIGO Scientific, VIRGO,
  KAGRA Scientific}),\ }\href@noop {} {\bibfield  {journal} {\bibinfo
  {journal} {arXiv e-prints}\ ,\ \bibinfo {eid} {arXiv:2111.03634}} (\bibinfo
  {year} {2021}{\natexlab{b}})},\ \Eprint {https://arxiv.org/abs/2111.03634}
  {arXiv:2111.03634 [astro-ph.HE]} \BibitemShut {NoStop}%
\bibitem [{\citenamefont {Zevin}\ \emph {et~al.}(2021)\citenamefont {Zevin},
  \citenamefont {Bavera}, \citenamefont {Berry}, \citenamefont {Kalogera},
  \citenamefont {Fragos}, \citenamefont {Marchant}, \citenamefont {Rodriguez},
  \citenamefont {Antonini}, \citenamefont {Holz},\ and\ \citenamefont
  {Pankow}}]{Zevin:2020gbd}%
  \BibitemOpen
  \bibfield  {author} {\bibinfo {author} {\bibfnamefont {M.}~\bibnamefont
  {Zevin}}, \bibinfo {author} {\bibfnamefont {S.~S.}\ \bibnamefont {Bavera}},
  \bibinfo {author} {\bibfnamefont {C.~P.~L.}\ \bibnamefont {Berry}}, \bibinfo
  {author} {\bibfnamefont {V.}~\bibnamefont {Kalogera}}, \bibinfo {author}
  {\bibfnamefont {T.}~\bibnamefont {Fragos}}, \bibinfo {author} {\bibfnamefont
  {P.}~\bibnamefont {Marchant}}, \bibinfo {author} {\bibfnamefont {C.~L.}\
  \bibnamefont {Rodriguez}}, \bibinfo {author} {\bibfnamefont {F.}~\bibnamefont
  {Antonini}}, \bibinfo {author} {\bibfnamefont {D.~E.}\ \bibnamefont {Holz}},\
  and\ \bibinfo {author} {\bibfnamefont {C.}~\bibnamefont {Pankow}},\ }\href
  {https://doi.org/10.3847/1538-4357/abe40e} {\bibfield  {journal} {\bibinfo
  {journal} {Astrophys. J.}\ }\textbf {\bibinfo {volume} {910}},\ \bibinfo
  {pages} {152} (\bibinfo {year} {2021})},\ \Eprint
  {https://arxiv.org/abs/2011.10057} {arXiv:2011.10057 [astro-ph.HE]}
  \BibitemShut {NoStop}%
\bibitem [{\citenamefont {Wong}\ \emph {et~al.}(2021)\citenamefont {Wong},
  \citenamefont {Breivik}, \citenamefont {Kremer},\ and\ \citenamefont
  {Callister}}]{Wong:2020ise}%
  \BibitemOpen
  \bibfield  {author} {\bibinfo {author} {\bibfnamefont {K.~W.~K.}\
  \bibnamefont {Wong}}, \bibinfo {author} {\bibfnamefont {K.}~\bibnamefont
  {Breivik}}, \bibinfo {author} {\bibfnamefont {K.}~\bibnamefont {Kremer}},\
  and\ \bibinfo {author} {\bibfnamefont {T.}~\bibnamefont {Callister}},\ }\href
  {https://doi.org/10.1103/PhysRevD.103.083021} {\bibfield  {journal} {\bibinfo
   {journal} {Phys. Rev. D}\ }\textbf {\bibinfo {volume} {103}},\ \bibinfo
  {pages} {083021} (\bibinfo {year} {2021})},\ \Eprint
  {https://arxiv.org/abs/2011.03564} {arXiv:2011.03564 [astro-ph.HE]}
  \BibitemShut {NoStop}%
\bibitem [{\citenamefont {Wong}\ and\ \citenamefont
  {Gerosa}(2019)}]{Wong:2019uni}%
  \BibitemOpen
  \bibfield  {author} {\bibinfo {author} {\bibfnamefont {K.~W.~K.}\
  \bibnamefont {Wong}}\ and\ \bibinfo {author} {\bibfnamefont {D.}~\bibnamefont
  {Gerosa}},\ }\href {https://doi.org/10.1103/PhysRevD.100.083015} {\bibfield
  {journal} {\bibinfo  {journal} {Phys. Rev. D}\ }\textbf {\bibinfo {volume}
  {100}},\ \bibinfo {pages} {083015} (\bibinfo {year} {2019})},\ \Eprint
  {https://arxiv.org/abs/1909.06373} {arXiv:1909.06373 [astro-ph.HE]}
  \BibitemShut {NoStop}%
\bibitem [{\citenamefont {Colloms}\ \emph {et~al.}(2025)\citenamefont
  {Colloms}, \citenamefont {Berry}, \citenamefont {Veitch},\ and\ \citenamefont
  {Zevin}}]{Colloms:2025hib}%
  \BibitemOpen
  \bibfield  {author} {\bibinfo {author} {\bibfnamefont {S.}~\bibnamefont
  {Colloms}}, \bibinfo {author} {\bibfnamefont {C.~P.~L.}\ \bibnamefont
  {Berry}}, \bibinfo {author} {\bibfnamefont {J.}~\bibnamefont {Veitch}},\ and\
  \bibinfo {author} {\bibfnamefont {M.}~\bibnamefont {Zevin}},\ }\href@noop {}
  {\bibfield  {journal} {\bibinfo  {journal} {Arxiv}\ } (\bibinfo {year}
  {2025})},\ \Eprint {https://arxiv.org/abs/2503.03819} {arXiv:2503.03819
  [astro-ph.HE]} \BibitemShut {NoStop}%
\bibitem [{\citenamefont {Plunkett}\ \emph {et~al.}(2025)\citenamefont
  {Plunkett}, \citenamefont {Mould},\ and\ \citenamefont
  {Vitale}}]{Plunkett:2025mjr}%
  \BibitemOpen
  \bibfield  {author} {\bibinfo {author} {\bibfnamefont {C.}~\bibnamefont
  {Plunkett}}, \bibinfo {author} {\bibfnamefont {M.}~\bibnamefont {Mould}},\
  and\ \bibinfo {author} {\bibfnamefont {S.}~\bibnamefont {Vitale}},\
  }\href@noop {} {\bibfield  {journal} {\bibinfo  {journal} {arxiv}\ }
  (\bibinfo {year} {2025})},\ \Eprint {https://arxiv.org/abs/2504.18615}
  {arXiv:2504.18615 [gr-qc]} \BibitemShut {NoStop}%
\bibitem [{\citenamefont {Franciolini}\ \emph {et~al.}(2022)\citenamefont
  {Franciolini}, \citenamefont {Baibhav}, \citenamefont {De~Luca},
  \citenamefont {Ng}, \citenamefont {Wong}, \citenamefont {Berti},
  \citenamefont {Pani}, \citenamefont {Riotto},\ and\ \citenamefont
  {Vitale}}]{Franciolini:2021tla}%
  \BibitemOpen
  \bibfield  {author} {\bibinfo {author} {\bibfnamefont {G.}~\bibnamefont
  {Franciolini}}, \bibinfo {author} {\bibfnamefont {V.}~\bibnamefont
  {Baibhav}}, \bibinfo {author} {\bibfnamefont {V.}~\bibnamefont {De~Luca}},
  \bibinfo {author} {\bibfnamefont {K.~K.~Y.}\ \bibnamefont {Ng}}, \bibinfo
  {author} {\bibfnamefont {K.~W.~K.}\ \bibnamefont {Wong}}, \bibinfo {author}
  {\bibfnamefont {E.}~\bibnamefont {Berti}}, \bibinfo {author} {\bibfnamefont
  {P.}~\bibnamefont {Pani}}, \bibinfo {author} {\bibfnamefont {A.}~\bibnamefont
  {Riotto}},\ and\ \bibinfo {author} {\bibfnamefont {S.}~\bibnamefont
  {Vitale}},\ }\href {https://doi.org/10.1103/PhysRevD.105.083526} {\bibfield
  {journal} {\bibinfo  {journal} {Phys. Rev. D}\ }\textbf {\bibinfo {volume}
  {105}},\ \bibinfo {pages} {083526} (\bibinfo {year} {2022})},\ \Eprint
  {https://arxiv.org/abs/2105.03349} {arXiv:2105.03349 [gr-qc]} \BibitemShut
  {NoStop}%
\bibitem [{\citenamefont {Cheng}\ \emph {et~al.}(2023)\citenamefont {Cheng},
  \citenamefont {Zevin},\ and\ \citenamefont {Vitale}}]{Cheng:2023ddt}%
  \BibitemOpen
  \bibfield  {author} {\bibinfo {author} {\bibfnamefont {A.~Q.}\ \bibnamefont
  {Cheng}}, \bibinfo {author} {\bibfnamefont {M.}~\bibnamefont {Zevin}},\ and\
  \bibinfo {author} {\bibfnamefont {S.}~\bibnamefont {Vitale}},\ }\href
  {https://doi.org/10.3847/1538-4357/aced98} {\bibfield  {journal} {\bibinfo
  {journal} {Astrophys. J.}\ }\textbf {\bibinfo {volume} {955}},\ \bibinfo
  {pages} {127} (\bibinfo {year} {2023})},\ \Eprint
  {https://arxiv.org/abs/2307.03129} {arXiv:2307.03129 [astro-ph.HE]}
  \BibitemShut {NoStop}%
\bibitem [{\citenamefont {Heinzel}\ \emph {et~al.}(2024)\citenamefont
  {Heinzel}, \citenamefont {Biscoveanu},\ and\ \citenamefont
  {Vitale}}]{Heinzel:2023hlb}%
  \BibitemOpen
  \bibfield  {author} {\bibinfo {author} {\bibfnamefont {J.}~\bibnamefont
  {Heinzel}}, \bibinfo {author} {\bibfnamefont {S.}~\bibnamefont
  {Biscoveanu}},\ and\ \bibinfo {author} {\bibfnamefont {S.}~\bibnamefont
  {Vitale}},\ }\href {https://doi.org/10.1103/PhysRevD.109.103006} {\bibfield
  {journal} {\bibinfo  {journal} {Phys. Rev. D}\ }\textbf {\bibinfo {volume}
  {109}},\ \bibinfo {pages} {103006} (\bibinfo {year} {2024})},\ \Eprint
  {https://arxiv.org/abs/2312.00993} {arXiv:2312.00993 [astro-ph.HE]}
  \BibitemShut {NoStop}%
\bibitem [{\citenamefont {Gerosa}\ \emph {et~al.}(2013)\citenamefont {Gerosa},
  \citenamefont {Kesden}, \citenamefont {Berti}, \citenamefont
  {O'Shaughnessy},\ and\ \citenamefont {Sperhake}}]{Gerosa:2013laa}%
  \BibitemOpen
  \bibfield  {author} {\bibinfo {author} {\bibfnamefont {D.}~\bibnamefont
  {Gerosa}}, \bibinfo {author} {\bibfnamefont {M.}~\bibnamefont {Kesden}},
  \bibinfo {author} {\bibfnamefont {E.}~\bibnamefont {Berti}}, \bibinfo
  {author} {\bibfnamefont {R.}~\bibnamefont {O'Shaughnessy}},\ and\ \bibinfo
  {author} {\bibfnamefont {U.}~\bibnamefont {Sperhake}},\ }\href
  {https://doi.org/10.1103/PhysRevD.87.104028} {\bibfield  {journal} {\bibinfo
  {journal} {Phys. Rev. D}\ }\textbf {\bibinfo {volume} {87}},\ \bibinfo
  {pages} {104028} (\bibinfo {year} {2013})},\ \Eprint
  {https://arxiv.org/abs/1302.4442} {arXiv:1302.4442 [gr-qc]} \BibitemShut
  {NoStop}%
\bibitem [{\citenamefont {Gerosa}\ \emph {et~al.}(2014)\citenamefont {Gerosa},
  \citenamefont {O'Shaughnessy}, \citenamefont {Kesden}, \citenamefont
  {Berti},\ and\ \citenamefont {Sperhake}}]{Gerosa:2014kta}%
  \BibitemOpen
  \bibfield  {author} {\bibinfo {author} {\bibfnamefont {D.}~\bibnamefont
  {Gerosa}}, \bibinfo {author} {\bibfnamefont {R.}~\bibnamefont
  {O'Shaughnessy}}, \bibinfo {author} {\bibfnamefont {M.}~\bibnamefont
  {Kesden}}, \bibinfo {author} {\bibfnamefont {E.}~\bibnamefont {Berti}},\ and\
  \bibinfo {author} {\bibfnamefont {U.}~\bibnamefont {Sperhake}},\ }\href
  {https://doi.org/10.1103/PhysRevD.89.124025} {\bibfield  {journal} {\bibinfo
  {journal} {Phys. Rev. D}\ }\textbf {\bibinfo {volume} {89}},\ \bibinfo
  {pages} {124025} (\bibinfo {year} {2014})},\ \Eprint
  {https://arxiv.org/abs/1403.7147} {arXiv:1403.7147 [gr-qc]} \BibitemShut
  {NoStop}%
\bibitem [{\citenamefont {Farr}\ \emph {et~al.}(2017)\citenamefont {Farr},
  \citenamefont {Stevenson}, \citenamefont {Coleman~Miller}, \citenamefont
  {Mandel}, \citenamefont {Farr},\ and\ \citenamefont
  {Vecchio}}]{Farr:2017uvj}%
  \BibitemOpen
  \bibfield  {author} {\bibinfo {author} {\bibfnamefont {W.~M.}\ \bibnamefont
  {Farr}}, \bibinfo {author} {\bibfnamefont {S.}~\bibnamefont {Stevenson}},
  \bibinfo {author} {\bibfnamefont {M.}~\bibnamefont {Coleman~Miller}},
  \bibinfo {author} {\bibfnamefont {I.}~\bibnamefont {Mandel}}, \bibinfo
  {author} {\bibfnamefont {B.}~\bibnamefont {Farr}},\ and\ \bibinfo {author}
  {\bibfnamefont {A.}~\bibnamefont {Vecchio}},\ }\href
  {https://doi.org/10.1038/nature23453} {\bibfield  {journal} {\bibinfo
  {journal} {Nature}\ }\textbf {\bibinfo {volume} {548}},\ \bibinfo {pages}
  {426} (\bibinfo {year} {2017})},\ \Eprint {https://arxiv.org/abs/1706.01385}
  {arXiv:1706.01385 [astro-ph.HE]} \BibitemShut {NoStop}%
\bibitem [{\citenamefont {Mould}\ and\ \citenamefont
  {Gerosa}(2022)}]{Mould:2021xst}%
  \BibitemOpen
  \bibfield  {author} {\bibinfo {author} {\bibfnamefont {M.}~\bibnamefont
  {Mould}}\ and\ \bibinfo {author} {\bibfnamefont {D.}~\bibnamefont {Gerosa}},\
  }\href {https://doi.org/10.1103/PhysRevD.105.024076} {\bibfield  {journal}
  {\bibinfo  {journal} {Phys. Rev. D}\ }\textbf {\bibinfo {volume} {105}},\
  \bibinfo {pages} {024076} (\bibinfo {year} {2022})},\ \Eprint
  {https://arxiv.org/abs/2110.05507} {arXiv:2110.05507 [astro-ph.HE]}
  \BibitemShut {NoStop}%
\bibitem [{\citenamefont {Varma}\ \emph
  {et~al.}(2022{\natexlab{a}})\citenamefont {Varma}, \citenamefont
  {Biscoveanu}, \citenamefont {Isi}, \citenamefont {Farr},\ and\ \citenamefont
  {Vitale}}]{Varma:2021xbh}%
  \BibitemOpen
  \bibfield  {author} {\bibinfo {author} {\bibfnamefont {V.}~\bibnamefont
  {Varma}}, \bibinfo {author} {\bibfnamefont {S.}~\bibnamefont {Biscoveanu}},
  \bibinfo {author} {\bibfnamefont {M.}~\bibnamefont {Isi}}, \bibinfo {author}
  {\bibfnamefont {W.~M.}\ \bibnamefont {Farr}},\ and\ \bibinfo {author}
  {\bibfnamefont {S.}~\bibnamefont {Vitale}},\ }\href
  {https://doi.org/10.1103/PhysRevLett.128.031101} {\bibfield  {journal}
  {\bibinfo  {journal} {Phys. Rev. Lett.}\ }\textbf {\bibinfo {volume} {128}},\
  \bibinfo {pages} {031101} (\bibinfo {year} {2022}{\natexlab{a}})},\ \Eprint
  {https://arxiv.org/abs/2107.09693} {arXiv:2107.09693 [astro-ph.HE]}
  \BibitemShut {NoStop}%
\bibitem [{\citenamefont {Varma}\ \emph
  {et~al.}(2022{\natexlab{b}})\citenamefont {Varma}, \citenamefont {Isi},
  \citenamefont {Biscoveanu}, \citenamefont {Farr},\ and\ \citenamefont
  {Vitale}}]{Varma:2021csh}%
  \BibitemOpen
  \bibfield  {author} {\bibinfo {author} {\bibfnamefont {V.}~\bibnamefont
  {Varma}}, \bibinfo {author} {\bibfnamefont {M.}~\bibnamefont {Isi}}, \bibinfo
  {author} {\bibfnamefont {S.}~\bibnamefont {Biscoveanu}}, \bibinfo {author}
  {\bibfnamefont {W.~M.}\ \bibnamefont {Farr}},\ and\ \bibinfo {author}
  {\bibfnamefont {S.}~\bibnamefont {Vitale}},\ }\href
  {https://doi.org/10.1103/PhysRevD.105.024045} {\bibfield  {journal} {\bibinfo
   {journal} {Phys. Rev. D}\ }\textbf {\bibinfo {volume} {105}},\ \bibinfo
  {pages} {024045} (\bibinfo {year} {2022}{\natexlab{b}})},\ \Eprint
  {https://arxiv.org/abs/2107.09692} {arXiv:2107.09692 [astro-ph.HE]}
  \BibitemShut {NoStop}%
\bibitem [{\citenamefont {Biscoveanu}(2025)}]{Biscoveanu:2025jpc}%
  \BibitemOpen
  \bibfield  {author} {\bibinfo {author} {\bibfnamefont {S.}~\bibnamefont
  {Biscoveanu}},\ }\href@noop {} {\bibfield  {journal} {\bibinfo  {journal}
  {Arxiv}\ } (\bibinfo {year} {2025})},\ \Eprint
  {https://arxiv.org/abs/2502.04278} {arXiv:2502.04278 [astro-ph.HE]}
  \BibitemShut {NoStop}%
\bibitem [{\citenamefont {Roulet}\ \emph {et~al.}(2021)\citenamefont {Roulet},
  \citenamefont {Chia}, \citenamefont {Olsen}, \citenamefont {Dai},
  \citenamefont {Venumadhav}, \citenamefont {Zackay},\ and\ \citenamefont
  {Zaldarriaga}}]{Roulet:2021hcu}%
  \BibitemOpen
  \bibfield  {author} {\bibinfo {author} {\bibfnamefont {J.}~\bibnamefont
  {Roulet}}, \bibinfo {author} {\bibfnamefont {H.~S.}\ \bibnamefont {Chia}},
  \bibinfo {author} {\bibfnamefont {S.}~\bibnamefont {Olsen}}, \bibinfo
  {author} {\bibfnamefont {L.}~\bibnamefont {Dai}}, \bibinfo {author}
  {\bibfnamefont {T.}~\bibnamefont {Venumadhav}}, \bibinfo {author}
  {\bibfnamefont {B.}~\bibnamefont {Zackay}},\ and\ \bibinfo {author}
  {\bibfnamefont {M.}~\bibnamefont {Zaldarriaga}},\ }\href
  {https://doi.org/10.1103/PhysRevD.104.083010} {\bibfield  {journal} {\bibinfo
   {journal} {Phys. Rev. D}\ }\textbf {\bibinfo {volume} {104}},\ \bibinfo
  {pages} {083010} (\bibinfo {year} {2021})},\ \Eprint
  {https://arxiv.org/abs/2105.10580} {arXiv:2105.10580 [astro-ph.HE]}
  \BibitemShut {NoStop}%
\bibitem [{\citenamefont {Stevenson}\ \emph {et~al.}(2017)\citenamefont
  {Stevenson}, \citenamefont {Berry},\ and\ \citenamefont
  {Mandel}}]{Stevenson:2017dlk}%
  \BibitemOpen
  \bibfield  {author} {\bibinfo {author} {\bibfnamefont {S.}~\bibnamefont
  {Stevenson}}, \bibinfo {author} {\bibfnamefont {C.~P.~L.}\ \bibnamefont
  {Berry}},\ and\ \bibinfo {author} {\bibfnamefont {I.}~\bibnamefont
  {Mandel}},\ }\href {https://doi.org/10.1093/mnras/stx1764} {\bibfield
  {journal} {\bibinfo  {journal} {Mon. Not. Roy. Astron. Soc.}\ }\textbf
  {\bibinfo {volume} {471}},\ \bibinfo {pages} {2801} (\bibinfo {year}
  {2017})},\ \Eprint {https://arxiv.org/abs/1703.06873} {arXiv:1703.06873
  [astro-ph.HE]} \BibitemShut {NoStop}%
\bibitem [{\citenamefont {P\'erigois}\ \emph {et~al.}(2023)\citenamefont
  {P\'erigois}, \citenamefont {Mapelli}, \citenamefont {Santoliquido},
  \citenamefont {Bouffanais},\ and\ \citenamefont {Rufolo}}]{Perigois:2023ihi}%
  \BibitemOpen
  \bibfield  {author} {\bibinfo {author} {\bibfnamefont {C.}~\bibnamefont
  {P\'erigois}}, \bibinfo {author} {\bibfnamefont {M.}~\bibnamefont {Mapelli}},
  \bibinfo {author} {\bibfnamefont {F.}~\bibnamefont {Santoliquido}}, \bibinfo
  {author} {\bibfnamefont {Y.}~\bibnamefont {Bouffanais}},\ and\ \bibinfo
  {author} {\bibfnamefont {R.}~\bibnamefont {Rufolo}},\ }\href
  {https://doi.org/10.3390/universe9120507} {\bibfield  {journal} {\bibinfo
  {journal} {Universe}\ }\textbf {\bibinfo {volume} {9}},\ \bibinfo {pages}
  {507} (\bibinfo {year} {2023})},\ \Eprint {https://arxiv.org/abs/2301.01312}
  {arXiv:2301.01312 [astro-ph.HE]} \BibitemShut {NoStop}%
\bibitem [{\citenamefont {Rodriguez}\ \emph {et~al.}(2016)\citenamefont
  {Rodriguez}, \citenamefont {Zevin}, \citenamefont {Pankow}, \citenamefont
  {Kalogera},\ and\ \citenamefont {Rasio}}]{Rodriguez:2016vmx}%
  \BibitemOpen
  \bibfield  {author} {\bibinfo {author} {\bibfnamefont {C.~L.}\ \bibnamefont
  {Rodriguez}}, \bibinfo {author} {\bibfnamefont {M.}~\bibnamefont {Zevin}},
  \bibinfo {author} {\bibfnamefont {C.}~\bibnamefont {Pankow}}, \bibinfo
  {author} {\bibfnamefont {V.}~\bibnamefont {Kalogera}},\ and\ \bibinfo
  {author} {\bibfnamefont {F.~A.}\ \bibnamefont {Rasio}},\ }\href
  {https://doi.org/10.3847/2041-8205/832/1/L2} {\bibfield  {journal} {\bibinfo
  {journal} {Astrophys. J. Lett.}\ }\textbf {\bibinfo {volume} {832}},\
  \bibinfo {pages} {L2} (\bibinfo {year} {2016})},\ \Eprint
  {https://arxiv.org/abs/1609.05916} {arXiv:1609.05916 [astro-ph.HE]}
  \BibitemShut {NoStop}%
\bibitem [{\citenamefont {Marchant}\ \emph {et~al.}(2021)\citenamefont
  {Marchant}, \citenamefont {Pappas}, \citenamefont {Gallegos-Garcia},
  \citenamefont {Berry}, \citenamefont {Taam}, \citenamefont {Kalogera},\ and\
  \citenamefont {Podsiadlowski}}]{Marchant:2021hiv}%
  \BibitemOpen
  \bibfield  {author} {\bibinfo {author} {\bibfnamefont {P.}~\bibnamefont
  {Marchant}}, \bibinfo {author} {\bibfnamefont {K.~M.~W.}\ \bibnamefont
  {Pappas}}, \bibinfo {author} {\bibfnamefont {M.}~\bibnamefont
  {Gallegos-Garcia}}, \bibinfo {author} {\bibfnamefont {C.~P.~L.}\ \bibnamefont
  {Berry}}, \bibinfo {author} {\bibfnamefont {R.~E.}\ \bibnamefont {Taam}},
  \bibinfo {author} {\bibfnamefont {V.}~\bibnamefont {Kalogera}},\ and\
  \bibinfo {author} {\bibfnamefont {P.}~\bibnamefont {Podsiadlowski}},\ }\href
  {https://doi.org/10.1051/0004-6361/202039992} {\bibfield  {journal} {\bibinfo
   {journal} {Astron. Astrophys.}\ }\textbf {\bibinfo {volume} {650}},\
  \bibinfo {pages} {A107} (\bibinfo {year} {2021})},\ \Eprint
  {https://arxiv.org/abs/2103.09243} {arXiv:2103.09243 [astro-ph.SR]}
  \BibitemShut {NoStop}%
\bibitem [{\citenamefont {Fryer}\ and\ \citenamefont
  {Kalogera}(2001)}]{Fryer:1999ht}%
  \BibitemOpen
  \bibfield  {author} {\bibinfo {author} {\bibfnamefont {C.~L.}\ \bibnamefont
  {Fryer}}\ and\ \bibinfo {author} {\bibfnamefont {V.}~\bibnamefont
  {Kalogera}},\ }\href {https://doi.org/10.1086/321359} {\bibfield  {journal}
  {\bibinfo  {journal} {Astrophys. J.}\ }\textbf {\bibinfo {volume} {554}},\
  \bibinfo {pages} {548} (\bibinfo {year} {2001})},\ \Eprint
  {https://arxiv.org/abs/astro-ph/9911312} {arXiv:astro-ph/9911312}
  \BibitemShut {NoStop}%
\bibitem [{\citenamefont {Kalogera}(2000)}]{Kalogera:1999tq}%
  \BibitemOpen
  \bibfield  {author} {\bibinfo {author} {\bibfnamefont {V.}~\bibnamefont
  {Kalogera}},\ }\href {https://doi.org/10.1086/309400} {\bibfield  {journal}
  {\bibinfo  {journal} {Astrophys. J.}\ }\textbf {\bibinfo {volume} {541}},\
  \bibinfo {pages} {319} (\bibinfo {year} {2000})},\ \Eprint
  {https://arxiv.org/abs/astro-ph/9911417} {arXiv:astro-ph/9911417}
  \BibitemShut {NoStop}%
\bibitem [{\citenamefont {{Fryer}}\ \emph {et~al.}(2012)\citenamefont
  {{Fryer}}, \citenamefont {{Belczynski}}, \citenamefont {{Wiktorowicz}},
  \citenamefont {{Dominik}}, \citenamefont {{Kalogera}},\ and\ \citenamefont
  {{Holz}}}]{2012ApJ...749...91F}%
  \BibitemOpen
  \bibfield  {author} {\bibinfo {author} {\bibfnamefont {C.~L.}\ \bibnamefont
  {{Fryer}}}, \bibinfo {author} {\bibfnamefont {K.}~\bibnamefont
  {{Belczynski}}}, \bibinfo {author} {\bibfnamefont {G.}~\bibnamefont
  {{Wiktorowicz}}}, \bibinfo {author} {\bibfnamefont {M.}~\bibnamefont
  {{Dominik}}}, \bibinfo {author} {\bibfnamefont {V.}~\bibnamefont
  {{Kalogera}}},\ and\ \bibinfo {author} {\bibfnamefont {D.~E.}\ \bibnamefont
  {{Holz}}},\ }\href {https://doi.org/10.1088/0004-637X/749/1/91} {\bibfield
  {journal} {\bibinfo  {journal} {Astrophysics Journal}\ }\textbf {\bibinfo
  {volume} {749}},\ \bibinfo {eid} {91} (\bibinfo {year} {2012})},\ \Eprint
  {https://arxiv.org/abs/1110.1726} {arXiv:1110.1726 [astro-ph.SR]}
  \BibitemShut {NoStop}%
\bibitem [{\citenamefont {Vitale}\ \emph {et~al.}(2022)\citenamefont {Vitale},
  \citenamefont {Biscoveanu},\ and\ \citenamefont {Talbot}}]{Vitale:2022dpa}%
  \BibitemOpen
  \bibfield  {author} {\bibinfo {author} {\bibfnamefont {S.}~\bibnamefont
  {Vitale}}, \bibinfo {author} {\bibfnamefont {S.}~\bibnamefont {Biscoveanu}},\
  and\ \bibinfo {author} {\bibfnamefont {C.}~\bibnamefont {Talbot}},\ }\href
  {https://doi.org/10.1051/0004-6361/202245084} {\bibfield  {journal} {\bibinfo
   {journal} {Astron. Astrophys.}\ }\textbf {\bibinfo {volume} {668}},\
  \bibinfo {pages} {L2} (\bibinfo {year} {2022})},\ \Eprint
  {https://arxiv.org/abs/2209.06978} {arXiv:2209.06978 [astro-ph.HE]}
  \BibitemShut {NoStop}%
\bibitem [{\citenamefont {Li}\ \emph {et~al.}(2024)\citenamefont {Li},
  \citenamefont {Wang}, \citenamefont {Tang},\ and\ \citenamefont
  {Fan}}]{Li:2023yyt}%
  \BibitemOpen
  \bibfield  {author} {\bibinfo {author} {\bibfnamefont {Y.-J.}\ \bibnamefont
  {Li}}, \bibinfo {author} {\bibfnamefont {Y.-Z.}\ \bibnamefont {Wang}},
  \bibinfo {author} {\bibfnamefont {S.-P.}\ \bibnamefont {Tang}},\ and\
  \bibinfo {author} {\bibfnamefont {Y.-Z.}\ \bibnamefont {Fan}},\ }\href
  {https://doi.org/10.1103/PhysRevLett.133.051401} {\bibfield  {journal}
  {\bibinfo  {journal} {Phys. Rev. Lett.}\ }\textbf {\bibinfo {volume} {133}},\
  \bibinfo {pages} {051401} (\bibinfo {year} {2024})},\ \Eprint
  {https://arxiv.org/abs/2303.02973} {arXiv:2303.02973 [astro-ph.HE]}
  \BibitemShut {NoStop}%
\bibitem [{\citenamefont {van~der Sluys}\ \emph {et~al.}(2008)\citenamefont
  {van~der Sluys}, \citenamefont {R\"over}, \citenamefont {Stroeer},
  \citenamefont {Raymond}, \citenamefont {Mandel}, \citenamefont {Christensen},
  \citenamefont {Kalogera}, \citenamefont {Meyer},\ and\ \citenamefont
  {Vecchio}}]{vanderSluys:2007st}%
  \BibitemOpen
  \bibfield  {author} {\bibinfo {author} {\bibfnamefont {M.~V.}\ \bibnamefont
  {van~der Sluys}}, \bibinfo {author} {\bibfnamefont {C.}~\bibnamefont
  {R\"over}}, \bibinfo {author} {\bibfnamefont {A.}~\bibnamefont {Stroeer}},
  \bibinfo {author} {\bibfnamefont {V.}~\bibnamefont {Raymond}}, \bibinfo
  {author} {\bibfnamefont {I.}~\bibnamefont {Mandel}}, \bibinfo {author}
  {\bibfnamefont {N.}~\bibnamefont {Christensen}}, \bibinfo {author}
  {\bibfnamefont {V.}~\bibnamefont {Kalogera}}, \bibinfo {author}
  {\bibfnamefont {R.}~\bibnamefont {Meyer}},\ and\ \bibinfo {author}
  {\bibfnamefont {A.}~\bibnamefont {Vecchio}},\ }\href
  {https://doi.org/10.1086/595279} {\bibfield  {journal} {\bibinfo  {journal}
  {Astrophys. J. Lett.}\ }\textbf {\bibinfo {volume} {688}},\ \bibinfo {pages}
  {L61} (\bibinfo {year} {2008})},\ \Eprint {https://arxiv.org/abs/0710.1897}
  {arXiv:0710.1897 [astro-ph]} \BibitemShut {NoStop}%
\bibitem [{\citenamefont {van~der Sluys}\ \emph {et~al.}(2009)\citenamefont
  {van~der Sluys}, \citenamefont {Mandel}, \citenamefont {Raymond},
  \citenamefont {Kalogera}, \citenamefont {Rover},\ and\ \citenamefont
  {Christensen}}]{vanderSluys:2009bf}%
  \BibitemOpen
  \bibfield  {author} {\bibinfo {author} {\bibfnamefont {M.}~\bibnamefont
  {van~der Sluys}}, \bibinfo {author} {\bibfnamefont {I.}~\bibnamefont
  {Mandel}}, \bibinfo {author} {\bibfnamefont {V.}~\bibnamefont {Raymond}},
  \bibinfo {author} {\bibfnamefont {V.}~\bibnamefont {Kalogera}}, \bibinfo
  {author} {\bibfnamefont {C.}~\bibnamefont {Rover}},\ and\ \bibinfo {author}
  {\bibfnamefont {N.}~\bibnamefont {Christensen}},\ }\href
  {https://doi.org/10.1088/0264-9381/26/20/204010} {\bibfield  {journal}
  {\bibinfo  {journal} {Class. Quant. Grav.}\ }\textbf {\bibinfo {volume}
  {26}},\ \bibinfo {pages} {204010} (\bibinfo {year} {2009})},\ \Eprint
  {https://arxiv.org/abs/0905.1323} {arXiv:0905.1323 [gr-qc]} \BibitemShut
  {NoStop}%
\bibitem [{\citenamefont {Vitale}\ \emph {et~al.}(2014)\citenamefont {Vitale},
  \citenamefont {Lynch}, \citenamefont {Veitch}, \citenamefont {Raymond},\ and\
  \citenamefont {Sturani}}]{Vitale:2014mka}%
  \BibitemOpen
  \bibfield  {author} {\bibinfo {author} {\bibfnamefont {S.}~\bibnamefont
  {Vitale}}, \bibinfo {author} {\bibfnamefont {R.}~\bibnamefont {Lynch}},
  \bibinfo {author} {\bibfnamefont {J.}~\bibnamefont {Veitch}}, \bibinfo
  {author} {\bibfnamefont {V.}~\bibnamefont {Raymond}},\ and\ \bibinfo {author}
  {\bibfnamefont {R.}~\bibnamefont {Sturani}},\ }\href
  {https://doi.org/10.1103/PhysRevLett.112.251101} {\bibfield  {journal}
  {\bibinfo  {journal} {Phys. Rev. Lett.}\ }\textbf {\bibinfo {volume} {112}},\
  \bibinfo {pages} {251101} (\bibinfo {year} {2014})},\ \Eprint
  {https://arxiv.org/abs/1403.0129} {arXiv:1403.0129 [gr-qc]} \BibitemShut
  {NoStop}%
\bibitem [{\citenamefont {P\"urrer}\ \emph {et~al.}(2016)\citenamefont
  {P\"urrer}, \citenamefont {Hannam},\ and\ \citenamefont
  {Ohme}}]{Purrer:2015nkh}%
  \BibitemOpen
  \bibfield  {author} {\bibinfo {author} {\bibfnamefont {M.}~\bibnamefont
  {P\"urrer}}, \bibinfo {author} {\bibfnamefont {M.}~\bibnamefont {Hannam}},\
  and\ \bibinfo {author} {\bibfnamefont {F.}~\bibnamefont {Ohme}},\ }\href
  {https://doi.org/10.1103/PhysRevD.93.084042} {\bibfield  {journal} {\bibinfo
  {journal} {Phys. Rev. D}\ }\textbf {\bibinfo {volume} {93}},\ \bibinfo
  {pages} {084042} (\bibinfo {year} {2016})},\ \Eprint
  {https://arxiv.org/abs/1512.04955} {arXiv:1512.04955 [gr-qc]} \BibitemShut
  {NoStop}%
\bibitem [{\citenamefont {Vitale}\ \emph
  {et~al.}(2017{\natexlab{b}})\citenamefont {Vitale}, \citenamefont {Lynch},
  \citenamefont {Raymond}, \citenamefont {Sturani}, \citenamefont {Veitch},\
  and\ \citenamefont {Graff}}]{Vitale:2016avz}%
  \BibitemOpen
  \bibfield  {author} {\bibinfo {author} {\bibfnamefont {S.}~\bibnamefont
  {Vitale}}, \bibinfo {author} {\bibfnamefont {R.}~\bibnamefont {Lynch}},
  \bibinfo {author} {\bibfnamefont {V.}~\bibnamefont {Raymond}}, \bibinfo
  {author} {\bibfnamefont {R.}~\bibnamefont {Sturani}}, \bibinfo {author}
  {\bibfnamefont {J.}~\bibnamefont {Veitch}},\ and\ \bibinfo {author}
  {\bibfnamefont {P.}~\bibnamefont {Graff}},\ }\href
  {https://doi.org/10.1103/PhysRevD.95.064053} {\bibfield  {journal} {\bibinfo
  {journal} {Phys. Rev. D}\ }\textbf {\bibinfo {volume} {95}},\ \bibinfo
  {pages} {064053} (\bibinfo {year} {2017}{\natexlab{b}})},\ \Eprint
  {https://arxiv.org/abs/1611.01122} {arXiv:1611.01122 [gr-qc]} \BibitemShut
  {NoStop}%
\bibitem [{\citenamefont {Dhani}\ \emph {et~al.}(2024)\citenamefont {Dhani},
  \citenamefont {V\"olkel}, \citenamefont {Buonanno}, \citenamefont {Estelles},
  \citenamefont {Gair}, \citenamefont {Pfeiffer}, \citenamefont {Pompili},\
  and\ \citenamefont {Toubiana}}]{Dhani:2024jja}%
  \BibitemOpen
  \bibfield  {author} {\bibinfo {author} {\bibfnamefont {A.}~\bibnamefont
  {Dhani}}, \bibinfo {author} {\bibfnamefont {S.}~\bibnamefont {V\"olkel}},
  \bibinfo {author} {\bibfnamefont {A.}~\bibnamefont {Buonanno}}, \bibinfo
  {author} {\bibfnamefont {H.}~\bibnamefont {Estelles}}, \bibinfo {author}
  {\bibfnamefont {J.}~\bibnamefont {Gair}}, \bibinfo {author} {\bibfnamefont
  {H.~P.}\ \bibnamefont {Pfeiffer}}, \bibinfo {author} {\bibfnamefont
  {L.}~\bibnamefont {Pompili}},\ and\ \bibinfo {author} {\bibfnamefont
  {A.}~\bibnamefont {Toubiana}},\ }\href@noop {} {\bibfield  {journal}
  {\bibinfo  {journal} {Arxiv}\ } (\bibinfo {year} {2024})},\ \Eprint
  {https://arxiv.org/abs/2404.05811} {arXiv:2404.05811 [gr-qc]} \BibitemShut
  {NoStop}%
\bibitem [{\citenamefont {Miller}\ \emph {et~al.}(2024)\citenamefont {Miller},
  \citenamefont {Ko}, \citenamefont {Callister},\ and\ \citenamefont
  {Chatziioannou}}]{Miller:2024sui}%
  \BibitemOpen
  \bibfield  {author} {\bibinfo {author} {\bibfnamefont {S.~J.}\ \bibnamefont
  {Miller}}, \bibinfo {author} {\bibfnamefont {Z.}~\bibnamefont {Ko}}, \bibinfo
  {author} {\bibfnamefont {T.}~\bibnamefont {Callister}},\ and\ \bibinfo
  {author} {\bibfnamefont {K.}~\bibnamefont {Chatziioannou}},\ }\href
  {https://doi.org/10.1103/PhysRevD.109.104036} {\bibfield  {journal} {\bibinfo
   {journal} {Phys. Rev. D}\ }\textbf {\bibinfo {volume} {109}},\ \bibinfo
  {pages} {104036} (\bibinfo {year} {2024})},\ \Eprint
  {https://arxiv.org/abs/2401.05613} {arXiv:2401.05613 [gr-qc]} \BibitemShut
  {NoStop}%
\bibitem [{\citenamefont {Wysocki}\ \emph {et~al.}(2019)\citenamefont
  {Wysocki}, \citenamefont {Lange},\ and\ \citenamefont
  {O'Shaughnessy}}]{Wysocki:2018}%
  \BibitemOpen
  \bibfield  {author} {\bibinfo {author} {\bibfnamefont {D.}~\bibnamefont
  {Wysocki}}, \bibinfo {author} {\bibfnamefont {J.}~\bibnamefont {Lange}},\
  and\ \bibinfo {author} {\bibfnamefont {R.}~\bibnamefont {O'Shaughnessy}},\
  }\href {https://doi.org/10.1103/PhysRevD.100.043012} {\bibfield  {journal}
  {\bibinfo  {journal} {Phys. Rev. D}\ }\textbf {\bibinfo {volume} {100}},\
  \bibinfo {pages} {043012} (\bibinfo {year} {2019})}\BibitemShut {NoStop}%
\bibitem [{\citenamefont {{LIGO Scientific Collaboration}}()}]{LIGOT2000012}%
  \BibitemOpen
  \bibfield  {author} {\bibinfo {author} {\bibnamefont {{LIGO Scientific
  Collaboration}}},\ }\href@noop {} {\bibinfo {title} {{Noise curves used for
  Simulations in the update of the Observing Scenarios Paper}, howpublished =
  {\url{https://dcc.ligo.org/ligo-t2000012/public}}, note = {LIGO-T2000012-v5},
  year = {2020}}}\BibitemShut {NoStop}%
\bibitem [{\citenamefont {Essick}\ and\ \citenamefont
  {Fishbach}(2024)}]{Essick:2023upv}%
  \BibitemOpen
  \bibfield  {author} {\bibinfo {author} {\bibfnamefont {R.}~\bibnamefont
  {Essick}}\ and\ \bibinfo {author} {\bibfnamefont {M.}~\bibnamefont
  {Fishbach}},\ }\href {https://doi.org/10.3847/1538-4357/ad1604} {\bibfield
  {journal} {\bibinfo  {journal} {Astrophys. J.}\ }\textbf {\bibinfo {volume}
  {962}},\ \bibinfo {pages} {169} (\bibinfo {year} {2024})},\ \Eprint
  {https://arxiv.org/abs/2310.02017} {arXiv:2310.02017 [gr-qc]} \BibitemShut
  {NoStop}%
\bibitem [{\citenamefont {Ashton}\ \emph {et~al.}(2019)\citenamefont {Ashton}
  \emph {et~al.}}]{Ashton:2018jfp}%
  \BibitemOpen
  \bibfield  {author} {\bibinfo {author} {\bibfnamefont {G.}~\bibnamefont
  {Ashton}} \emph {et~al.},\ }\href {https://doi.org/10.3847/1538-4365/ab06fc}
  {\bibfield  {journal} {\bibinfo  {journal} {Astrophys. J. Suppl.}\ }\textbf
  {\bibinfo {volume} {241}},\ \bibinfo {pages} {27} (\bibinfo {year} {2019})},\
  \Eprint {https://arxiv.org/abs/1811.02042} {arXiv:1811.02042 [astro-ph.IM]}
  \BibitemShut {NoStop}%
\bibitem [{\citenamefont {Romero-Shaw}\ \emph {et~al.}(2020)\citenamefont
  {Romero-Shaw} \emph {et~al.}}]{Romero-Shaw:2020owr}%
  \BibitemOpen
  \bibfield  {author} {\bibinfo {author} {\bibfnamefont {I.~M.}\ \bibnamefont
  {Romero-Shaw}} \emph {et~al.},\ }\href
  {https://doi.org/10.1093/mnras/staa2850} {\bibfield  {journal} {\bibinfo
  {journal} {Mon. Not. Roy. Astron. Soc.}\ }\textbf {\bibinfo {volume} {499}},\
  \bibinfo {pages} {3295} (\bibinfo {year} {2020})},\ \Eprint
  {https://arxiv.org/abs/2006.00714} {arXiv:2006.00714 [astro-ph.IM]}
  \BibitemShut {NoStop}%
\bibitem [{\citenamefont {Pratten}\ \emph {et~al.}(2020)\citenamefont
  {Pratten}, \citenamefont {Husa}, \citenamefont {Garcia-Quiros}, \citenamefont
  {Colleoni}, \citenamefont {Ramos-Buades}, \citenamefont {Estelles},\ and\
  \citenamefont {Jaume}}]{Pratten:2020fqn}%
  \BibitemOpen
  \bibfield  {author} {\bibinfo {author} {\bibfnamefont {G.}~\bibnamefont
  {Pratten}}, \bibinfo {author} {\bibfnamefont {S.}~\bibnamefont {Husa}},
  \bibinfo {author} {\bibfnamefont {C.}~\bibnamefont {Garcia-Quiros}}, \bibinfo
  {author} {\bibfnamefont {M.}~\bibnamefont {Colleoni}}, \bibinfo {author}
  {\bibfnamefont {A.}~\bibnamefont {Ramos-Buades}}, \bibinfo {author}
  {\bibfnamefont {H.}~\bibnamefont {Estelles}},\ and\ \bibinfo {author}
  {\bibfnamefont {R.}~\bibnamefont {Jaume}},\ }\href
  {https://doi.org/10.1103/PhysRevD.102.064001} {\bibfield  {journal} {\bibinfo
   {journal} {Phys. Rev. D}\ }\textbf {\bibinfo {volume} {102}},\ \bibinfo
  {pages} {064001} (\bibinfo {year} {2020})},\ \Eprint
  {https://arxiv.org/abs/2001.11412} {arXiv:2001.11412 [gr-qc]} \BibitemShut
  {NoStop}%
\bibitem [{\citenamefont {Krishna}\ \emph {et~al.}(2023)\citenamefont
  {Krishna}, \citenamefont {Vijaykumar}, \citenamefont {Ganguly}, \citenamefont
  {Talbot}, \citenamefont {Biscoveanu}, \citenamefont {George}, \citenamefont
  {Williams},\ and\ \citenamefont {Zimmerman}}]{Krishna:2023bug}%
  \BibitemOpen
  \bibfield  {author} {\bibinfo {author} {\bibfnamefont {K.}~\bibnamefont
  {Krishna}}, \bibinfo {author} {\bibfnamefont {A.}~\bibnamefont {Vijaykumar}},
  \bibinfo {author} {\bibfnamefont {A.}~\bibnamefont {Ganguly}}, \bibinfo
  {author} {\bibfnamefont {C.}~\bibnamefont {Talbot}}, \bibinfo {author}
  {\bibfnamefont {S.}~\bibnamefont {Biscoveanu}}, \bibinfo {author}
  {\bibfnamefont {R.~N.}\ \bibnamefont {George}}, \bibinfo {author}
  {\bibfnamefont {N.}~\bibnamefont {Williams}},\ and\ \bibinfo {author}
  {\bibfnamefont {A.}~\bibnamefont {Zimmerman}},\ }\href@noop {} {\bibfield
  {journal} {\bibinfo  {journal} {Arxiv}\ } (\bibinfo {year} {2023})},\ \Eprint
  {https://arxiv.org/abs/2312.06009} {arXiv:2312.06009 [gr-qc]} \BibitemShut
  {NoStop}%
\bibitem [{\citenamefont {Talbot}\ \emph {et~al.}(2024)\citenamefont {Talbot},
  \citenamefont {Farah}, \citenamefont {Galaudage}, \citenamefont {Golomb},\
  and\ \citenamefont {Tong}}]{Talbot:2024yqw}%
  \BibitemOpen
  \bibfield  {author} {\bibinfo {author} {\bibfnamefont {C.}~\bibnamefont
  {Talbot}}, \bibinfo {author} {\bibfnamefont {A.}~\bibnamefont {Farah}},
  \bibinfo {author} {\bibfnamefont {S.}~\bibnamefont {Galaudage}}, \bibinfo
  {author} {\bibfnamefont {J.}~\bibnamefont {Golomb}},\ and\ \bibinfo {author}
  {\bibfnamefont {H.}~\bibnamefont {Tong}},\ }\href@noop {} {\bibfield
  {journal} {\bibinfo  {journal} {Arxiv}\ } (\bibinfo {year} {2024})},\ \Eprint
  {https://arxiv.org/abs/2409.14143} {arXiv:2409.14143 [astro-ph.IM]}
  \BibitemShut {NoStop}%
\bibitem [{\citenamefont {{Loredo}}\ and\ \citenamefont
  {{Wasserman}}(1995)}]{1995ApJS...96..261L}%
  \BibitemOpen
  \bibfield  {author} {\bibinfo {author} {\bibfnamefont {T.~J.}\ \bibnamefont
  {{Loredo}}}\ and\ \bibinfo {author} {\bibfnamefont {I.~M.}\ \bibnamefont
  {{Wasserman}}},\ }\href {https://doi.org/10.1086/192119} {\bibfield
  {journal} {\bibinfo  {journal} {Astrophysical Journal Supplement Series}\
  }\textbf {\bibinfo {volume} {96}},\ \bibinfo {pages} {261} (\bibinfo {year}
  {1995})}\BibitemShut {NoStop}%
\bibitem [{\citenamefont {Loredo}\ and\ \citenamefont
  {Wasserman}(1998)}]{Loredo:1997qf}%
  \BibitemOpen
  \bibfield  {author} {\bibinfo {author} {\bibfnamefont {T.~J.}\ \bibnamefont
  {Loredo}}\ and\ \bibinfo {author} {\bibfnamefont {I.~M.}\ \bibnamefont
  {Wasserman}},\ }\href {https://doi.org/10.1086/305870} {\bibfield  {journal}
  {\bibinfo  {journal} {Astrophys. J.}\ }\textbf {\bibinfo {volume} {502}},\
  \bibinfo {pages} {75} (\bibinfo {year} {1998})},\ \Eprint
  {https://arxiv.org/abs/astro-ph/9701111} {arXiv:astro-ph/9701111}
  \BibitemShut {NoStop}%
\bibitem [{\citenamefont {Loredo}\ and\ \citenamefont
  {Lamb}(2002)}]{Loredo:2001rx}%
  \BibitemOpen
  \bibfield  {author} {\bibinfo {author} {\bibfnamefont {T.~J.}\ \bibnamefont
  {Loredo}}\ and\ \bibinfo {author} {\bibfnamefont {D.~Q.}\ \bibnamefont
  {Lamb}},\ }\href {https://doi.org/10.1103/PhysRevD.65.063002} {\bibfield
  {journal} {\bibinfo  {journal} {Phys. Rev. D}\ }\textbf {\bibinfo {volume}
  {65}},\ \bibinfo {pages} {063002} (\bibinfo {year} {2002})},\ \Eprint
  {https://arxiv.org/abs/astro-ph/0107260} {arXiv:astro-ph/0107260}
  \BibitemShut {NoStop}%
\bibitem [{\citenamefont {Mandel}\ \emph {et~al.}(2019)\citenamefont {Mandel},
  \citenamefont {Farr},\ and\ \citenamefont {Gair}}]{Mandel:2018mve}%
  \BibitemOpen
  \bibfield  {author} {\bibinfo {author} {\bibfnamefont {I.}~\bibnamefont
  {Mandel}}, \bibinfo {author} {\bibfnamefont {W.~M.}\ \bibnamefont {Farr}},\
  and\ \bibinfo {author} {\bibfnamefont {J.~R.}\ \bibnamefont {Gair}},\ }\href
  {https://doi.org/10.1093/mnras/stz896} {\bibfield  {journal} {\bibinfo
  {journal} {Mon. Not. Roy. Astron. Soc.}\ }\textbf {\bibinfo {volume} {486}},\
  \bibinfo {pages} {1086} (\bibinfo {year} {2019})},\ \Eprint
  {https://arxiv.org/abs/1809.02063} {arXiv:1809.02063 [physics.data-an]}
  \BibitemShut {NoStop}%
\bibitem [{\citenamefont {Vitale}\ \emph {et~al.}(2020)\citenamefont {Vitale},
  \citenamefont {Gerosa}, \citenamefont {Farr},\ and\ \citenamefont
  {Taylor}}]{Vitale:2020aaz}%
  \BibitemOpen
  \bibfield  {author} {\bibinfo {author} {\bibfnamefont {S.}~\bibnamefont
  {Vitale}}, \bibinfo {author} {\bibfnamefont {D.}~\bibnamefont {Gerosa}},
  \bibinfo {author} {\bibfnamefont {W.~M.}\ \bibnamefont {Farr}},\ and\
  \bibinfo {author} {\bibfnamefont {S.~R.}\ \bibnamefont {Taylor}},\ }\bibfield
   {journal} {\bibinfo  {journal} {Handbook of Gravitational Wave Astronomy}\
  }\href {https://doi.org/"10.1007/978-981-15-4702-7\_45-1"}
  {"10.1007/978-981-15-4702-7\_45-1"} (\bibinfo {year} {2020}),\ \Eprint
  {https://arxiv.org/abs/2007.05579} {arXiv:2007.05579 [astro-ph.IM]}
  \BibitemShut {NoStop}%
\bibitem [{\citenamefont {{Thrane}}\ and\ \citenamefont
  {{Talbot}}(2019)}]{2019PASA...36...10T}%
  \BibitemOpen
  \bibfield  {author} {\bibinfo {author} {\bibfnamefont {E.}~\bibnamefont
  {{Thrane}}}\ and\ \bibinfo {author} {\bibfnamefont {C.}~\bibnamefont
  {{Talbot}}},\ }\href {https://doi.org/10.1017/pasa.2019.2} {\bibfield
  {journal} {\bibinfo  {journal} {Publications of the Astronomical Society of
  Australia}\ }\textbf {\bibinfo {volume} {36}},\ \bibinfo {eid} {e010}
  (\bibinfo {year} {2019})},\ \Eprint {https://arxiv.org/abs/1809.02293}
  {arXiv:1809.02293 [astro-ph.IM]} \BibitemShut {NoStop}%
\bibitem [{\citenamefont {Speagle}(2020)}]{Speagle_2020}%
  \BibitemOpen
  \bibfield  {author} {\bibinfo {author} {\bibfnamefont {J.~S.}\ \bibnamefont
  {Speagle}},\ }\href {https://doi.org/10.1093/mnras/staa278} {\bibfield
  {journal} {\bibinfo  {journal} {MNRAS}\ }\textbf {\bibinfo {volume} {493}},\
  \bibinfo {pages} {3132} (\bibinfo {year} {2020})},\ \Eprint
  {https://arxiv.org/abs/https://academic.oup.com/mnras/article-pdf/493/3/3132/32890730/staa278.pdf}
  {https://academic.oup.com/mnras/article-pdf/493/3/3132/32890730/staa278.pdf}
  \BibitemShut {NoStop}%
\bibitem [{\citenamefont {Tiwari}(2018)}]{Tiwari:2017ndi}%
  \BibitemOpen
  \bibfield  {author} {\bibinfo {author} {\bibfnamefont {V.}~\bibnamefont
  {Tiwari}},\ }\href {https://doi.org/10.1088/1361-6382/aac89d} {\bibfield
  {journal} {\bibinfo  {journal} {Class. Quant. Grav.}\ }\textbf {\bibinfo
  {volume} {35}},\ \bibinfo {pages} {145009} (\bibinfo {year} {2018})},\
  \Eprint {https://arxiv.org/abs/1712.00482} {arXiv:1712.00482 [astro-ph.HE]}
  \BibitemShut {NoStop}%
\bibitem [{\citenamefont {{Farr}}(2019)}]{2019RNAAS...3...66F}%
  \BibitemOpen
  \bibfield  {author} {\bibinfo {author} {\bibfnamefont {W.~M.}\ \bibnamefont
  {{Farr}}},\ }\href {https://doi.org/10.3847/2515-5172/ab1d5f} {\bibfield
  {journal} {\bibinfo  {journal} {Research Notes of the American Astronomical
  Society}\ }\textbf {\bibinfo {volume} {3}},\ \bibinfo {eid} {66} (\bibinfo
  {year} {2019})},\ \Eprint {https://arxiv.org/abs/1904.10879}
  {arXiv:1904.10879 [astro-ph.IM]} \BibitemShut {NoStop}%
\bibitem [{\citenamefont {Talbot}\ and\ \citenamefont
  {Golomb}(2023)}]{Talbot:2023pex}%
  \BibitemOpen
  \bibfield  {author} {\bibinfo {author} {\bibfnamefont {C.}~\bibnamefont
  {Talbot}}\ and\ \bibinfo {author} {\bibfnamefont {J.}~\bibnamefont
  {Golomb}},\ }\href {https://doi.org/10.1093/mnras/stad2968} {\bibfield
  {journal} {\bibinfo  {journal} {Mon. Not. Roy. Astron. Soc.}\ }\textbf
  {\bibinfo {volume} {526}},\ \bibinfo {pages} {3495} (\bibinfo {year}
  {2023})},\ \Eprint {https://arxiv.org/abs/2304.06138} {arXiv:2304.06138
  [astro-ph.IM]} \BibitemShut {NoStop}%
\bibitem [{\citenamefont {Essick}(2021)}]{Essick_2021}%
  \BibitemOpen
  \bibfield  {author} {\bibinfo {author} {\bibfnamefont {R.}~\bibnamefont
  {Essick}},\ }\href {https://doi.org/10.3847/2515-5172/ac2ba7} {\bibfield
  {journal} {\bibinfo  {journal} {Research Notes of the AAS}\ }\textbf
  {\bibinfo {volume} {5}},\ \bibinfo {pages} {220} (\bibinfo {year}
  {2021})}\BibitemShut {NoStop}%
\bibitem [{\citenamefont {Essick}(2023)}]{Essick:2023toz}%
  \BibitemOpen
  \bibfield  {author} {\bibinfo {author} {\bibfnamefont {R.}~\bibnamefont
  {Essick}},\ }\href {https://doi.org/10.1103/PhysRevD.108.043011} {\bibfield
  {journal} {\bibinfo  {journal} {Phys. Rev. D}\ }\textbf {\bibinfo {volume}
  {108}},\ \bibinfo {pages} {043011} (\bibinfo {year} {2023})},\ \Eprint
  {https://arxiv.org/abs/2307.02765} {arXiv:2307.02765 [gr-qc]} \BibitemShut
  {NoStop}%
\bibitem [{\citenamefont {Lorenzo-Medina}\ and\ \citenamefont
  {Dent}(2025)}]{Lorenzo-Medina:2024opt}%
  \BibitemOpen
  \bibfield  {author} {\bibinfo {author} {\bibfnamefont {A.}~\bibnamefont
  {Lorenzo-Medina}}\ and\ \bibinfo {author} {\bibfnamefont {T.}~\bibnamefont
  {Dent}},\ }\href {https://doi.org/10.1088/1361-6382/ad9c0e} {\bibfield
  {journal} {\bibinfo  {journal} {Class. Quant. Grav.}\ }\textbf {\bibinfo
  {volume} {42}},\ \bibinfo {pages} {045008} (\bibinfo {year} {2025})},\
  \Eprint {https://arxiv.org/abs/2408.13383} {arXiv:2408.13383 [gr-qc]}
  \BibitemShut {NoStop}%
\bibitem [{\citenamefont {Islam}\ \emph {et~al.}(2023)\citenamefont {Islam},
  \citenamefont {Vajpeyi}, \citenamefont {Shaik}, \citenamefont {Haster},
  \citenamefont {Varma}, \citenamefont {Field}, \citenamefont {Lange},
  \citenamefont {O'Shaughnessy},\ and\ \citenamefont {Smith}}]{Islam:2023zzj}%
  \BibitemOpen
  \bibfield  {author} {\bibinfo {author} {\bibfnamefont {T.}~\bibnamefont
  {Islam}}, \bibinfo {author} {\bibfnamefont {A.}~\bibnamefont {Vajpeyi}},
  \bibinfo {author} {\bibfnamefont {F.~H.}\ \bibnamefont {Shaik}}, \bibinfo
  {author} {\bibfnamefont {C.-J.}\ \bibnamefont {Haster}}, \bibinfo {author}
  {\bibfnamefont {V.}~\bibnamefont {Varma}}, \bibinfo {author} {\bibfnamefont
  {S.~E.}\ \bibnamefont {Field}}, \bibinfo {author} {\bibfnamefont
  {J.}~\bibnamefont {Lange}}, \bibinfo {author} {\bibfnamefont
  {R.}~\bibnamefont {O'Shaughnessy}},\ and\ \bibinfo {author} {\bibfnamefont
  {R.}~\bibnamefont {Smith}},\ }\href@noop {} {\bibfield  {journal} {\bibinfo
  {journal} {Arxiv}\ } (\bibinfo {year} {2023})},\ \Eprint
  {https://arxiv.org/abs/2309.14473} {arXiv:2309.14473 [gr-qc]} \BibitemShut
  {NoStop}%
\bibitem [{\citenamefont {Kiendrebeogo}\ \emph {et~al.}(2023)\citenamefont
  {Kiendrebeogo} \emph {et~al.}}]{Kiendrebeogo:2023hzf}%
  \BibitemOpen
  \bibfield  {author} {\bibinfo {author} {\bibfnamefont {R.~W.}\ \bibnamefont
  {Kiendrebeogo}} \emph {et~al.},\ }\href
  {https://doi.org/10.3847/1538-4357/acfcb1} {\bibfield  {journal} {\bibinfo
  {journal} {Astrophys. J.}\ }\textbf {\bibinfo {volume} {958}},\ \bibinfo
  {pages} {158} (\bibinfo {year} {2023})},\ \Eprint
  {https://arxiv.org/abs/2306.09234} {arXiv:2306.09234 [astro-ph.HE]}
  \BibitemShut {NoStop}%
\bibitem [{\citenamefont {Damour}(2001)}]{Damour:2001tu}%
  \BibitemOpen
  \bibfield  {author} {\bibinfo {author} {\bibfnamefont {T.}~\bibnamefont
  {Damour}},\ }\href {https://doi.org/10.1103/PhysRevD.64.124013} {\bibfield
  {journal} {\bibinfo  {journal} {Phys. Rev. D}\ }\textbf {\bibinfo {volume}
  {64}},\ \bibinfo {pages} {124013} (\bibinfo {year} {2001})},\ \Eprint
  {https://arxiv.org/abs/gr-qc/0103018} {arXiv:gr-qc/0103018} \BibitemShut
  {NoStop}%
\bibitem [{\citenamefont {Racine}(2008)}]{Racine:2008qv}%
  \BibitemOpen
  \bibfield  {author} {\bibinfo {author} {\bibfnamefont {E.}~\bibnamefont
  {Racine}},\ }\href {https://doi.org/10.1103/PhysRevD.78.044021} {\bibfield
  {journal} {\bibinfo  {journal} {Phys. Rev. D}\ }\textbf {\bibinfo {volume}
  {78}},\ \bibinfo {pages} {044021} (\bibinfo {year} {2008})},\ \Eprint
  {https://arxiv.org/abs/0803.1820} {arXiv:0803.1820 [gr-qc]} \BibitemShut
  {NoStop}%
\bibitem [{\citenamefont {Santamaria}\ \emph {et~al.}(2010)\citenamefont
  {Santamaria} \emph {et~al.}}]{Santamaria:2010yb}%
  \BibitemOpen
  \bibfield  {author} {\bibinfo {author} {\bibfnamefont {L.}~\bibnamefont
  {Santamaria}} \emph {et~al.},\ }\href
  {https://doi.org/10.1103/PhysRevD.82.064016} {\bibfield  {journal} {\bibinfo
  {journal} {Phys. Rev. D}\ }\textbf {\bibinfo {volume} {82}},\ \bibinfo
  {pages} {064016} (\bibinfo {year} {2010})},\ \Eprint
  {https://arxiv.org/abs/1005.3306} {arXiv:1005.3306 [gr-qc]} \BibitemShut
  {NoStop}%
\bibitem [{\citenamefont {Abbott}\ \emph
  {et~al.}(2021{\natexlab{c}})\citenamefont {Abbott} \emph
  {et~al.}}]{LIGOScientific:2021djp}%
  \BibitemOpen
  \bibfield  {author} {\bibinfo {author} {\bibfnamefont {R.}~\bibnamefont
  {Abbott}} \emph {et~al.} (\bibinfo {collaboration} {LIGO Scientific, VIRGO,
  KAGRA}),\ }\href@noop {} {\bibfield  {journal} {\bibinfo  {journal} {arXiv
  e-prints}\ ,\ \bibinfo {eid} {arXiv:2111.03606}} (\bibinfo {year}
  {2021}{\natexlab{c}})},\ \Eprint {https://arxiv.org/abs/2111.03606}
  {arXiv:2111.03606 [gr-qc]} \BibitemShut {NoStop}%
\bibitem [{\citenamefont {Varma}\ \emph
  {et~al.}(2019{\natexlab{a}})\citenamefont {Varma}, \citenamefont {Field},
  \citenamefont {Scheel}, \citenamefont {Blackman}, \citenamefont {Gerosa},
  \citenamefont {Stein}, \citenamefont {Kidder},\ and\ \citenamefont
  {Pfeiffer}}]{Varma:2019csw}%
  \BibitemOpen
  \bibfield  {author} {\bibinfo {author} {\bibfnamefont {V.}~\bibnamefont
  {Varma}}, \bibinfo {author} {\bibfnamefont {S.~E.}\ \bibnamefont {Field}},
  \bibinfo {author} {\bibfnamefont {M.~A.}\ \bibnamefont {Scheel}}, \bibinfo
  {author} {\bibfnamefont {J.}~\bibnamefont {Blackman}}, \bibinfo {author}
  {\bibfnamefont {D.}~\bibnamefont {Gerosa}}, \bibinfo {author} {\bibfnamefont
  {L.~C.}\ \bibnamefont {Stein}}, \bibinfo {author} {\bibfnamefont {L.~E.}\
  \bibnamefont {Kidder}},\ and\ \bibinfo {author} {\bibfnamefont {H.~P.}\
  \bibnamefont {Pfeiffer}},\ }\href
  {https://doi.org/10.1103/PhysRevResearch.1.033015} {\bibfield  {journal}
  {\bibinfo  {journal} {Phys. Rev. Research.}\ }\textbf {\bibinfo {volume}
  {1}},\ \bibinfo {pages} {033015} (\bibinfo {year} {2019}{\natexlab{a}})},\
  \Eprint {https://arxiv.org/abs/1905.09300} {arXiv:1905.09300 [gr-qc]}
  \BibitemShut {NoStop}%
\bibitem [{\citenamefont {Varma}\ \emph
  {et~al.}(2019{\natexlab{b}})\citenamefont {Varma}, \citenamefont {Field},
  \citenamefont {Scheel}, \citenamefont {Blackman}, \citenamefont {Kidder},\
  and\ \citenamefont {Pfeiffer}}]{Varma:2018mmi}%
  \BibitemOpen
  \bibfield  {author} {\bibinfo {author} {\bibfnamefont {V.}~\bibnamefont
  {Varma}}, \bibinfo {author} {\bibfnamefont {S.~E.}\ \bibnamefont {Field}},
  \bibinfo {author} {\bibfnamefont {M.~A.}\ \bibnamefont {Scheel}}, \bibinfo
  {author} {\bibfnamefont {J.}~\bibnamefont {Blackman}}, \bibinfo {author}
  {\bibfnamefont {L.~E.}\ \bibnamefont {Kidder}},\ and\ \bibinfo {author}
  {\bibfnamefont {H.~P.}\ \bibnamefont {Pfeiffer}},\ }\href
  {https://doi.org/10.1103/PhysRevD.99.064045} {\bibfield  {journal} {\bibinfo
  {journal} {Phys. Rev. D}\ }\textbf {\bibinfo {volume} {99}},\ \bibinfo
  {pages} {064045} (\bibinfo {year} {2019}{\natexlab{b}})},\ \Eprint
  {https://arxiv.org/abs/1812.07865} {arXiv:1812.07865 [gr-qc]} \BibitemShut
  {NoStop}%
\bibitem [{\citenamefont {{Vitale}}\ and\ \citenamefont
  {{Mould}}(2025)}]{vitale_data_2025}%
  \BibitemOpen
  \bibfield  {author} {\bibinfo {author} {\bibfnamefont {S.}~\bibnamefont
  {{Vitale}}}\ and\ \bibinfo {author} {\bibfnamefont {M.}~\bibnamefont
  {{Mould}}},\ }\href {https://doi.org/10.5281/zenodo.15492970} {\bibinfo
  {title} {Data for: The long road to alignment: Measuring black hole spin
  orientation with expanding gravitational-wave datasets}} (\bibinfo {year}
  {2025})\BibitemShut {NoStop}%
\end{thebibliography}%

\clearpage
\appendix

\onecolumngrid
\section{Simulated population and \gwtcthree results}\label{App.Truth}

As mentioned in the body of the paper, the BBH population we simulated is consistent with that measured using the \lvk model in \gwtcthree data~\cite{Vitale:2022dpa}. This is shown in Fig.~\ref{Fig.TrueCorner} below, where posteriors from \gwtcthree are shown in blue, and the \truthcolor marks the values used for our simulations. Notice that the fraction of sources in the non-isotropic tilt component, \fa, is mass-ratio dependent in our population. 

\begin{figure}[b]
\centering
\includegraphics[width=\textwidth]{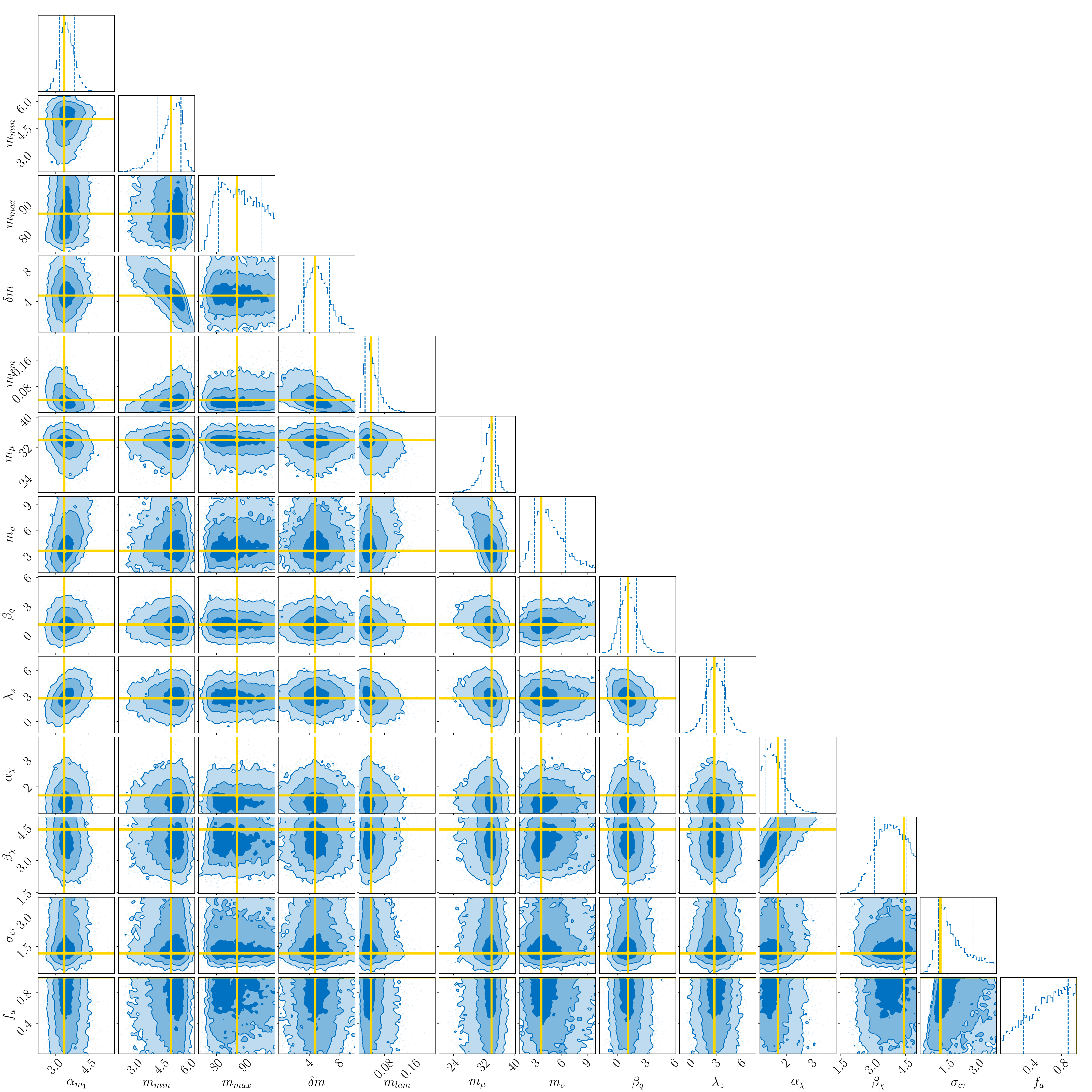}
\vspace{0.5em}
\caption{The corner plot shows the \gwtcthree hyper posterior of the \lvk model. The \truthcolor lines mark the values used to generate our synthetic population.}
\label{Fig.TrueCorner}
\end{figure}
\vfill
\clearpage

\section{PPD for other hyper parameters}\label{App.PPD}

Even though the focus of this paper are the tilt angles, we should at least mention that the measurements of all other parameters -- primary mass $m_1$, mass ratio $q \in [0.1,1]$, redshift $z$ and spin magnitude $\chi$ -- are also unbiased for all models and catalog sizes. 

\begin{figure}[h]
\centering
\begin{minipage}[t]{0.48\textwidth}
  \includegraphics[width=\linewidth]{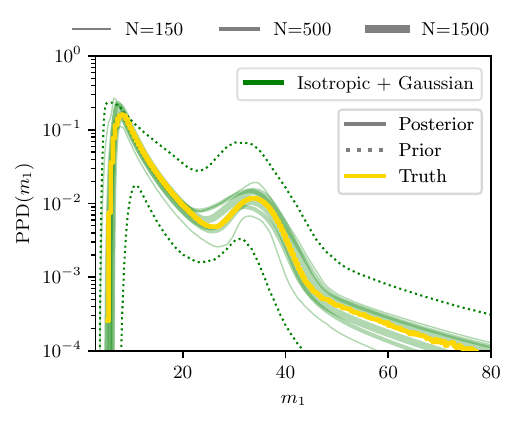}
\end{minipage}
\hfill
\begin{minipage}[t]{0.48\textwidth}
  \includegraphics[width=\linewidth]{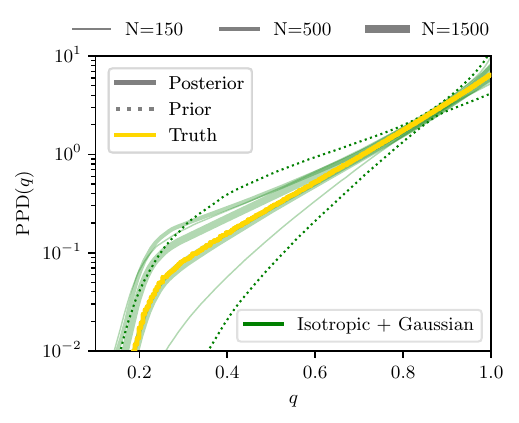}
\end{minipage}

\vspace{1em}

\begin{minipage}[t]{0.48\textwidth}
  \includegraphics[width=\linewidth]{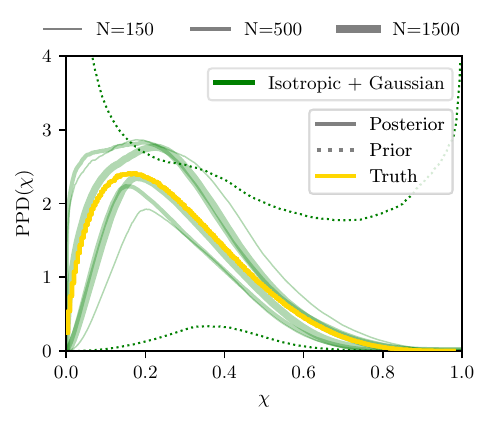}
\end{minipage}
\hfill
\begin{minipage}[t]{0.48\textwidth}
  \includegraphics[width=\linewidth]{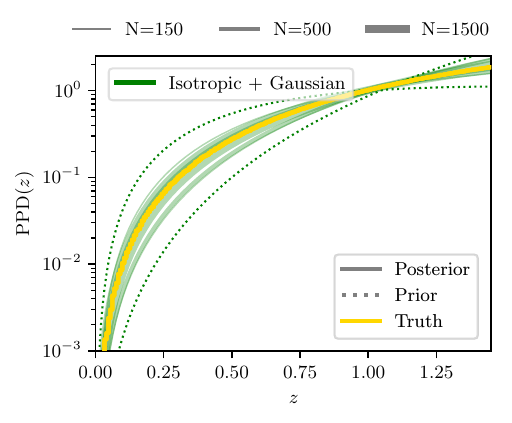}
\end{minipage}

\caption{PPD for the parameters $m_1$, $q$, $\chi$, and $z$ for the \isogaus model. Posteriors are shown for catalogs with 150, 500, and 1500 sources.}
\label{Fig.OtherParsPPDIsoGauss}
\end{figure}

In Fig.~\ref{Fig.OtherParsPPDIsoGauss} we show the PPDs for these parameters obtained with the \isogaus model and all three catalog sizes. The solid curves enclose 90\% of the posterior and the dotted lines 90\% of the prior. The \truthcolor curves are the true distributions of these parameters and are generally included within the 90\% credible intervals. No significant biases are seen anywhere. 
The PPDs for mass, mass ratio and redshift are nearly indistinguishable when our other models are used. For the \isobeta and \isocbeta models only, a small difference is visible in $\chi$'s PPD: the upper edge of the credible region moves up by $\sim10\%$ at $\chi~\in[0-0.2]$ compared to the \isogaus and all other models. 

\vspace{1em}

\vfill
\clearpage

\section{Tables}\label{App.Tables}

We provide some tabulated values for median and 90\% uncertainties for posterior and priors for the PPD sliced at $-1,-0.5,0,0.5,1$ in Tab.~\ref{tab:cosTau_slices_uncorr} (uncorrelated models) and Tab.~\ref{tab:cosTau_slices_corr} (correlated models). Tab.~\ref{Tab.FractionsUncorr} and Tab.~\ref{Tab.FractionsCorr} report median and 90\% uncertainties for posterior and priors for the fraction of sources with $\ct\leq-0.98$, $\ct\leq 0$ and $\ct \geq 0.98$. The true values are given in all tables for every quantity.

\begin{table}[ht]
\centering
\small
\begin{tabular}{c c l c c c c}
\toprule
$\cos\tau$ & Truth & Model & 150 events & 500 events & 1500 events & Prior \\
\midrule
\multirow{4}{*}{-1}&\multirow{4}{*}{$0.28$}
  & \lvk  & $0.28^{+0.10}_{-0.11}$ & $0.34^{+0.08}_{-0.07}$ & $0.30^{+0.05}_{-0.04}$ & $0.34^{+0.14}_{-0.25}$ \\
&  & \isogaus     & $0.25^{+0.13}_{-0.12}$ & $0.32^{+0.08}_{-0.09}$ & $0.28^{+0.06}_{-0.06}$ & $0.47^{+0.20}_{-0.28}$ \\
 & & \isobeta     & $0.30^{+0.11}_{-0.18}$ & $0.36^{+0.07}_{-0.14}$ & $0.31^{+0.07}_{-0.19}$ & $0.29^{+7.61}_{-0.26}$ \\
 & & \isotuk      & $0.27^{+0.14}_{-0.20}$ & $0.32^{+0.10}_{-0.15}$ & $0.26^{+0.10}_{-0.10}$ & $0.50^{+0.21}_{-0.40}$ \\
\midrule
\multirow{4}{*}{-0.5}&\multirow{4}{*}{$0.39$}
  & \lvk         & $0.36^{+0.07}_{-0.09}$ & $0.41^{+0.04}_{-0.05}$ & $0.40^{+0.02}_{-0.03}$ & $0.42^{+0.07}_{-0.27}$ \\
&  & \isogaus     & $0.36^{+0.08}_{-0.09}$ & $0.41^{+0.04}_{-0.05}$ & $0.39^{+0.03}_{-0.03}$ & $0.50^{+0.19}_{-0.18}$ \\
&  & \isobeta     & $0.34^{+0.09}_{-0.09}$ & $0.40^{+0.05}_{-0.06}$ & $0.38^{+0.04}_{-0.04}$ & $0.47^{+0.46}_{-0.34}$ \\
&  & \isotuk      & $0.35^{+0.18}_{-0.13}$ & $0.41^{+0.11}_{-0.09}$ & $0.40^{+0.08}_{-0.08}$ & $0.50^{+0.30}_{-0.33}$ \\
\midrule
\multirow{4}{*}{0}&\multirow{4}{*}{$0.51$}
  & \lvk         & $0.50^{+0.02}_{-0.12}$ & $0.51^{+0.01}_{-0.06}$ & $0.51^{+0.01}_{-0.02}$ & $0.50^{+0.02}_{-0.17}$ \\
&  & \isogaus     & $0.52^{+0.08}_{-0.09}$ & $0.52^{+0.05}_{-0.05}$ & $0.53^{+0.05}_{-0.03}$ & $0.52^{+0.09}_{-0.05}$ \\
&  & \isobeta     & $0.44^{+0.19}_{-0.10}$ & $0.48^{+0.12}_{-0.06}$ & $0.51^{+0.06}_{-0.06}$ & $0.52^{+0.50}_{-0.31}$ \\
&  & \isotuk      & $0.55^{+0.11}_{-0.25}$ & $0.55^{+0.06}_{-0.16}$ & $0.57^{+0.04}_{-0.07}$ & $0.50^{+0.20}_{-0.30}$ \\
\midrule
\multirow{4}{*}{0.5}&\multirow{4}{*}{$0.62$}
  & \lvk         & $0.65^{+0.09}_{-0.07}$ & $0.59^{+0.05}_{-0.04}$ & $0.61^{+0.03}_{-0.02}$ & $0.59^{+0.26}_{-0.08}$ \\
&  & \isogaus     & $0.65^{+0.11}_{-0.09}$ & $0.60^{+0.06}_{-0.05}$ & $0.62^{+0.03}_{-0.03}$ & $0.50^{+0.20}_{-0.17}$ \\
&  & \isobeta     & $0.65^{+0.21}_{-0.14}$ & $0.62^{+0.13}_{-0.10}$ & $0.65^{+0.09}_{-0.07}$ & $0.47^{+0.46}_{-0.34}$ \\
&  & \isotuk      & $0.64^{+0.22}_{-0.19}$ & $0.59^{+0.10}_{-0.05}$ & $0.60^{+0.04}_{-0.03}$ & $0.50^{+0.29}_{-0.34}$ \\
\midrule
\multirow{4}{*}{1}&\multirow{4}{*}{$0.67$}
  & \lvk         & $0.72^{+0.24}_{-0.12}$ & $0.63^{+0.11}_{-0.06}$ & $0.65^{+0.04}_{-0.04}$ & $0.62^{+0.59}_{-0.11}$ \\
&  & \isogaus     & $0.63^{+0.18}_{-0.19}$ & $0.58^{+0.10}_{-0.11}$ & $0.61^{+0.06}_{-0.10}$ & $0.47^{+0.20}_{-0.27}$ \\
&  & $\isobeta^{(1)}$     & $0.67^{+1.63}_{-0.44}$ & $0.46^{+0.94}_{-0.16}$ & $0.43^{+0.38}_{-0.11}$ & $0.32^{+3.26}_{-0.29}$ \\
&  & \isotuk      & $0.62^{+0.37}_{-0.27}$ & $0.58^{+0.10}_{-0.15}$ & $0.59^{+0.05}_{-0.09}$ & $0.50^{+0.18}_{-0.40}$ \\
\bottomrule
\end{tabular}
\caption{PPD (median and 90\% credible interval) of $\cos\tau$ at fixed values of \ct (first column) for the uncorrelated models. The prior and true value are also reported. $^{(1)}$ For the \isobeta only, the slice is actually taken at $\ct=0.99$  numerical issues with the large values that singular beta posteriors can take in the last bin.}
\label{tab:cosTau_slices_uncorr}
\end{table}
\clearpage

\begin{table}
\begin{tabular}{c c l c c c c}
\toprule
$\cos\tau$ &Truth & Model & 150 events & 500 events & 1500 events & Prior \\
\midrule
\multirow{4}{*}{-1}&\multirow{4}{*}{$0.28$}
  & \lvkc & $0.33^{+0.10}_{-0.11}$ & $0.37^{+0.06}_{-0.08}$ & $0.30^{+0.05}_{-0.04}$ & $0.43^{+0.06}_{-0.15}$ \\
 & & \isocgaus & $0.32^{+0.14}_{-0.12}$ & $0.36^{+0.07}_{-0.09}$ & $0.29^{+0.06}_{-0.06}$ & $0.49^{+0.10}_{-0.13}$ \\
 &  & \isocbeta & $0.33^{+0.09}_{-0.12}$ & $0.38^{+0.06}_{-0.09}$ & $0.31^{+0.07}_{-0.16}$ & $0.41^{+3.73}_{-0.19}$ \\
 &  & \isoctuk & $0.33^{+0.11}_{-0.15}$ & $0.37^{+0.07}_{-0.12}$ & $0.27^{+0.09}_{-0.09}$ & $0.50^{+0.09}_{-0.19}$ \\
\midrule
\multirow{4}{*}{-0.5}&\multirow{4}{*}{$0.39$}
  & \lvkc & $0.37^{+0.09}_{-0.09}$ & $0.41^{+0.05}_{-0.05}$ & $0.39^{+0.02}_{-0.03}$ & $0.46^{+0.04}_{-0.14}$ \\
 &  & \isocgaus & $0.38^{+0.10}_{-0.11}$ & $0.41^{+0.05}_{-0.05}$ & $0.39^{+0.03}_{-0.03}$ & $0.50^{+0.10}_{-0.08}$ \\
 &  & \isocbeta & $0.35^{+0.08}_{-0.09}$ & $0.40^{+0.05}_{-0.05}$ & $0.38^{+0.04}_{-0.04}$ & $0.49^{+0.22}_{-0.17}$ \\
 &  & \isoctuk & $0.36^{+0.14}_{-0.11}$ & $0.40^{+0.08}_{-0.07}$ & $0.39^{+0.10}_{-0.09}$ & $0.50^{+0.14}_{-0.15}$ \\
\midrule
\multirow{4}{*}{0}&\multirow{4}{*}{$0.51$}
  & \lvkc & $0.48^{+0.04}_{-0.11}$ & $0.49^{+0.02}_{-0.07}$ & $0.51^{+0.01}_{-0.03}$ & $0.50^{+0.01}_{-0.09}$ \\
 &  & \isocgaus & $0.51^{+0.07}_{-0.10}$ & $0.51^{+0.05}_{-0.06}$ & $0.52^{+0.05}_{-0.03}$ & $0.51^{+0.04}_{-0.02}$ \\
 &  & \isocbeta & $0.43^{+0.15}_{-0.09}$ & $0.46^{+0.10}_{-0.05}$ & $0.51^{+0.07}_{-0.06}$ & $0.51^{+0.25}_{-0.14}$ \\
 &  & \isoctuk & $0.48^{+0.16}_{-0.19}$ & $0.52^{+0.08}_{-0.14}$ & $0.57^{+0.04}_{-0.07}$ & $0.50^{+0.10}_{-0.13}$ \\
\midrule
\multirow{4}{*}{0.5}&\multirow{4}{*}{$0.62$}
  & \lvkc & $0.63^{+0.09}_{-0.08}$ & $0.59^{+0.05}_{-0.05}$ & $0.61^{+0.03}_{-0.02}$ & $0.54^{+0.13}_{-0.04}$ \\
 &  & \isocgaus & $0.64^{+0.12}_{-0.11}$ & $0.60^{+0.06}_{-0.05}$ & $0.62^{+0.04}_{-0.03}$ & $0.50^{+0.09}_{-0.09}$ \\
 &  & \isocbeta & $0.61^{+0.21}_{-0.11}$ & $0.59^{+0.13}_{-0.09}$ & $0.65^{+0.09}_{-0.07}$ & $0.49^{+0.22}_{-0.17}$ \\
 &  & \isoctuk & $0.64^{+0.23}_{-0.25}$ & $0.60^{+0.13}_{-0.08}$ & $0.60^{+0.05}_{-0.03}$ & $0.50^{+0.13}_{-0.15}$ \\
\midrule
\multirow{4}{*}{1}&\multirow{4}{*}{$0.67$}
  & \lvkc & $0.73^{+0.27}_{-0.16}$ & $0.65^{+0.14}_{-0.08}$ & $0.66^{+0.05}_{-0.04}$ & $0.56^{+0.30}_{-0.05}$ \\
&   & \isocgaus & $0.62^{+0.23}_{-0.15}$ & $0.59^{+0.11}_{-0.11}$ & $0.61^{+0.06}_{-0.11}$ & $0.49^{+0.09}_{-0.15}$ \\
 &  & $\isocbeta^{(1)}$ & $0.75^{+1.52}_{-0.47}$ & $0.56^{+0.89}_{-0.21}$ & $0.43^{+0.41}_{-0.11}$ & $0.42^{+1.44}_{-0.19}$ \\
 &  & \isoctuk & $0.63^{+0.45}_{-0.27}$ & $0.58^{+0.16}_{-0.17}$ & $0.59^{+0.05}_{-0.10}$ & $0.50^{+0.08}_{-0.20}$ \\
\midrule
\bottomrule
\end{tabular}
\caption{PPD (median and 90\% credible interval) of $\cos\tau$ at fixed values of \ct (first column) for the correlated models. The prior and true value are also reported. $^{(1)}$ For the \isobeta only, the slice is actually taken at $\ct=0.99$  numerical issues with the large values that singular beta posteriors can take in the last bin.}\label{tab:cosTau_slices_corr}

\end{table}

\clearpage

\begin{table}
\begin{tabular}{c c l c c c c}
\toprule
$\cos\tau$ & Truth & Model & 150 events & 500 events & 1500 events & Prior \\
\midrule
\multirow{4}{*}{$\cos\tau \leq -0.98$} & \multirow{4}{*}{0.4} & \lvk & $0.4^{+0.2}_{-0.2}$ & $0.5^{+0.1}_{-0.1}$ & $0.4^{+0.1}_{-0.0}$ & $0.5^{+0.2}_{-0.4}$ \\
  & & \isogaus & $0.4^{+0.2}_{-0.2}$ & $0.5^{+0.2}_{-0.1}$ & $0.5^{+0.0}_{-0.1}$ & $0.7^{+0.3}_{-0.4}$ \\
  & & \isobeta & $0.5^{+0.2}_{-0.3}$ & $0.5^{+0.2}_{-0.2}$ & $0.4^{+0.1}_{-0.3}$ & $0.5^{+6.6}_{-0.4}$ \\
  & & \isotuk & $0.4^{+0.2}_{-0.3}$ & $0.5^{+0.2}_{-0.2}$ & $0.4^{+0.2}_{-0.1}$ & $0.7^{+0.4}_{-0.6}$ \\
\midrule
\multirow{4}{*}{$\cos\tau \leq 0.00$} & \multirow{4}{*}{38.9} & \lvk & $36.5^{+6.6}_{-7.9}$ & $41.3^{+3.7}_{-4.2}$ & $40.2^{+2.6}_{-1.2}$ & $42.0^{+7.2}_{-24.3}$ \\
  & & \isogaus & $37.2^{+7.4}_{-7.7}$ & $41.7^{+4.0}_{-4.5}$ & $40.4^{+3.1}_{-1.6}$ & $50.0^{+15.1}_{-15.8}$ \\
  & & \isobeta & $35.0^{+7.6}_{-7.9}$ & $40.7^{+4.3}_{-4.9}$ & $40.4^{+2.3}_{-1.4}$ & $49.9^{+33.1}_{-33.1}$ \\
  & & \isotuk & $37.1^{+8.2}_{-9.5}$ & $41.9^{+4.4}_{-5.1}$ & $41.0^{+1.9}_{-1.5}$ & $50.0^{+26.0}_{-25.3}$ \\
\midrule
\multirow{4}{*}{$\cos\tau \geq 0.98$} & \multirow{4}{*}{1.0} & \lvk & $1.1^{+0.4}_{-0.2}$ & $1.0^{+0.2}_{-0.1}$ & $1.0^{+0.2}_{-0.2}$ & $1.0^{+0.9}_{-0.2}$ \\
  & & \isogaus & $1.0^{+0.3}_{-0.3}$ & $0.9^{+0.2}_{-0.2}$ & $0.9^{+0.1}_{-0.2}$ & $0.7^{+0.3}_{-0.4}$ \\
  & & \isobeta & $0.9^{+3.4}_{-0.6}$ & $0.7^{+1.6}_{-0.3}$ & $0.6^{+0.2}_{-0.2}$ & $0.5^{+6.7}_{-0.4}$ \\
  & & \isotuk & $0.9^{+0.6}_{-0.4}$ & $0.9^{+0.2}_{-0.2}$ & $0.9^{+0.1}_{-0.1}$ & $0.7^{+0.3}_{-0.6}$ \\
\midrule
\bottomrule
\end{tabular}
\caption{Median and 90\% credible interval on the percentage of sources with \ct in the range given in the first column for the uncorrelated models. Truth and priors are also reported.}\label{Tab.FractionsUncorr}
\end{table}

\begin{table}
\begin{tabular}{c c l c c c c}
\toprule
$\cos\tau$ & Truth & Model & 150 events & 500 events & 1500 events & Prior \\
\midrule
\multirow{4}{*}{$\cos\tau \leq -0.98$} & \multirow{4}{*}{0.4} & \lvkc & $0.5^{+0.2}_{-0.2}$ & $0.6^{+0.1}_{-0.1}$ & $0.5^{+0.1}_{-0.1}$ & $0.6^{+0.2}_{-0.2}$ \\
  & & \isocgaus & $0.5^{+0.2}_{-0.2}$ & $0.5^{+0.1}_{-0.2}$ & $0.4^{+0.1}_{-0.1}$ & $0.7^{+0.2}_{-0.2}$ \\
  & & \isocbeta & $0.5^{+0.2}_{-0.2}$ & $0.6^{+0.1}_{-0.1}$ & $0.4^{+0.1}_{-0.1}$ & $0.6^{+3.1}_{-0.3}$ \\
  & & \isoctuk & $0.5^{+0.2}_{-0.2}$ & $0.6^{+0.1}_{-0.2}$ & $0.4^{+0.1}_{-0.2}$ & $0.7^{+0.2}_{-0.3}$ \\
\midrule
\multirow{4}{*}{$\cos\tau \leq 0.00$} & \multirow{4}{*}{38.9} & \lvkc & $37.5^{+7.9}_{-8.5}$ & $41.4^{+4.5}_{-4.6}$ & $40.4^{+2.2}_{-1.4}$ & $46.2^{+3.5}_{-12.7}$ \\
  & & \isocgaus & $38.9^{+9.4}_{-9.2}$ & $41.8^{+4.8}_{-4.6}$ & $39.9^{+1.8}_{-3.1}$ & $50.0^{+8.4}_{-7.4}$ \\
  & & \isocbeta & $36.4^{+7.6}_{-8.1}$ & $40.9^{+4.9}_{-4.7}$ & $40.0^{+2.0}_{-1.0}$ & $50.1^{+15.8}_{-15.4}$ \\
  & & \isoctuk & $37.1^{+10.9}_{-9.4}$ & $41.3^{+5.2}_{-5.4}$ & $40.8^{+2.1}_{-2.1}$ & $50.0^{+11.5}_{-10.9}$ \\
\midrule
\multirow{4}{*}{$\cos\tau \geq 0.98$} & \multirow{4}{*}{1.0} & \lvkc & $1.1^{+0.4}_{-0.3}$ & $1.0^{+0.2}_{-0.2}$ & $1.0^{+0.1}_{-0.1}$ & $0.9^{+0.5}_{-0.1}$ \\
  & & \isocgaus & $0.9^{+0.4}_{-0.3}$ & $0.9^{+0.2}_{-0.2}$ & $0.9^{+0.2}_{-0.1}$ & $0.7^{+0.2}_{-0.2}$ \\
  & & \isocbeta & $1.1^{+3.1}_{-0.7}$ & $0.8^{+1.5}_{-0.3}$ & $0.6^{+0.2}_{-0.1}$ & $0.6^{+3.2}_{-0.3}$ \\
  & & \isoctuk & $1.0^{+0.7}_{-0.4}$ & $0.9^{+0.2}_{-0.3}$ & $0.9^{+0.1}_{-0.2}$ & $0.7^{+0.2}_{-0.3}$ \\
\midrule
\bottomrule
\end{tabular}
\caption{Median and 90\% credible interval on the percentage of sources with \ct in the range given in the first column for the correlated models. Truth and priors are also reported.}\label{Tab.FractionsCorr}
\end{table}

\end{document}